\documentclass[11pt]{article}
\usepackage[a4paper,margin=27mm]{geometry}
\usepackage{amsmath,amssymb,amsthm,mathtools,microtype,mathrsfs}
\usepackage[colorlinks=true,linkcolor=blue,citecolor=blue,urlcolor=blue]{hyperref}
\usepackage{tikz}
\usetikzlibrary{arrows.meta}
\hypersetup{pdftitle={The Maxwell Bethe–Sommerfeld problem I},pdfauthor={Habib Ammari, Yu Gao, Jiayu Qiu}}
\usepackage{enumitem}
\newtheorem{theorem}{Theorem}[section]
\newtheorem{lemma}[theorem]{Lemma}
\newtheorem{proposition}[theorem]{Proposition}
\newtheorem{remark}[theorem]{Remark}

\newtheorem{definition}{Definition}[section]
\newtheorem*{conjecture}{Conjecture}
\numberwithin{equation}{section}

\newcommand{\R}{\mathbb R}\newcommand{\Z}{\mathbb Z}
\newcommand{\T}{\mathbb T}\newcommand{\C}{\mathbb C}
\newcommand{\ii}{\mathrm i}
\newcommand{\one}{\mathbf 1}
\newcommand{\spec}{\sigma}
\newcommand{\dd}{\,\mathrm dx}
\DeclareMathOperator{\dist}{dist}
\DeclareMathOperator{\Dom}{Dom}
\DeclareMathOperator{\diver}{div}
\newcommand{\norm}[1]{\left\|#1\right\|}
\title{The Maxwell Bethe–Sommerfeld problem I. Thin-layer induction and results for planar TE modes}

\author{Habib Ammari$^{1}$ \qquad Yu Gao$^{2}$ \qquad Jiayu Qiu$^{1}$\\[5pt]
\small $^{1}$Department of Mathematics, ETH Z\"urich, Switzerland\\
\small $^{2}$Department of Mathematics, The Hong Kong University of Science and Technology,\\[-2pt]
\small Clear Water Bay, Kowloon, Hong Kong}

\date{}

\begin{document}
\maketitle
\begin{abstract}
We present a framework for constructing a single medium that supports infinitely many complete spectral gaps for periodic divergence-form operators, which arise from the transverse electric (TE) polarization of Maxwell's equations. The construction proceeds by inductively inserting a seed medium with a prescribed spectral gap into a thin layer of the unit cell. As a result, we obtain densely many counterexamples to the Bethe-Sommerfeld Conjecture (BSC) for divergence-form operators. Moreover, we prove that the thin-layer construction controls the gap topology: by carefully selecting the seed medium, one can realize a single operator with any prescribed configuration of gap Chern numbers. This is the first of two papers that contains all the fundamental ingredients for the thin-layer induction. In the second paper, we will apply this framework to disprove the BSC for the full three-dimensional Maxwell operator.
\end{abstract}
\begin{quotation}\small
\noindent\textit{Keywords.} Bethe-Sommerfeld conjecture, periodic divergence-form operators, complete spectral gaps, thin-layer construction, gap Chern numbers, Maxwell's equations.

\smallskip
\noindent\textit{2020 Mathematics Subject Classification.} Primary 35P05. Secondary 35B27, 35Q61, 81Q70.
\end{quotation}

\begingroup
\small
\makeatletter
\let\originaladdvspace\addvspace
\def\addvspace#1{\originaladdvspace{.03em}}
\tableofcontents
\makeatother
\endgroup

\section{Introduction and main result}

We consider periodic elliptic divergence-form operators, which arise as a two-dimensional reduction of the three-dimensional Maxwell operator under the TE polarization; see Appendix \ref{app:maxwell-reduction}. More precisely, for a measurable $\Z^2$-periodic Hermitian matrix field $A=A(x)$ satisfying $0<\alpha I_2\leq A\leq\beta I_2<\infty$, let $H_A$ be the nonnegative self-adjoint operator associated with the closed form
\begin{equation} \label{eq:sesquilinear_form}
q_A[u]:=\int_{\R^2}(A\nabla u)\cdot\overline{\nabla u}\dd,\quad
 \Dom (q_A)=H^1(\R^2).
\end{equation}
In other words, $H_A=-\diver(A\nabla)$ in the weak sense. 

The main result of this paper is as follows. We construct (infinitely many) bounded measurable coefficient fields $A=A(x)$ for which $H_{A}$ has infinitely many complete spectral gaps that tend to infinity. As a result, they provide counterexamples to the Bethe-Sommerfeld Conjecture (BSC) for this class of divergence-form operators; see below for the statement of the conjecture and an account of its development.

\begin{theorem}\label{thm:main}
There exist $\alpha,\beta>0$ such that the collection
\[
 \mathcal X=\{A=A^{\mathsf T}\text{ real and periodic}:
                       \alpha I_2\leq A\leq\beta I_2\text{ a.e.}\}
\]
satisfies the following properties. First, there exists $A\in\mathcal X$ for which $\spec(H_A)$ has infinitely many bounded open gap components tending to infinity. In other words, there exist open intervals $G_j$ such that
\[
G_j\cap \spec(H_A)=\varnothing, \quad \text{with}\quad \sup G_j\leq\inf G_{j+1},\quad \inf G_j\to \infty .
\]
Moreover, the collection of such $A$ forms a dense $G_\delta$ subset of $\mathcal X$ in the $L^1$ metric.
\end{theorem}

The proof is based on a thin-layer induction, as sketched in Section \ref{sec:proof-idea}. Having proved Theorem \ref{thm:main}, we ask whether the construction provides information beyond the existence of infinitely many gaps. To this end, we consider \textit{the topological properties} of the constructed spectral gaps. In fact, the spectral projections onto the occupied bands below a spectral gap $G$ define a bundle parametrized by the quasi-momentum, whose topology is characterized by \textit{the (first) Chern number $\operatorname{Ch}(A;G)$}, defined in \eqref{eq:chern-number}. This invariant is a central object in modern condensed matter theory and topological photonics; we refer the reader to \cite{Ozawa_2019} and the references therein for an overview of the Chern number and its applications. We prove that the thin-layer construction also controls the Chern numbers of the constructed gaps. In particular, by carefully selecting the seed media, one can realize a single operator with \textit{any prescribed configuration of gap Chern numbers}.

\begin{theorem}
\label{thm:chern-main}
For every $\eta>0$, the following statements hold.
\begin{enumerate}[label=\textup{(\roman*)}]
\item There exist $\alpha,\beta>0$ such that, for every integer $m$, there is a smooth $\Z^2$-periodic Hermitian matrix field $A_m$ satisfying
\begin{equation}\label{eq:common-chern-bounds}
 \alpha I_2\leq A_m(x)\leq \beta I_2,
 \quad \norm{\operatorname{Im}A_m}_{L^\infty}\leq\eta,
\end{equation}
such that $H_{A_m}$ has a complete positive gap $G$ with $\operatorname{Ch}(A_m;G)=m$.
\item For every sequence $(m_j)_{j\geq1}$ of integers, there is a bounded measurable $\Z^2$-periodic Hermitian matrix field $A_\infty$ satisfying the bounds in \eqref{eq:common-chern-bounds} and having distinct bounded complete gap components $G_j$ with
\[
 \sup G_j\leq\inf G_{j+1},\quad
 \inf G_j\to \infty,\quad
 \operatorname{Ch}(A_\infty;G_j)=m_j.
\]
\end{enumerate}
\end{theorem}
The real symmetric coefficients in Theorem~\ref{thm:main} make $H_{A}$ time-reversal invariant; consequently, all its gaps have zero Chern number (see the discussion following \eqref{eq:chern-number}). Nonzero Chern numbers require Hermitian coefficients that are not real-valued, as allowed in Theorem \ref{thm:chern-main}. The bound $\eta$ on the imaginary part has a physical interpretation: this part is usually introduced through magneto-optic effects induced by external magnetic fields \cite{lee2019elliptic,qiu2024mathematical}. Hence, Theorem \ref{thm:chern-main} indicates that one can realize any prescribed Chern-number configuration in a single medium \textit{with arbitrarily small external magnetic fields}, at the cost of increasingly complex spatial configurations during the induction.

\subsection{Background}

The Bethe-Sommerfeld conjecture originates in the electronic theory of metals developed by Sommerfeld and Bethe in the 1930s \cite{SommerfeldBethe1933}. In its classical mathematical formulation, it states that a periodic Schr\"odinger operator $-\Delta+V$ on $\R^d$ ($d\geq2$) has only finitely many spectral gaps, or equivalently, that its spectrum contains a half-line $[E_*,\infty)$. This differs from the one-dimensional situation, where smooth periodic potentials generically have infinitely many open gaps \cite{Simon1976}. See \cite[Section~6.1.3]{kuchment2016overview} for an overview of this dimensional contrast. The physical intuition underlying the conjecture is that a periodic electric potential cannot block the propagation of electrons at sufficiently high energies. For classical wave systems, such as electromagnetic waves governed by Maxwell's equations, the analogous question is whether a periodic medium can exclude wave propagation at arbitrarily high frequencies. This is the main focus of this paper.

After several decades of work, the original conjecture for the Schrödinger operator has been proved in its full generality; we refer the reader to \cite{PopovSkriganov1981,DahlbergTrubowitz1982,Skriganov1985Bands,SkriganovSobolev2006,HelfferMohamed1998,Parnovski2008,Karpeshina1997,Veliev2015} and, in particular, the review \cite[Section~6.1.3]{kuchment2016overview} for a comprehensive account. The theory also extends beyond the bounded electric perturbations of the Laplacian. For example, finite-gap results have been obtained for magnetic Schrödinger and periodic Landau operators \cite{Mohamed1997,Karpeshina2004,Elton2004}, higher-order operators with bounded periodic potentials \cite{Shen2003}, and pseudodifferential perturbations \cite{BarbatisParnovski2009}; see also \cite{Sobolev2007}. A general theorem in \cite{ParnovskiSobolev2010} proves, in particular, the conjecture for periodic magnetic Schr\"odinger operators in all dimensions $d\geq2$. Related developments concern high-energy spectral asymptotics for Schrödinger operators \cite{ParnovskiShterenberg2009,ParnovskiShterenberg2012}, Dirac operators \cite{LagaceEtAl2023}, and quasi-periodic settings \cite{KarpeshinaShterenberg2019,KarpeshinaParnovskiShterenberg2026}.


Despite these advances for Schrödinger operators and their generalizations, the BSC for Maxwell operators remains largely unexplored, which is formulated and explicitly proposed in \cite[Conjecture~7.8]{Kuchment2001}. From a practical aspect, the interest in the Maxwell BSC is rooted in the rapid development of metamaterials and, in particular, photonic cyrstals, for which a comprehensive understanding of the availability of high-frequency gaps is fundamental for applications \cite{Ozawa_2019}. However, the mathematical difficulty to solve the Maxwell BSC is that the principal symbol of the Maxwell operator has spatially varying coefficients, unlike that of the Schrödinger operator; see \cite[Section~6.1.3, Remark~6.6]{kuchment2016overview}. This feature explicitly distinguishes the general elliptic and Maxwell questions from established results on lower-order perturbations of Schrödinger operators. A positive result for the operator $-\epsilon^{-1}\Delta$ is proved in \cite{Vorobets}, where the dielectric function $\epsilon=\epsilon(x)$ has a special separated form $\epsilon(x)=\epsilon_1(x_1)+\epsilon_2(x_2)$. We note that there has also been considerable work on the construction and analysis of spectral gaps in classical wave systems and on related homogenization methods; see \cite{FigotinKuchment1996a,FigotinKuchment1996b,HempelLienau2000,HempelPost2003,Zhikov2000,Khrabustovskyi2013} and, in particular, \cite{FigotinKuchment1998,BorisovExner2021}, where ideas related to the thin-layer construction used in the current paper have appeared. These works provide ways to create gaps and approximate prescribed spectral configurations, but they do not construct a single operator with infinitely many gaps and hence do not disprove the BSC for classical wave systems. For example, \cite{FigotinKuchment1998} studies thin layers with increasing material contrast, while \cite{BorisovExner2021} uses increasingly high, narrow potential walls together with a small differential perturbation to open spectral gaps. Remarkably, in our thin-layer construction, \textit{we replace the growth of material coefficients with rapid oscillations of the medium within the layers}. Together with the scaling law of the Maxwell operator and its divergence-form reduction, this replacement allows us to obtain a single limiting medium with infinitely many spectral gaps. See Section \ref{sec:proof-idea} for an overview of the proof.

The physical interest in the gap Chern numbers considered in Theorem~\ref{thm:chern-main} can be traced back to the work of Haldane and Raghu \cite{HaldaneRaghu2008}, who proposed a photonic analogue of the quantum Hall effect. Since then, \textit{topological photonics} has attracted considerable interest in both the physics and mathematics communities; see \cite{Ozawa_2019,lee2019elliptic,qiu23waveguide} and the references therein. Photonic crystals provide a platform for the realization of topological phenomena because their material structures can be designed and fabricated, offering greater flexibility than many condensed matter systems. These developments, together with the constructions of large Chern numbers in photonic and lattice models \cite{SkirloEtAl2015,SticletPiechon,Devescovi}, motivate the question of how much topological information can be prescribed within a single periodic medium. Theorem~\ref{thm:chern-main} realizes any prescribed sequence of integers as the Chern numbers of selected gap components tending to infinity in a single medium with uniformly bounded coefficients. Through the bulk-edge correspondence (BEC), this result suggests a way to control directed interface transport in arbitrarily high frequency ranges within a fixed material configuration. We refer the reader to \cite{Drouot2021BulkEdge,qiu2025bulk} and the references therein for rigorous results on BEC and further discussion of its photonic interpretation and condensed matter origins.

\subsection{Main idea of the proof and outline of paper}\label{sec:proof-idea}

The proof of Theorem~\ref{thm:main} is presented in Section \ref{sec:infinite-gap-construction}. Here, we sketch its main idea. To prove the theorem, we first find a medium with a prescribed spectral gap $I$ by a standard variational argument; see Lemma \ref{lem:seed}. We denote its coefficient matrix by $S=S(x)$ and call it \textit{the seed medium}. We then begin the thin-layer induction: at each step $j$, we insert a rapidly rescaled copy of the seed, denoted by $S_{N_j}(x):=S(N_{j}x)$ ($N_j\in\Z$ is large), into a thin layer surrounding each period cell. In particular, we let the layer occupy an asymptotically vanishing volume fraction, denoted as $w_j\to 0$, while the oscillations become increasingly rapid, i.e., $N_j\to\infty$; see Figure \ref{fig:layer_construction}. The construction relies on three observations:
\begin{itemize}
    \item[(i)] By the scaling property of the divergence-form operator (see Remark \ref{rmk:scaling_covariance}), the rapidly rescaled seed has a spectral gap $N_j^2 I$ at increasingly high frequencies, while $S_{N_j}$ satisfies the same coefficient bounds as the original seed $S$.
    \item[(ii)] By choosing the layer thickness $w_j$ proportional to $(\log N_j)/N_j$, we ensure that the composite medium inherits at least part of the gap $N_j^2 I$ of $S_{N_j}$ after the $j$\textsuperscript{th} insertion; see Proposition \ref{prop:insert}. The physical idea is that, at this thickness, the rapid repetition of the gapped seed within the surrounding layer suppresses coupling between adjacent period cells. The argument is given in detail in Section \ref{sec:layer-insertion}.
    \item[(iii)] Since the layer thickness $w_j$ tends to zero while the material coefficients remain uniformly bounded as the induction proceeds, a limiting medium $A_{\infty}$ exists. The limiting medium is sufficiently close in $L^1$ norm to each composite medium, which guaranties that it inherits the increasingly large gaps constructed during induction. As a consequence, the spectrum of the limiting operator has infinitely many spectral gaps, as claimed in Theorem \ref{thm:main}.
\end{itemize}

Readers interested only in Theorem \ref{thm:main} may stop at Section \ref{sec:infinite-gap-construction}, where its proof is completed.

To prove Theorem~\ref{thm:chern-main} in Section \ref{sec:chern-layers}, we use the same thin-layer induction. At the $j$\textsuperscript{th} step, we insert a seed medium $S_j$ with a gap of prescribed Chern number $m_j$. This requires two additional inputs:
\begin{itemize}
    \item[(i)] After the $j$\textsuperscript{th} insertion, the composite medium inherits both part of the rescaled gap $N_{j}^2 I_j$ and its gap Chern number $m_j$; see Proposition \ref{prop:chern-insert}. The key point is that any eigenbundle of the composite medium in $N_{j}^2 I_j$ is topologically trivial, where the underlying physical idea is exactly the same as in point (ii) above: the layer of appropriate thickness suppresses the intercell connection, and hence prevents a nontrivial Chern number. A familiar illustration of this mechanism is found by the eigenbundles admitting an exponentially localized Wannier basis, corresponding to exponentially small intercell coupling; in particular, it is well known that such bundles have zero Chern numbers (see, e.g., \cite{monaco2018optimal,bols2026localization}).
    \item[(ii)] For every integer $m_j$, there exists a medium with a spectral gap of Chern number $m_j$; see Proposition \ref{prop:uniform-seeds}. This supplies the seed required at each step. The proof of Proposition \ref{prop:uniform-seeds} is the most technical part of the paper and accounts for a large part of its length. The availability of a gap with an arbitrary Chern number is indeed standard for the Landau operator, which can be realized by simply choosing a sufficiently high gap. Nevertheless, the corresponding question for divergence-form operators has received much less attention. To prove Proposition \ref{prop:uniform-seeds}, we use the Landau operator as a starting point. More detailedly, we construct a periodic target operator with any prescribed Chern number by removing a small disk from the domain of the Landau operator, i.e., the magnetic torus, and then trivializing the magnetic line bundle on the complement; see Lemma \ref{lem:seed-magnetic-target}. We then use high-frequency homogenization to realize this target operator near a band edge of a host operator. In the high-frequency limit, the effective operator is precisely the target in Lemma \ref{lem:seed-magnetic-target}, so the resulting operator inherits its prescribed gap and Chern number (see Lemma \ref{lem:seed-transfer}). Details are given in Sections \ref{sec:chern_roadmap} and \ref{sec:chern_auxiliary_proof}.
\end{itemize}

\subsection{Perspectives} \label{sec:perspective}

The thin-layer construction in Section \ref{sec:proof-idea} applies beyond the planar divergence-form operators. In the second paper, we will apply it to the full three-dimensional Maxwell operator with magnetic permeability fixed at one, namely,
\begin{equation} \label{eq:maxwell_def}
\mathcal{M}_\epsilon H:=\operatorname{curl}
       (\epsilon^{-1}\operatorname{curl}H),
 \quad \operatorname{div}H=0,
\end{equation}
where $\epsilon$ is a real scalar $\Z^3$-periodic dielectric function that satisfies $0<\epsilon_-\leq\epsilon(x)\leq\epsilon_+<\infty$ almost everywhere. To carry out the construction for the full Maxwell operator, we need two additional inputs:
\begin{itemize}
    \item[(i)] A seed medium with a complete gap for the full Maxwell operator;
    \item[(ii)] An adaptation of the layer-localization argument in Section \ref{sec:layer-insertion} to incorporate the constraint $\operatorname{div}H=0$.
\end{itemize}
Once these inputs are established, the insertion and limiting arguments follow the same scheme as in the present paper. In fact, the proofs of these inputs have been obtained by the authors and are under polishing. We therefore postpone the full three-dimensional counterexample to the second paper.

Another question is whether the BSC holds for Maxwell or divergence-form operators \textit{under stronger regularity assumptions on the coefficients}. Since the thin-layer construction inserts rapidly oscillating coefficients, the limiting medium in Theorem \ref{thm:main} can be discontinuous. Even though the regularity of coefficients appears less restrictive for Schrödinger operators, as the BSC has been proved for $L^{\infty}$ potentials in dimension two \cite{ParnovskiSobolev2001} (although this is not known in general dimensions), it is natural to ask whether the BSC for Maxwell or divergence-form operators is more closely tied to coefficient regularity. This motivates the following conjecture.

\begin{conjecture}
\begin{enumerate}[label=\textup{(\roman*)}]
\item Let $A$ be a continuous, real symmetric, positive-definite $\Z^2$-periodic matrix field. Then $\spec(H_A)$ contains a half-line.
\item Let $\epsilon$ be a continuous, real, positive $\Z^3$-periodic scalar dielectric function. Then the spectrum of the Maxwell operator \eqref{eq:maxwell_def} contains a half-line.
\end{enumerate}
\end{conjecture}

\subsection{Notation and preliminaries}
\label{sec:notations-preliminaries}

In this paper, function spaces may have real or complex scalar, vector, or matrix values. We use $|\cdot|$ for the absolute value of scalars, the Euclidean norm of vectors, and the induced operator norm of matrices, $|B|:=\sup_{|z|=1}|Bz|$. When the value space is clear, we write $L^p(U)$ for the corresponding Lebesgue space. All integrals use Lebesgue measure, and
\[
 \begin{aligned}
 \|f\|_{L^p(U)}
 :=\left(\int_U|f(x)|^p\,dx\right)^{1/p},
 1\le p<\infty,\qquad
 \|f\|_{L^\infty(U)}
 :=\operatorname*{ess\,sup}_{x\in U}|f(x)|.
 \end{aligned}
\]
For scalar and vector fields in $L^2(U)$, we take
\[
 \langle f,g\rangle
 :=\int_U f(x)\cdot\overline{g(x)}\,dx.
\]
For Banach spaces $X,Y$, we denote by $\mathcal L(X,Y)$ the bounded
linear operators, with norm
\[
 \|T\|:=\sup_{\|u\|_X=1}\|Tu\|_Y.
\]
The symbol ${}^*$ denotes the Hilbert-space adjoint, or conjugate
transpose for matrices. Note that the subscripts of norms that indicate the corresponding spaces will be omitted if no confusion arises. 

Let the spatial torus $\T^2$ and its centered unit cell $\Omega$
be defined by
\[
 \T^2:=\R^2/\Z^2,\qquad
 \Omega:=(-\tfrac12,\tfrac12)^2,
\]
with opposite faces of $\overline\Omega$ identified.
The inner square $Q$ and its complement $Q^c$ in $\T^2$ are
\[
 Q:=\tfrac12\Omega=(-\tfrac14,\tfrac14)^2,\qquad
 Q^c:=\T^2\setminus Q.
\]
For a Lipschitz domain $U\subset\R^2$, we denote by $H^1(U)$ the
Sobolev space of $L^2$ functions with $L^2$ weak first derivatives,
and by $H_0^1(U)$ its subspace with zero trace on $\partial U$. When taking into account the periodic/quasi-periodic boundary conditions, we write
\[
 H_{\mathrm{per}}^1(\Omega):=H^1(\T^2),\qquad
 H_{\mathrm{qp}}^1(\Omega;k)
 :=e^{\ii k\cdot x}H_{\mathrm{per}}^1(\Omega),
 \quad k\in\R^2.
\]
In particular, since the quasi-periodic condition holds on opposite faces in the trace sense, we have
\begin{equation}
\label{eq:space-inclusions}
    H_0^1(\Omega)\subset H_{\mathrm{qp}}^1(\Omega;k)\subset H^1(\Omega).
\end{equation}

For $0<\alpha\le\beta<\infty$, define the Hermitian and real symmetric
coefficient classes by
\[
 \begin{aligned}
 \mathcal X^{\mathrm H}_{\alpha,\beta}
 &:=\left\{
 A\in L^\infty(\T^2;\C^{2\times2}):
 A=A^*,\ \alpha I_2\le A\le\beta I_2\ \text{a.e.}
 \right\},\\
 \mathcal X^{\mathrm R}_{\alpha,\beta}
 &:=\left\{
 A\in\mathcal X^{\mathrm H}_{\alpha,\beta}:
 A=\overline A\ \text{a.e.}
 \right\}.
 \end{aligned}
\]
Here, the matrix inequalities are understood in the sense of sesquilinear forms. Both the two classes are equipped with the $L^1$ metric. For a contrast ratio $\rho\ge1$, the binary scalar medium defined as follows, which lies in the class $mathcal X^{\mathrm R}_{\alpha,\beta}$, will be frequently applied in the proof:
\[
 a_\rho:=\one_Q+\rho\,\one_{Q^c},\qquad
 a_\rho I_2\in\mathcal X^{\mathrm R}_{1,\rho},
\]
where $\one_U$ denotes the indicator of $U$.

For $A\in\mathcal X^{\mathrm H}_{\alpha,\beta}$, let
$H_A=-\nabla\cdot(A\nabla)$ be the periodic operator defined through \eqref{eq:sesquilinear_form}. Its Bloch fibers $H_A(k)$,
$k\in\mathcal B:=[-\pi,\pi]^2$, are associated with the closed forms
\[
 q_{A,k}[u]:=\int_\Omega
 (A\nabla_k u)\cdot\overline{\nabla_k u}\,dx,\qquad
 \nabla_k:=\nabla+\ii k,\qquad
 \Dom (q_{A,k})=H_{\mathrm{per}}^1(\Omega).
\]
Thus, $H_A(k)=-\nabla_k\cdot A\nabla_k$ in the weak sense.
Each fiber has a compact resolvent, and we write its energy eigenvalues
and corresponding frequencies as
\[
 0\le\lambda_1(A;k)\le\lambda_2(A;k)\le\cdots,\qquad
 \omega_n(A;k):=\sqrt{\lambda_n(A;k)},
\]
with eigenvalues counted with multiplicity.
Floquet decomposition gives
\[
 \sigma(H_A)=\bigcup_{k\in\mathcal B}\sigma(H_A(k)).
\]
A nonempty open interval $G\subset(0,\infty)$ is a complete spectral
gap if
\[
 G\cap\sigma(H_A)=\varnothing,
 \quad\text{equivalently}\quad
 G\cap\sigma(H_A(k))=\varnothing
 \quad\text{for every }k\in\mathcal B.
\]
A maximal such interval is called a gap component. For fixed $A$ and $E\in G$, define the occupied projection
\[
 P(k):=\one_{(-\infty,E)}\bigl(H_A(k)\bigr).
\]
Then the finite-rank projection $P(k)$ is smooth in $k$ and satisfies
\[
 P(k+2\pi e_j)=\tau_jP(k)\tau_j^*,\qquad
 (\tau_j u)(x):=e^{-2\pi\ii x_j}u(x).
\]
Hence, it defines a Bloch bundle over $\R^2/(2\pi\Z)^2$.
With $\partial_j:=\partial/\partial k_j$ and orientation
$\mathrm dk_1\wedge\mathrm dk_2$, the (first) Chern number of this Bloch bundle is defined using the Kubo formula (ses e.g. \cite{Avron1994charge}):
\begin{equation}\label{eq:chern-number}
 \operatorname{Ch}(A;E)
 :=\frac{1}{2\pi\ii}\int_{\mathcal B}
 \operatorname{tr}\!\left(P[\partial_1P,\partial_2P]\right)
 \,\mathrm dk_1\,\mathrm dk_2
 \in\Z.
\end{equation}
This integer is independent of $E\in G$, and hence we sometimes write it as $\operatorname{Ch}(A;G)$. In particular, $\operatorname{Ch}(A;E)=0$ for real $A$: complex conjugation sends $P(k)$ to $P(-k)$ and thus changes the sign of the curvature.

Finally, the periodic cell boundary and its $w$-neighborhood are defined by
\[
 \begin{aligned}
 \Sigma
 :=\bigcup_{j=1}^2
 \{x\in\T^2:x_j\equiv\tfrac12\pmod1\},\quad
 \Sigma_w
 :=\{x\in\T^2:\dist_{\T^2}(x,\Sigma)<w\},
 \end{aligned}
\]
where $0<w<1/8$. These notions are important for the thin-layer construction used in this paper.

\section{A binary planar seed from a variational argument} \label{sec:gap_seed_variational}

The thin-layer construction starts with a periodic medium for which one complete spectral gap is known, which is realized using the binary medium $a_\rho$. Specifically, we set $a_\rho$ equal to $1$ on the inner square $Q=(-\tfrac14,\tfrac14)^2$ and to $\rho$ in $Q^{c}=\T^2\backslash Q$. Then, the Dirichlet-Neumann bracketing reduces the gap estimate to separating the first Dirichlet eigenvalue from the second Neumann eigenvalue on the period cell. We obtain this separation by a compactness argument at large contrast (cf.~\cite{HempelLienau2000,CassierWeinstein2021}) and then show that the same energy interval remains a gap of $\spec(H_{a_\rho})$.

\begin{lemma}\label{lem:seed} 
There exists $\rho_0>1$ such that, if $\rho\ge\rho_0$, it holds that
\begin{equation}\label{eq:parameterized-square-gap}
 \left(8\pi^2,9\pi^2\right)
 \cap\sigma(H_{a_\rho})=\varnothing.
\end{equation}
\end{lemma}

\begin{proof}
{\color{blue}Step 1.} Let $\lambda_j^{\mathrm N}(\rho)$ and $\lambda_j^{\mathrm D}(\rho)$ denote the Neumann and Dirichlet eigenvalues of $-\nabla\cdot(a_\rho\nabla)$ on $\Omega$, respectively, and let $\mu_j$ denote the Dirichlet eigenvalues of $-\Delta$ on $Q$. The eigenvalues are listed in nondecreasing order and counted with multiplicity.  In the quasi-periodic representation for the Bloch fibers, we obtain the Dirichlet, Bloch, and Neumann eigenvalues by applying the min-max
principle to the quadratic form
\[
 u\longmapsto \int_\Omega a_\rho|\nabla u|^2\,dx
\]
restricted to the corresponding domains $H_0^1(\Omega)$, $H_{\mathrm{qp}}^1(\Omega;k)$ and $H^1(\Omega)$, respectively. The form-domain inclusions
in \eqref{eq:space-inclusions} then give
\[
 \lambda_j^{\mathrm N}(\rho)
 \le \lambda_j(a_\rho;k)
 \le \lambda_j^{\mathrm D}(\rho),
 \qquad j\ge 1,\quad k\in\mathcal B,
\]
where $\lambda_j(a_\rho;k)$ denote the Bloch eigenvalues. The Floquet decomposition and these bounds reduce the proof to showing that, for all sufficiently large $\rho$,
\begin{equation} \label{eq:binary_seed}
\lambda_1^{\mathrm D}(\rho)\le \mu_1,
 \qquad
 \lambda_2^{\mathrm N}(\rho)>\mu_1+\pi^2.
\end{equation}

{\color{blue}Step 2.1.} To prove the first estimate in \eqref{eq:binary_seed}, we let $\phi$ be the positive, $L^2(Q)$-normalized first Dirichlet eigenfunction of $-\Delta$ on $Q$, given by
\[
 \phi(x)=4\cos(2\pi x_1)\cos(2\pi x_2).
\]
Its zero extension $\widetilde\phi\in H_0^1(\Omega)$ has unit $L^2(\Omega)$ norm. Using $\widetilde\phi$ as a trial function and
noting that $a_\rho=1$ on $Q$, we obtain
\begin{equation}\label{eq:parameterized-square-dirichlet}
 \lambda_1^{\mathrm D}(\rho)
 \le \int_\Omega a_\rho|\nabla\widetilde\phi|^2\,dx
 = \int_Q|\nabla\phi|^2\,dx
 = \mu_1,
 \qquad \rho\ge 1.
\end{equation}

{\color{blue}Step 2.2.} To obtain the second estimate of \eqref{eq:binary_seed}, we first prove an upper bound of $\lambda_2^{\mathrm N}(\rho)$, which will provide the compactness needed to estimate the lower bound. First, the Neumann kernel consists of constants, so $\lambda_2^{\mathrm N}(\rho)$ is the minimum of the Rayleigh quotient over nonzero mean-zero
functions in $H^1(\Omega)$. Set $m:=\int_Q\phi\,dx=4/\pi^2$. As $|\Omega|=1$, the function $\widetilde\phi-m$ has zero mean and
squared $L^2(\Omega)$ norm $1-m^2$. Its gradient is equal to that of $\widetilde\phi$, so using it as a trial function gives
\[
 \lambda_2^{\mathrm N}(\rho)
 \le \frac{\int_\Omega a_\rho|\nabla\widetilde\phi|^2\,dx}{1-m^2}
 =\frac{\int_Q |\nabla \phi|^2\,dx}{1-m^2}
 = \frac{\mu_1}{1-m^2}.
\]
By the min--max principle, $\lambda_2^{\mathrm N}(\rho)$ is
nondecreasing in $\rho$, since $a_\rho$ is pointwise nondecreasing.
The above upper bound therefore gives a finite limit
\[
L:=\lim_{\rho\to\infty}\lambda_2^{\mathrm N}(\rho)<\infty.
\]
It then suffices to show that $L>9\pi^2$.

{\color{blue}Step 2.3.} Let $\rho_j\to\infty$ and choose $L^2(\Omega)$-normalized, mean-zero
eigenfunctions $u_j$ associated with $\lambda_2^{\mathrm N}(\rho_j)$.
Then
\[
\int_\Omega |\nabla u_j|^2\,dx
\le \int_\Omega a_{\rho_j}|\nabla u_j|^2\,dx
 = \lambda_2^{\mathrm N}(\rho_j)
 \le \frac{\mu_1}{1-m^2}.
\]
This bound and the normalization imply that $\{u_j\}$ is bounded in $H^1(\Omega)$. Note that $a_{\rho_j}=\rho_j$ in $\Omega\setminus\overline Q$, the same bound gives
\begin{equation}\label{eq:exterior-energy-estimate}
 \int_{\Omega\setminus\overline Q}|\nabla u_j|^2\,dx
 \le \frac{\mu_1}{\rho_j(1-m^2)}.
\end{equation}
By weak compactness in $H^1(\Omega)$ and the Rellich compactness
theorem, there exists $u\in H^1(\Omega)$ such that, along a subsequence,
\[
 u_j\rightharpoonup u\quad\text{in }H^1(\Omega),
 \qquad
 u_j\to u\quad\text{in }L^2(\Omega).
\]
The exterior energy estimate \eqref{eq:exterior-energy-estimate} and the weak $H^1$ convergence imply that $\nabla u=0$ on the connected
exterior $\Omega\setminus\overline Q$. Thus $u=c$ in $\Omega\setminus\overline Q$ for some constant $c$. The traces of $u$ agree across $\partial Q$, so $v:=u|_Q-c$ has zero trace and belongs to $H_0^1(Q)$. Hence $u=c+\widetilde v$, where $\widetilde v$ is the zero extension of $v$. Strong $L^2$ convergence gives $\int_\Omega u\,dx=0$ and $\|u\|_{L^2(\Omega)}=1$. Using $u=c+\widetilde v$ and $|\Omega|=1$,
we obtain
\[
 c=-\int_Qv\,dx,\qquad
 1=\|v\|_{L^2(Q)}^2-|c|^2
  =\|v\|_{L^2(Q)}^2-\left|\int_Qv\,dx\right|^2.
\]
The weak convergence of the gradients and lower semicontinuity of the $L^2(Q)$ norm give
\[
 \|\nabla v\|_{L^2(Q)}^2
 =\|\nabla u\|_{L^2(Q)}^2
 \le\liminf_{j\to\infty}\|\nabla u_j\|_{L^2(Q)}^2
 \le L,
\]
where the last inequality follows from $a_{\rho_j}=1$ on $Q$ and
$\lambda_2^{\mathrm N}(\rho_j)\to L$.

{\color{blue}Step 2.4.} Now we bound the Dirichlet energy of $v$ from below using its normalization, with the result providing the desired lower bound of $L$. To this end, we decompose $v$ into the ground state and its orthogonal complement. Set $\gamma:=\langle v,\phi\rangle_{L^2(Q)}$ and write
$v=\gamma\phi+v_\perp$, where $v_\perp\perp\phi$ in $L^2(Q)$. By the eigenfunction equation for $\phi$, the two components are
also orthogonal in the Dirichlet energy. Since $\mu_2=20\pi^2$, we have
\begin{equation} \label{eq:binary_seed_proof_1}
\|\nabla v\|_{L^2(Q)}^2
 =\mu_1|\gamma|^2+\|\nabla v_\perp\|_{L^2(Q)}^2
 \ge 8\pi^2|\gamma|^2+20\pi^2\|v_\perp\|_{L^2(Q)}^2.
\end{equation}
For normalization, the same decomposition gives
\[
 \|v\|_{L^2(Q)}^2=|\gamma|^2+\|v_\perp\|_{L^2(Q)}^2,
 \qquad
 \int_Qv\,dx=m\gamma+\int_Qv_\perp\,dx.
\]
A lower bound on the squared mean gives an upper bound on the
normalization expression. Applying $|a+b|^2\ge\tfrac34|a|^2-3|b|^2$
to the second identity and using Cauchy--Schwarz with $|Q|=\tfrac14$,
we obtain
\[
 \left|\int_Qv\,dx\right|^2
 \ge \frac34m^2|\gamma|^2-3\left|\int_Qv_\perp\,dx\right|^2
 \ge \frac34m^2|\gamma|^2-\frac34\|v_\perp\|_{L^2(Q)}^2.
\]
Subtracting this lower bound from the squared norm gives
\begin{equation}\label{eq:parameterized-square-rankone}
 1=\|v\|_{L^2(Q)}^2-\left|\int_Qv\,dx\right|^2
 \le \left(1-\frac{3m^2}{4}\right)|\gamma|^2
    +\frac74\|v_\perp\|_{L^2(Q)}^2.
\end{equation}
By direct calculation, one sees that $9(1-3m^2/4)<8$, and consequently, for $v\ne0$,
\[
 9\pi^2
 <\pi^2\bigl(8|\gamma|^2+20\|v_\perp\|_{L^2(Q)}^2\bigr)
 \le\|\nabla v\|_{L^2(Q)}^2
 \le L,
\]
where \eqref{eq:binary_seed_proof_1} is applied to derive the second inequality. This completes the proof of $\lambda_2^{\mathrm N}(\rho)>9\pi^2$ for all sufficiently large $\rho$.
\end{proof}

The binary medium constructed above provides a seed for Theorem~\ref{thm:main}. To prepare for the thin-layer construction in Section~\ref{sec:layer-insertion}, we introduce the following notation for gapped seeds and their frequency windows.

\begin{definition}[Seed and frequency windows]\label{def:binary-seed}
A coefficient \(S \in\mathcal X^{\mathrm H}_{\alpha,\beta}\) is called a \emph{seed} if there exists an open frequency interval
\(G=(\omega_-,\omega_+)\), with \(0<\omega_-<\omega_+\), such that
\[
  (\omega_-^2,\omega_+^2)\cap\sigma(H_S)=\varnothing.
\]
For such a seed, fix \(G\) as above and choose nonempty open intervals
\[
  G_j=(\omega_{j,-},\omega_{j,+}),\qquad j=0,1,
  \qquad G_1\Subset G_0\Subset G.
\]
We then define the distance \(\delta:=\operatorname{dist}(\overline{G_0},\R\setminus G)\).
\end{definition}

By Lemma~\ref{lem:seed}, for any fixed $\rho\ge\rho_0$, the coefficient $S=a_\rho I_2$ is a seed coefficient in $\mathcal X^{\mathrm R}_{1,\rho}$ with frequency interval $G=(2\sqrt{2}\pi,3\pi)$.  Consequently, in the construction for Theorem~\ref{thm:main}, we may choose this binary seed as our seed coefficient.

\section{Insertion of high-frequency gaps by thin layers} \label{sec:layer-insertion}

The thin-layer construction outlined in Subsection~\ref{sec:proof-idea} creates a high-frequency gap by inserting a rapidly rescaled copy of a gapped seed near the boundary of each period cell. The layer width controls the $L^1$ change in the coefficient. By Lemma~\ref{lem:persistence}, this allows compact subintervals of previously opened gaps to remain free of spectrum as new gaps are added. The construction thus provides the inductive step for producing a medium with infinitely many gaps in Section~\ref{sec:infinite-gap-construction}.

\begin{figure}[tbp]
\centering
\begin{tikzpicture}[
  x=6.2cm,y=6.2cm,
  line cap=round,line join=round,
  dimension/.style={
    {Latex[length=1.7pt,width=1.8pt]}-{Latex[length=1.7pt,width=1.8pt]},
    draw=layerink,line width=.5pt
  },
  guide/.style={draw=layerguide,line width=.4pt},
  smalllabel/.style={font=\footnotesize,text=layerink,inner sep=2pt},
  legendtext/.style={anchor=west,text=layerink,inner sep=0pt},
  swatch/.style={line width=.5pt,rounded corners=1pt}
]
  \definecolor{layerink}{HTML}{19314B}
  \definecolor{layerguide}{HTML}{728399}
  \definecolor{layerbase}{HTML}{F2F6FC}
  \definecolor{layerbaseedge}{HTML}{8297AE}
  \definecolor{layersky}{HTML}{69CEF5}
  \definecolor{layerskyedge}{HTML}{247FAB}
  \definecolor{layerorange}{HTML}{FFAA46}
  \definecolor{layerorangeedge}{HTML}{BA6117}

  \pgfmathsetmacro{\drawwidth}{0.12}
  \pgfmathsetmacro{\drawsubwidth}{\drawwidth/4}
  \pgfmathsetmacro{\baseedge}{.5-\drawwidth}
  \pgfmathsetmacro{\subedge}{.5-\drawsubwidth}

  \fill[layerorange] (-.5,-.5) rectangle (.5,.5);
  \fill[layersky] (-\subedge,-\subedge) rectangle (\subedge,\subedge);
  \fill[layerbase] (-\baseedge,-\baseedge) rectangle (\baseedge,\baseedge);
  \draw[layerorangeedge,line width=.55pt]
    (-\subedge,-\subedge) rectangle (\subedge,\subedge);
  \draw[layerskyedge,line width=.6pt]
    (-\baseedge,-\baseedge) rectangle (\baseedge,\baseedge);
  \draw[layerink,line width=.8pt] (-.5,-.5) rectangle (.5,.5);

  \node[text=layerink,font=\large] at (0,.600) {$\Omega$};
  \node[text=layerink,font=\large] at (0,0) {$C(x)$};

  \foreach \a/\lab in {-.5/-\frac12,0/0,.5/\frac12} {
    \draw[guide] (\a,-.5) -- (\a,-.518);
    \node[smalllabel,below] at (\a,-.522) {$\lab$};
    \draw[guide] (-.5,\a) -- (-.518,\a);
    \node[smalllabel,left] at (-.522,\a) {$\lab$};
  }

  \draw[guide] (\baseedge,.510) -- (\baseedge,.678);
  \draw[guide] (\subedge,.510) -- (\subedge,.586);
  \draw[guide] (.5,.510) -- (.5,.678);
  \draw[dimension] (\baseedge,.654) -- (.5,.654);
  \node[smalllabel,above] at ({(\baseedge+.5)/2},.669) {$w$};
  \draw[dimension] (\subedge,.563) -- (.5,.563);
  \node[smalllabel,right] at (.520,.563) {$w/4$};

  \node[legendtext,font=\small\sffamily\bfseries]
    at (.67,.29) {Regions and coefficients};
  \draw[guide] (.67,.233) -- (1.23,.233);

  \path[swatch,fill=layerbase,draw=layerbaseedge]
    (.67,.135) rectangle (.728,.181);
  \node[legendtext] at (.77,.158) {$\Omega\setminus\Sigma_w$};
  \node[legendtext,font=\footnotesize]
    at (.77,.080) {Background $C(x)$};

  \path[swatch,fill=layersky,draw=layerskyedge]
    (.67,-.056) rectangle (.728,-.010);
  \node[legendtext] at (.77,-.033) {$\Sigma_w\setminus\Sigma_{w/4}$};
  \node[legendtext,font=\footnotesize] at (.77,-.111) {Seed $S(Nx)$};

  \path[swatch,fill=layerorange,draw=layerorangeedge]
    (.67,-.247) rectangle (.728,-.201);
  \node[legendtext] at (.77,-.224) {$\Sigma_{w/4}$};
  \node[legendtext,font=\footnotesize] at (.77,-.302) {Seed $S(Nx)$};
\end{tikzpicture}
\caption{Thin-layer insertion in $\Omega=(-1/2,1/2)^2$ with periodic
identification. The blue and orange strips form $\Sigma_w$, where
$D^S_{N,w}[C](x)=S(Nx)$.}
\label{fig:layer_construction}
\end{figure}
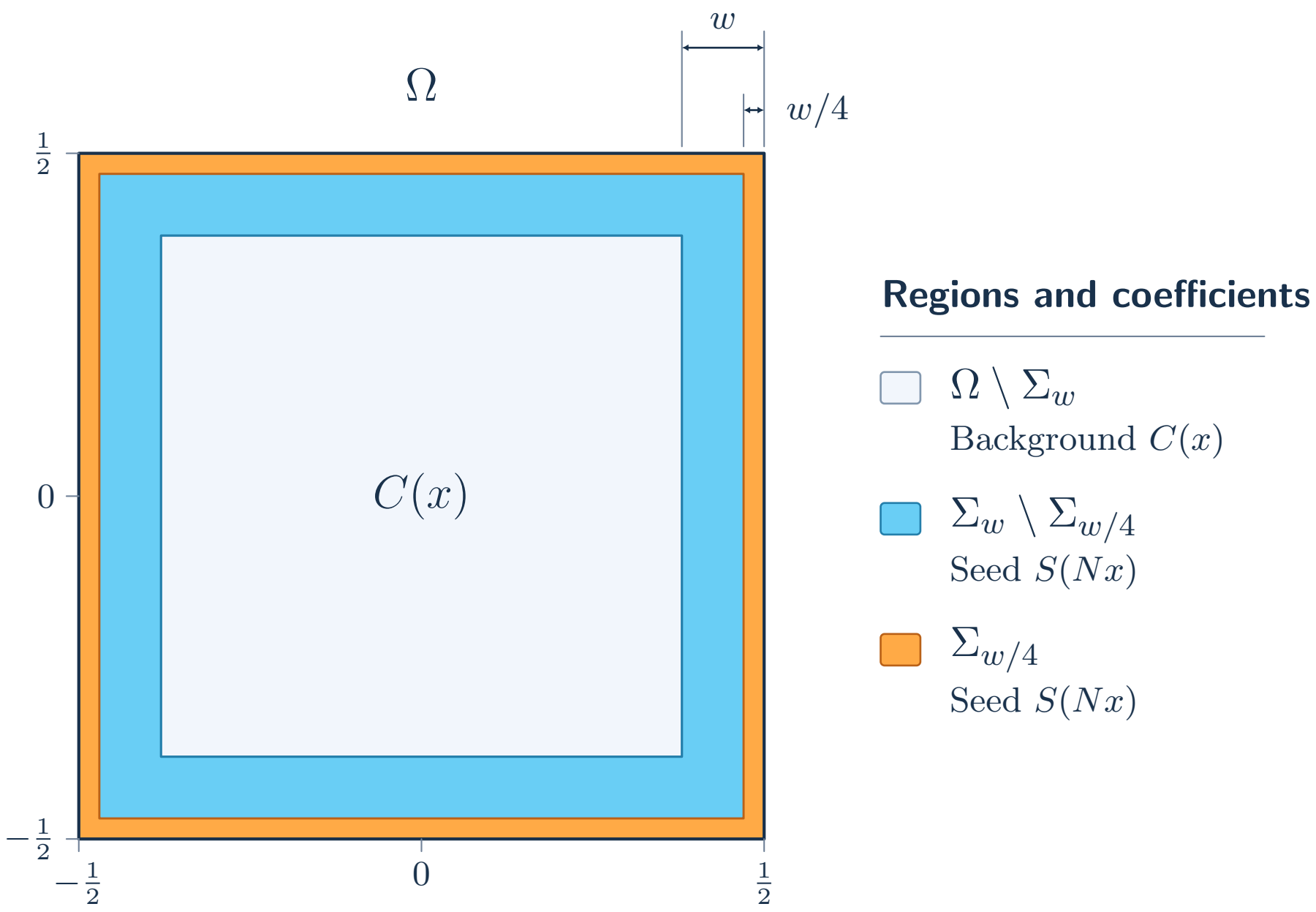

For a given background, the layer insertion replaces the coefficient by the rescaled seed only in the boundary layer $\Sigma_w$ shown in Figure~\ref{fig:layer_construction}.
\begin{definition}[Layer insertion]
For a background coefficient $C\in\mathcal X^{\mathrm H}_{\alpha,\beta}$ and a layer width $0<w<1/8$, the inserted coefficient at integer scale $N\ge 1$ is defined by
\begin{equation}\label{eq:insert}
 D^S_{N,w}[C](x):=
 \begin{cases}
  S(Nx), & x\in\Sigma_w,\\
  C(x), & x\notin\Sigma_w,
 \end{cases}
\end{equation}
where $S$ is the seed coefficient from Definition~\ref{def:binary-seed}.
\end{definition}

With these notions, we are now ready to present the main result of this section.
\begin{proposition}\label{prop:insert}
Fix a seed $S$, frequency windows $G_1\Subset G_0\Subset G$, and their separation $\delta>0$ as in Definition~\ref{def:binary-seed}. Set $K:=16\sqrt{\beta}/\delta$ and $w_N:=K(\log N)/N$. For $C\in\mathcal X^{\mathrm H}_{\alpha,\beta}$, set $D_N:=D^S_{N,w_N}[C]$. There exist $N_0\ge2$ and $c=c(\alpha,G_1)>0$ such that, for every $N\ge N_0$, one has $0<w_N<1/8$ and, there is an open energy interval $I_{C,N}$ satisfying
\begin{equation}\label{eq:properties-ICN}
   I_{C,N}\subset\{\omega^2:\omega\in NG_1\},
 \qquad
 |I_{C,N}|\ge\frac{c}{N^2},
 \qquad
 I_{C,N}\cap\sigma(H_{D_N})=\varnothing, 
\end{equation}
Moreover, the following estimate holds uniformly in $C\in\mathcal X^{\mathrm H}_{\alpha,\beta}$:
\begin{equation}\label{eq:induction_norm_difference}
 \|D_N-C\|_{L^1(\Omega)}\le 4(\beta-\alpha)w_N.
\end{equation}
\end{proposition}

\subsection{Roadmap of the proof}

First, we find a high-frequency gap for $D_N$ at zero quasimomentum, $k=0$, inside the frequency gap of the rescaled seed.
By the min--max principle and comparison with the periodic Laplacian, the number of Bloch frequencies in $NG_1$ is $O(N^2)$.
Since $|NG_1|=N|G_1|$, there is an open interval $(l,r)\subset NG_1$ such that
\[
 (l,r)\cap\{\omega_j(D_N;0):j\ge1\}=\varnothing,
 \qquad
 r-l\ge\frac{c_0}{N},
\]
where $c_0>0$ is independent of $C$ and $N$.

Second, we show that a central subinterval of $(l,r)$ is free of Bloch frequencies for every $k\in\mathcal B$. We control
the variation of each locally analytic eigenfrequency branch $\omega(t)$ along the segment $t\mapsto tk$ by
\[
 |\omega'(t)|\le V_{N,w_N}=o(N^{-1})
 \qquad\text{whenever }\omega(t)\in N\overline{G_0}.
\]
Because $G_1\Subset G_0$, a branch that meets the central subinterval can be followed back to $k=0$ within this frequency
window. Its variation is too small to leave $(l,r)$, contradicting the gap at $k=0$. The thin-layer construction is what makes the frequencies vary so slowly. The gapped seed suppresses the mass of eigenmodes in $\Sigma_{w_N/4}$, where the dependence on $k$ is localized by a gauge transformation. Physically, this corresponds to weak coupling between adjacent cells and nearly flat Bloch bands in the frequency window.

As seen above, to turn the intuitive argument into a rigorous proof, a key step is to control the eigenvalue motion. To this end, we introduce the following first-order differential operator, with the advantages 1) its positive eigenvalues are precisely the positive Bloch frequencies of the concerned elliptic operator and 2) it depends affinely on the quasi-momentum, and hence is convenient for estimating the momentum-derivative. We therefore use the standard first-order block construction (cf. \cite[Appendix~A]{GesztesyEtAl2012}).
\begin{definition}[First-order block operator]
For $C\in\mathcal X^{\mathrm H}_{\alpha,\beta}$ and $k\in\R^2$, regard $B_C(k)$ as an operator from $L^2(\T^2)$ to
$L^2(\T^2;\C^2)$ with domain $H^1(\T^2)$. The corresponding first-order block operator is defined by
\begin{equation}\label{eq:first-order-op}
 B_C(k):=C^{1/2}(\nabla_x+\ii k),
 \qquad
 M_C(k):=
 \begin{pmatrix}
  0 & B_C(k)^*\\
  B_C(k) & 0
 \end{pmatrix},
\end{equation}
where $B_C(k)^*$ denotes the $L^2$-adjoint of $B_C(k)$.
\end{definition}

The basic properties of $M_C(k)$ are summarized as the following lemma.
\begin{lemma}\label{lem:layer-block}
The operators $M_C(k)$ are self-adjoint and affine in $k$ on the common domain
\begin{equation}\label{eq:domain}
 \begin{aligned}
 \Dom (M_C(k))
 =\Dom (B_C(k))\oplus\Dom (B_C(k)^*)=H_{\mathrm{per}}^1(\Omega)
  \oplus C^{-1/2}H(\diver;\T^2),
 \end{aligned}
\end{equation}
where $C^{-1/2}H(\diver;\T^2)$ denotes the image of $H(\diver;\T^2)$ under multiplication by $C^{-1/2}$. The positive spectra of $M_C(k)$ are the square roots of the positive eigenvalues of $H_C(k)$:
\begin{equation}\label{eq:block-positive-spectrum}
 \sigma(M_C(k))\cap(0,\infty)
 =\bigl\{\sqrt{\lambda}:
     \lambda\in\sigma(H_C(k))\cap(0,\infty)\bigr\}.
\end{equation}
The correspondence $\lambda\mapsto\sqrt{\lambda}$ preserves the multiplicity of each positive eigenvalue.
\end{lemma}

The positive-spectrum correspondence in \eqref{eq:block-positive-spectrum} allows us to formulate the zero-quasimomentum estimate for $M_C(0)$. For the continuation of the frequency branches, we also need a quantitative lower bound on the distance from the central subinterval to $\sigma(M_C(0))$.

\begin{lemma}\label{lem:reference-hole}
Let $C\in\mathcal X^{\mathrm H}_{\alpha,\beta}$ and let $N\ge1$ be an integer. There exists an open interval $(l,r)\subseteq NG_1$ such that
\begin{equation}\label{eq:reference-hole-gap}
 (l,r)\cap\sigma(M_C(0))=\varnothing,
 \qquad
 \ell:=r-l\ge\frac{c}{N},
\end{equation}
where $c:=|G_1|/(C_{\mathrm{count}}+1)$ and $C_{\mathrm{count}}:=\bigl(1+\omega_{1,+}/(\pi\sqrt{\alpha})\bigr)^2$ depend only on $\alpha$ and $G_1$. The closed central half $J_N:=[l+\ell/4,r-\ell/4]$ satisfies
\begin{equation}\label{eq:reference-hole-separation}
 \operatorname{dist}\bigl(J_N,\sigma(M_C(0))\bigr)
 \ge\frac{\ell}{4}\ge\frac{c}{4N}.
\end{equation}
\end{lemma}

To control the variation of Bloch frequencies in $N\overline{G_0}$, we need to bound the mass of the corresponding eigenmodes in
$\Sigma_{w/4}$. The Combes--Thomas estimate for the rescaled seed gives this bound as follows.
\begin{lemma}
\label{lem:layer-decay}
For an integer $N\ge 1$, let $S_N(x):=S(Nx)$. For every $k\in\mathcal B$ and $\omega\in N\overline{G_0}$,
\begin{equation}\label{eq:seedgap}
 \dist\bigl(\omega,\spec(M_{S_N}(k))\bigr)\geq\delta N,\qquad k\in\mathcal B.
\end{equation}
The resolvent $(M_{S_N}(k)-\omega)^{-1}$ therefore exists and satisfies 
\begin{equation}\label{eq:CT}
 \norm{\one_X(M_{S_N}(k)-\omega)^{-1}\one_Y}
 \leq\frac2{\delta N}
 \exp\left(-\frac{\delta N}{2\sqrt\beta}\dist(X,Y)\right),
\end{equation}
for all measurable sets $X,Y\subset\T^2$. Moreover, for $C\in\mathcal X^{\mathrm H}_{\alpha,\beta}$ and $0<w<1/8$, let $D:=D^S_{N,w}[C]$. Every $L^2$-normalized eigenvector $F$ of $M_D(k)$ with eigenvalue $\omega\in N\overline{G_0}$ satisfies
\begin{equation}\label{eq:mass}
 \norm{\one_{\Sigma_{w/4}}F}_{L^2}
 \le \frac{8\sqrt{\beta}}{\delta Nw}
 \exp\left(-\frac{\delta Nw}{8\sqrt{\beta}}\right).
\end{equation}
\end{lemma}

We need one more technical ingredient to control the velocity of in-gap eigenvalues. We now introduce a gauge transform under which the quasimomentum derivative of the block operator for $D:=D^S_{N,w}[C]$ is supported solely in the thin layer $\Sigma_{w/4}$. Consequently, the eigenvalue derivative can be controlled by the mass of an eigenmode in that layer. This control is available thanks to our construction, in which the gapped medium is arranged in a layer surrounding the background medium.

\begin{definition}[Gauge transformation]
For $w\in (0,1/8)$, define the $1$-periodic function
\begin{equation}\label{eq:guage_transform_1}
 q_w(t):=\frac{2}{w}\,\one_{\{\operatorname{dist}(t,\tfrac12+\mathbb Z)<w/4\}},
 \qquad t\in\R.
\end{equation}
Since $\int_0^1 q_w(t)\,\mathrm dt=1$, we can choose a real, $1$-periodic Lipschitz function $f_w$ with $f_w'=q_w-1$ almost
everywhere. The gauge multiplier and the conjugated operator are given by
\begin{equation}\label{eq:guage_transform_2}
 U_w(k,x):=\exp\!\bigl(\ii k_1f_w(x_1)+\ii k_2f_w(x_2)\bigr),
 \qquad
 \widetilde M_C(k):=U_w(k)^*M_C(k)U_w(k).
\end{equation}
The scalar multiplier $U_w(k)$ acts on both components of the block space.
\end{definition}
Since $|U_w(k,x)|=1$, the operators $\widetilde M_C(k)$ and $M_C(k)$ are unitarily equivalent. If $F$ is a normalized eigenmode of
$M_D(k)$, then $U_w(k)^*F$ is a normalized eigenmode of
$\widetilde M_D(k)$ at the same eigenvalue, and
\[
 \bigl\|\one_{\Sigma_{w/4}}U_w(k)^*F\bigr\|_{L^2}
 =\bigl\|\one_{\Sigma_{w/4}}F\bigr\|_{L^2}.
\]
The layer mass estimate \eqref{eq:mass} therefore carries over to the transformed eigenmode. The next lemma identifies the
quasimomentum derivative of $\widetilde M_C(k)$ and makes its localization precise.

\begin{lemma}[Localized phase derivative and band motion]
\label{lem:layer-motion}
Fix $0<w<1/8$ and $C\in\mathcal X^{\mathrm H}_{\alpha,\beta}$. For $h\in\R^2$, define the directional derivative
\[
 \widetilde M_C'(k)[h]
 :=
 \left.\frac{\mathrm d}{\mathrm ds}
 \widetilde M_C(k+sh)\right|_{s=0}.
\]
This directional derivative is a bounded multiplication operator, independent of $k$, satisfying
\begin{equation}\label{eq:gauge}
 \begin{aligned}
 \widetilde M_C'(k)[h]
 =\one_{\Sigma_{w/4}}
 \bigl(\widetilde M_C'(k)[h]\bigr)
 \one_{\Sigma_{w/4}},\quad
 \norm{\widetilde M_C'(k)[h]}
 \le\frac{2\sqrt\beta\,|h|}{w}.
 \end{aligned}
\end{equation}

Let $D:=D^S_{N,w}[C]$, where $N\ge1$ is an integer. For each $k\in\mathcal B$, let $\omega(t)$ be any local real-analytic eigenfrequency branch of $\widetilde M_D(tk)$. Then
\begin{equation}\label{eq:velocity}
 |\omega'(t)|
 \le
 \frac{128\pi\sqrt2\,\beta^{3/2}}{\delta^2N^2w^3}
 \exp\!\left(-\frac{\delta Nw}{4\sqrt\beta}\right)
 =:v_{N,w},
\end{equation}
whenever the branch is defined at $t\in[0,1]$ and $\omega(t)\in N\overline{G_0}$. This bound is uniform in $C\in\mathcal X^{\mathrm H}_{\alpha,\beta}$ and $k\in\mathcal B$. Consequently, every eigenfrequency $\omega_*$ of $\widetilde M_D(k)$ satisfying
\[
 N\omega_{0,-}+v_{N,w}
 <\omega_*
 <N\omega_{0,+}-v_{N,w}
\]
can be continued along $t\mapsto tk$ to a locally real-analytic eigenfrequency branch $\omega(t)$ on $[0,1]$, with $\omega(1)=\omega_*$ and
\[
 |\omega(t)-\omega_*|\le(1-t)v_{N,w},
 \qquad
 \omega(t)\in NG_0,
 \qquad 0\le t\le1.
\]
\end{lemma}

The proofs of Lemmas~\ref{lem:layer-block}--\ref{lem:layer-motion} are provided in Appendix~\ref{app:prop-insert-auxiliary-lemmas}. With these preparations, we can now prove Proposition~\ref{prop:insert}.

\begin{proof}[Proof of Proposition~\ref{prop:insert}]
Take $w=w_N=\frac{16\sqrt\beta}{\delta}\frac{\log N}{N}$, with $N$ sufficiently large that $0<w_N<1/8$, and let $D_N:=D^S_{N,w_N}[C]$.
Lemma~\ref{lem:reference-hole}, applied to $D_N$, provides an interval $(l,r)\subseteq NG_1$ with $\ell:=r-l\ge c_0/N$, where $c_0>0$ depends only on $\alpha$ and $G_1$. Its closed central interval $J_N:=[l+\ell/4,r-\ell/4]$ satisfies
\begin{equation}\label{eq:insert-reference-margin}
 \dist\bigl(J_N,\sigma(\widetilde M_{D_N}(0))\bigr)
 \ge\frac{\ell}{4}\ge\frac{c_0}{4N},
\end{equation}
since $U_{w_N}(0)=I$ and hence $\widetilde M_{D_N}(0)=M_{D_N}(0)$. For the choice of width, the velocity bound in Lemma~\ref{lem:layer-motion} becomes
\[
 v_{N,w_N}
 =\frac{\pi\sqrt2\,\delta}{32N^3(\log N)^3}
 =o(N^{-1}).
\]
Increasing the lower bound on $N$ if necessary, we therefore have
\[
 v_{N,w_N}<\frac{c_0}{8N}\le\frac{\ell}{8},
 \qquad
 v_{N,w_N}
 <\frac N2\dist\bigl(\overline{G_1},\R\setminus G_0\bigr).
\]

Suppose that, for some $k\in\mathcal B$, the operator $\widetilde M_{D_N}(k)$ has an eigenfrequency $\omega\in J_N$. Since $J_N\subset NG_1$, the second inequality above gives $N\omega_{0,-}+v_{N,w_N}<\omega<N\omega_{0,+}-v_{N,w_N}$. The margin condition in Lemma~\ref{lem:layer-motion} is therefore satisfied, so the corresponding eigenfrequency branch can be continued back to $k=0$ within $NG_0$, we obtain
\[
 \dist\bigl(
 \omega,\sigma(\widetilde M_{D_N}(0))
 \bigr)
 \le v_{N,w_N}<\frac{\ell}{8}.
\]
This contradicts the reference estimate \eqref{eq:insert-reference-margin}. Hence no Bloch fiber has an eigenfrequency in $J_N$. Its interior lies in $NG_1$ and has length $\ell/2\ge c_0/(2N)$. By unitary equivalence, Lemma~\ref{lem:layer-block}, and Floquet decomposition, the energy interval
\[
 I_{C,N}:=
 \left(
 \left(l+\frac{\ell}{4}\right)^2,\,
 \left(r-\frac{\ell}{4}\right)^2
 \right)
\]
satisfies the properties \eqref{eq:properties-ICN}. Finally, $D_N=C$ outside $\Sigma_{w_N}$, while the common ellipticity bounds give $|S(Nx)-C(x)|\le\beta-\alpha$ almost everywhere. Using $|\Sigma_{w_N}|\le4w_N$, we obtain
\[
 \begin{aligned}
 \norm{D_N-C}_{L^1(\Omega)}
 =\int_{\Sigma_{w_N}}|S(Nx)-C(x)|\,\mathrm dx
 \le(\beta-\alpha)|\Sigma_{w_N}|
 \le4(\beta-\alpha)w_N.
 \end{aligned}
\]
This concludes the proof of \eqref{eq:induction_norm_difference}.
\end{proof}

\section{Stability of existing gaps in the coefficient topology}

The following lemma provides the stability needed for the iterative construction. More detailedly, we show that, within a fixed ellipticity class, every compact interval contained in a complete spectral gap persists under sufficiently small $L^1$ perturbations of the coefficients. This result will be applied to guarantee that a previously constructed gap is inherited while inserting new ones at higher energies.

\begin{lemma}\label{lem:persistence}
Let $A_n,A\in\mathcal X^{\mathrm H}_{\alpha,\beta}$ and suppose that $A_n\to A$ in $L^1(\Omega)$. Then
\begin{equation}\label{eq:normres}
 \sup_{k\in\mathcal B}
 \bigl\|(H_{A_n}(k)+1)^{-1}-(H_A(k)+1)^{-1}
 \bigr\|_{L^2(\Omega)\to L^2(\Omega)}
 \longrightarrow 0.
\end{equation}
Consequently, for every compact interval $I\subset\R\setminus\sigma(H_A)$, there exists $\eta=\eta(A,I)>0$ such that every
$\widetilde A\in\mathcal X^{\mathrm H}_{\alpha,\beta}$ with $\|\widetilde A-A\|_{L^1(\Omega)}<\eta$ satisfies
\[
 I\cap\sigma(H_{\widetilde A})=\varnothing.
\]
\end{lemma}

\begin{proof}[Proof of Lemma~\ref{lem:persistence}]
{\color{blue}Step 1.} We first show that the $L^1(\Omega)$ convergence of $A_n$ to $A$ implies the strong convergence $A_ng\to Ag$ in $L^2$ for every fixed $g\in L^2(\Omega;\mathbb C^2)$. To see this, note that by splitting the volume integral of $L^2$ norm into subdomains $\{|g|\le T\}$ and $\{|g|>T\}$, and using $|A_n-A|\le2\beta$ a.e., we obtain, for every $T>0$,
\[
 \|(A_n-A)g\|_{L^2(\Omega)}^2
 \le 2\beta T^2\|A_n-A\|_{L^1(\Omega)}
    +4\beta^2\int_{\{|g|>T\}}|g|^2\,dx.
\]
Taking the upper limit as $n\to\infty$ for fixed $T$, and then letting $T\to\infty$, yields
\begin{equation}\label{eq:persistence-multiplier}
 (A_n-A)g\to 0
 \quad\text{in }L^2(\Omega;\mathbb C^2).
\end{equation}
{\color{blue}Step 2.} To prove the resolvent convergence, let $R_C(k):=(H_C(k)+1)^{-1}$ and suppose that \eqref{eq:normres} fails. Then there exist $\varepsilon>0$ and a subsequence, still
indexed by $n$, such that
\[
 \sup_{k\in\mathcal B}\|R_{A_n}(k)-R_A(k)\|
 \ge 2\varepsilon,
\]
Thus, for each $n$, we can choose $k_n\in\mathcal B$ such that $\|R_{A_n}(k_n)-R_A(k_n)\|>\varepsilon$. By the definition of the operator norm, we can then choose $f_n\in L^2(\Omega)$ with $\|f_n\|_2=1$ so that
\begin{equation}\label{eq:persistence-contradiction}
 u_n:=R_{A_n}(k_n)f_n,\qquad
 v_n:=R_A(k_n)f_n,\qquad
 \|u_n-v_n\|_2\ge\varepsilon.
\end{equation}
For $(C_n,w_n)=(A_n,u_n)$ or $(A,v_n)$, the weak resolvent equation is
\begin{equation}\label{eq:persistence-weak}
 \int_\Omega (C_n\nabla_{k_n}w_n)\cdot
     \overline{\nabla_{k_n}\psi}\,dx
 +\int_\Omega w_n\overline{\psi}\,dx
 =\int_\Omega f_n\overline{\psi}\,dx,
 \qquad \psi\in H_{\mathrm{per}}^1(\Omega).
\end{equation}
Testing with $\psi=w_n$ and using uniform ellipticity gives
\[
 \alpha\|\nabla_{k_n}w_n\|_2^2+\|w_n\|_2^2
 \le |\langle f_n,w_n\rangle|
 \le \|w_n\|_2.
\]
Since $\|\nabla w_n\|_2\le\|\nabla_{k_n}w_n\|_2+|k_n|\|w_n\|_2$ and $\mathcal B$ is bounded, both $\{u_n\}$ and $\{v_n\}$ are bounded
in $H_{\mathrm{per}}^1(\Omega)$. By compactness of $\mathcal B$, weak compactness of bounded sequences in $L^2(\Omega)$ and $H_{\mathrm{per}}^1(\Omega)$, and the Rellich compact embedding $H_{\mathrm{per}}^1(\Omega)\hookrightarrow L^2(\Omega)$, we may pass to a common subsequence, still indexed by $n$, such that
\[
 \begin{gathered}
 k_n\to k\in\mathcal B,\qquad
 f_n\rightharpoonup f\quad\text{in }L^2(\Omega),\\
 w_n\rightharpoonup w\quad\text{in }H_{\mathrm{per}}^1(\Omega),
 \qquad
 w_n\to w\quad\text{in }L^2(\Omega),
 \end{gathered}
\]
where $(w_n,w)=(u_n,u)$ or $(v_n,v)$. By Hermitian symmetry, the gradient term in \eqref{eq:persistence-weak} is $\langle\nabla_{k_n}w_n,C_n\nabla_{k_n}\psi\rangle$.
Using $\nabla_{k_n}-\nabla_k=i(k_n-k)$ and $|C_n|\le\beta$, we obtain
\[
 \|C_n\nabla_{k_n}\psi-A\nabla_k\psi\|_2
 \le \|(C_n-A)\nabla_k\psi\|_2
    +\beta|k_n-k|\|\psi\|_2.
\]
The first term tends to zero by \eqref{eq:persistence-multiplier} with $g=\nabla_k\psi$ when $C_n=A_n$, and vanishes when $C_n=A$.
The second tends to zero since $k_n\to k$. Thus $C_n\nabla_{k_n}\psi\to A\nabla_k\psi$ strongly in
$L^2(\Omega;\mathbb C^2)$. Let $w$ denote the limit of $w_n$, so $w=u$ or $v$. Since $\nabla w_n\rightharpoonup\nabla w$ and $k_nw_n\to kw$
in $L^2$, we have $\nabla_{k_n}w_n\rightharpoonup\nabla_k w$. Together with the strong convergence of $C_n\nabla_{k_n}\psi$,
this allows us to pass to the limit in \eqref{eq:persistence-weak}. Hence
\[
 (H_A(k)+1)u=f=(H_A(k)+1)v.
\]
By uniqueness, $u=v$. Thus strong $L^2$ convergence gives $\|u_n-v_n\|_2\to0$, contradicting \eqref{eq:persistence-contradiction}.

{\color{blue}Step 3.} It remains to prove the spectral persistence using the resolvent convergence. By nonnegativity, it suffices to consider $J:=I\cap[0,\infty)$, and the conclusion is
immediate if $J=\varnothing$. For $J\ne\varnothing$, let $r(\lambda):=(1+\lambda)^{-1}$ for $\lambda\ge0$.
Since $J\cap\sigma(H_A)=\varnothing$, the sets
\[
 K:=r(J),\qquad K_A:=\{0\}\cup r(\sigma(H_A))
\]
are disjoint and compact. Spectral mapping and Floquet decomposition give $\sigma(R_A(k))\subset K_A$ for every $k\in\mathcal B$.
Hence $d:=\operatorname{dist}(K,K_A)>0$ is a uniform lower bound on the distance from $K$ to the reference resolvent spectra.
By \eqref{eq:normres}, there exists $\eta>0$ such that every $\widetilde A\in\mathcal X^{\mathrm H}_{\alpha,\beta}$ with
$\|\widetilde A-A\|_{L^1(\Omega)}<\eta$ satisfies
\[
 \sup_{k\in\mathcal B}
 \|R_{\widetilde A}(k)-R_A(k)\|<d/2.
\]
Spectral stability for bounded selfadjoint operators then gives, for every $k\in\mathcal B$,
\[
 \operatorname{dist}\bigl(K,\sigma(R_{\widetilde A}(k))\bigr)
 \ge d-\|R_{\widetilde A}(k)-R_A(k)\|
 >d/2.
\]
By spectral mapping, every perturbed fiber spectrum is disjoint from $J$. Floquet decomposition and nonnegativity therefore give $I\cap\sigma(H_{\widetilde A})=\varnothing$.
\end{proof}

\section{Construction of an infinite-gap medium: proof of Theorem~\ref{thm:main}} \label{sec:infinite-gap-construction}

We apply Proposition~\ref{prop:insert} repeatedly with a fixed gapped seed, inserting it into successively thinner layers to open gaps at energies tending to infinity. At stage $j$, we select a compact interval $I_j$ of positive length inside the new gap of $H_{C_j}$, with stability tolerance $\eta_j>0$ given by Lemma~\ref{lem:persistence}. The subsequent perturbations are chosen to satisfy
\[
 \sum_{n=j+1}^{\infty}
 \|C_n-C_{n-1}\|_{L^1(\Omega)}<\eta_j,
 \qquad j\ge1.
\]
These summable perturbations yield an $L^1$ limit $A_\infty$, while the tail bounds ensure that every $I_j$ remains as a gap of $\sigma(H_{A_\infty})$. Since the intervals $I_j$ tend to infinity by our construction, we conclude that $\spec(H_{A})$ has infinitely many gap components.

Finally, the spectral persistence and thin-layer insertion provide the open dense subsets of $\mathcal X$ needed to establish the dense $G_\delta$ conclusion by Baire's theorem.

\begin{proof}[Proof of Theorem~\ref{thm:main}]

For this seed, fix the frequency windows $G_1\Subset G_0\Subset G$ and their separation $\delta>0$ as in Definition~\ref{def:binary-seed}, and set
$K:=16\sqrt{\beta}/\delta$ as in Proposition~\ref{prop:insert}.

The class $\mathcal X$ is complete in the $L^1$ metric. Indeed, every $L^1$-convergent sequence in $\mathcal X$ has a subsequence converging almost everywhere to the same limit. The matrix bounds pass to this limit, so $\mathcal X$ is closed in the Banach space $L^1(\T^2;\R_{\mathrm{sym}}^{2\times2})$.

\medskip
\noindent{\color{blue}Step 1.}\enspace
\emph{Inductive construction.}
We initialize the iteration with the constant background coefficient $C_0$ and the reference width $w_0$, defined by
\[
 C_0:=2I_2\in\mathcal X,
 \qquad w_0:=1/8.
\]
For the first insertion, set $b_1:=1/2$ and choose an integer $N_1$ large enough for Proposition~\ref{prop:insert} to apply to $C_0$ and satisfying
\[
 N_1^2\omega_{1,-}^2>1,\qquad
 w_1:=K\frac{\log N_1}{N_1}<\frac{w_0}{2},\qquad
 4(\beta-\alpha)w_1<b_1.
\]
The resulting coefficient $C_1:=D^S_{N_1,w_1}[C_0]$ belongs to $\mathcal X$, and the Proposition~\ref{prop:insert} provides a nonempty open energy interval $J_1$ satisfying
\[
 J_1\subset
 (N_1^2\omega_{1,-}^2,N_1^2\omega_{1,+}^2),
 \qquad
 J_1\cap\sigma(H_{C_1})=\varnothing.
\]
We select a compact interval $I_1\Subset J_1$ of positive length. Applying Lemma~\ref{lem:persistence} to $C_1$ and $I_1$, there exists a stability tolerance $\eta_1>0$ such that, if $\|C-C_1\|_{L^1(\Omega)}<\eta_1$, we have
\[
 I_1\cap\sigma(H_C)=\varnothing,
 \qquad C\in\mathcal X
 .
\]

At each stage $j\ge2$, the tolerances from the preceding stages determine the bound
\[
 b_j:=2^{-j}\min\{1,\eta_1,\ldots,\eta_{j-1}\}.
\]
We choose an integer $N_j$ large enough for Proposition~\ref{prop:insert} to apply to $C_{j-1}$ and satisfying
\[
 N_j^2\omega_{1,-}^2>j,\quad
 N_j^2\omega_{1,-}^2>\sup I_{j-1},\quad
 w_j:=K\frac{\log N_j}{N_j}<\frac{w_{j-1}}2,\quad
 4(\beta-\alpha)w_j<b_j.
\]
We then define $C_j:=D^S_{N_j,w_j}[C_{j-1}]\in\mathcal X$ and select $I_j$ and $\eta_j$ as in the first step. This procedure constructs the sequences $C_j$, $I_j$, and
$\eta_j$ for all $j\ge1$.
By \eqref{eq:induction_norm_difference},
\begin{equation}\label{eq:budget}
 \|C_j-C_{j-1}\|_{L^1(\Omega)}
 \le4(\beta-\alpha)w_j<b_j.
\end{equation}
The location of each new gap and the choice of $N_j$ give
\[
 I_j\cap\sigma(H_{C_j})=\varnothing,\qquad
 \inf I_j>j,\qquad
 \inf I_j>\sup I_{j-1}\quad(j\ge2).
\]
Moreover, the stability tolerance $\eta_j$ ensures that, for every $C\in\mathcal X$,
\[
 I_j\cap\sigma(H_C)=\varnothing
 \qquad\text{whenever}\qquad
 \|C-C_j\|_{L^1(\Omega)}<\eta_j.
\]

\medskip
\noindent{\color{blue}Step 2.}\enspace
\emph{Passage to the limit and infinitely many gaps.}
Since $b_j\le2^{-j}$, estimate \eqref{eq:budget} gives
\[
 \sum_{j=1}^{\infty}
 \|C_j-C_{j-1}\|_{L^1(\Omega)}
 \le\sum_{j=1}^{\infty}b_j
 \le\sum_{j=1}^{\infty}2^{-j}<\infty.
\]
Thus $\{C_j\}$ is a Cauchy sequence in $\mathcal X$. By completeness, there exists $A_\infty\in\mathcal X$ such that
\[
 C_j\to A_\infty
 \quad\text{in }
 L^1(\T^2;\R_{\mathrm{sym}}^{2\times2}).
\]
For each fixed $j\ge1$, the subsequent bounds satisfy $b_n\le2^{-n}\eta_j$ for $n>j$. Consequently,
\[
 \begin{aligned}
 \|A_\infty-C_j\|_{L^1(\Omega)}
 \le\sum_{n=j+1}^{\infty}
      \|C_n-C_{n-1}\|_{L^1(\Omega)}
 \le\sum_{n=j+1}^{\infty}b_n
 \le2^{-j}\eta_j<\eta_j.
 \end{aligned}
\]
The stability property established in Step~1 therefore gives
\[
 I_j\cap\sigma(H_{A_\infty})=\varnothing,
 \qquad j\ge1.
\]
Since $\inf I_j>j\to\infty$, we conclude that there is no half-line of the form $[E,\infty)$ containing in $\sigma(H_{A_\infty})$, or equivalently, $\sigma(H_{A_\infty})$ contains infinitely many gap components.



\medskip
\noindent{\color{blue}Step 3.}\enspace
\emph{Density and the $G_\delta$ property.}
To describe gaps above a prescribed energy threshold, define, for each integer $m\ge1$,
\[
 \mathcal U_m:=
 \bigcup_{m<a<b}
 \left\{
 C\in\mathcal X:
 [a,b]\cap\sigma(H_C)=\varnothing
 \right\}.
\]
Each set in this union is open by Lemma~\ref{lem:persistence}, so $\mathcal U_m$ is open in $\mathcal X$.

To prove density, fix $C\in\mathcal X$ and $\varepsilon>0$. Since $w_N=K(\log N)/N\to0$ and $N^2\omega_{1,-}^2\to\infty$, we may choose an integer $N$ large enough for Proposition~\ref{prop:insert} to apply and such that
\[
 N^2\omega_{1,-}^2>m,\qquad
 4(\beta-\alpha)w_N<\varepsilon.
\]
For $D_N:=D^S_{N,w_N}[C]$, the proposition supplies a nonempty open interval above $m$ disjoint from $\sigma(H_{D_N})$.
This interval contains a compact subinterval $[a,b]$ with $m<a<b$. Hence
\[
 D_N\in\mathcal U_m,\qquad
 \|D_N-C\|_{L^1(\Omega)}
 \le4(\beta-\alpha)w_N<\varepsilon.
\]
Thus, $\mathcal U_m$ is dense in $\mathcal X$. Let $\mathcal G$ denote the class of coefficients described in the theorem. Compact subintervals of the gap components give $\mathcal G\subset\bigcap_{m\ge1}\mathcal U_m$. Conversely, for every $C\in\mathcal X$,
\[
 0\in\sigma(H_C),\qquad
 \sup\sigma(H_C)=\infty.
\]
The argument of Step~2 therefore gives the reverse inclusion, so
\[
 \mathcal G=\bigcap_{m=1}^{\infty}\mathcal U_m.
\]
Since each $\mathcal U_m$ is open and dense and $\mathcal X$ is complete, Baire's theorem shows that $\mathcal G$ is a dense $G_\delta$ subset of $\mathcal X$, proving the theorem.

\end{proof}

\section{Realization of prescribed Chern numbers: proof of Theorem \ref{thm:chern-main}}
\label{sec:chern-layers}

We apply the same thin-layer induction as in Section \ref{sec:infinite-gap-construction} to construct a single medium that supports an arbitrary Chern-number configuration, with a stronger condition that we prescribe the gap Chern number of each seed medium inserted during the induction. The availability of such medium is exactly stated in the claim (i) of Theorem \ref{thm:chern-main}, and is justified in the following proposition:
\begin{proposition} \label{prop:uniform-seeds}
There exist constants $0<\alpha<1$ and $\beta>2$ such that, for every $m\in\Z$ and $\eta>0$, there is a smooth $\Z^2$-periodic Hermitian matrix-valued field $A=A(x)$ such that
\begin{equation} \label{eq:Chern_seed_bound}
\alpha I_2\le A\le\beta I_2,\quad \|\operatorname{Im}A\|_{L^\infty}\le\eta,
\end{equation}
and the elliptic operator $H_A=-\operatorname{div}(A\nabla)$ (with the usual $H^1(\R^2)$ realization) has a complete positive spectral gap with the gap Chern number being $m$.
\end{proposition}
The proof of Proposition \ref{prop:uniform-seeds} constitutes the most technical part of this paper, and hence is postponed to Section \ref{sec:uniform-seeds}. With this result, we follow the construction in Section \ref{sec:infinite-gap-construction} by choosing a smooth seed medium $S_j$ with the prescribed gap Chern number $m_j$ in the $j$\textsuperscript{th} induction step. In particular, we require that the Chern number is inherited when we insert the seed. This is guaranteed by the following proposition:
\begin{proposition} \label{prop:chern-insert}
Let $S\in\mathcal X^{\mathrm H}_{\alpha,\beta}$, and let $G=(\omega_-,\omega_+)$, $G_1\Subset G_0\Subset G$, and $\delta$ be as in Definition~\ref{def:binary-seed}. Suppose $\operatorname{Ch}(S;(\omega_-^2,\omega_+^2))=c_S$. For $C\in\mathcal X^{\mathrm H}_{\alpha,\beta}$, set $D_N:=D^S_{N,w_N}[C]$, where
\[
K=\frac{40\sqrt\beta}{\delta},\quad w_N=K\frac{\log N}{N}.
\]
Then there exists $h>0$, independent of the background medium $C$, such that for all sufficiently large $N$ one can find $\omega_N\in NG_1$ with
\begin{equation} \label{eq:Chern_gap_estimate}
\inf_{k\in\mathcal B} \dist(\omega_N,\spec(M_{D_N}(k)))\geq h/N, \quad \operatorname{Ch}(D_N;\omega_N^2)=c_S.
\end{equation}
More generally, for every fixed $h'>0$, every $\omega\in N\overline{G_1}$ satisfying $\inf_{k\in\mathcal B} \dist(\omega,\spec(M_{D_N}(k)))\geq h'/N$ obeys $\operatorname{Ch}(D_N;\omega^2)=c_S$ once $N$ is sufficiently large.
\end{proposition}
The following property will be recalled to show the persistence of Chern numbers:
\begin{proposition} \label{lem:chern-persistence}
If $C\in\mathcal X^{\mathrm H}_{\alpha,\beta}$ and a compact interval $I$ lies inside a complete positive gap of $H_C$, then there is $r>0$ such that, for every $D\in\mathcal X^{\mathrm H}_{\alpha,\beta}$ with $\norm{D-C}_1<r$, $\spec(H_D)\cap I=\varnothing$ and $\operatorname{Ch}(D;E)=\operatorname{Ch}(C;E)$ for every $E\in I$.
\end{proposition}
With these ingredients, the construction in Section \ref{sec:infinite-gap-construction} yields a well-defined limiting medium, whose gap property and Chern numbers satisfy the requirement of Theorem \ref{thm:chern-main}:
\begin{proposition} \label{prop:chern-sequence}
Suppose $(S_j)_{j\geq1}\subset\mathcal X^{\mathrm H}_{\alpha,\beta}$ is a sequence of smooth periodic Hermitian matrix-valued fields, each with a complete positive gap of Chern number $m_j$. Suppose also that $\norm{\operatorname{Im}S_j}_\infty\leq\eta$ for every $j$. Then the limiting medium $A_\infty$, obtained by the induction in Section \ref{sec:infinite-gap-construction} with the seed $S_j$ in the $j$\textsuperscript{th} step, satisfies
\begin{equation} \label{eq:limiting_medium_bounded}
\alpha I_2\leq A_\infty\leq\beta I_2,
\quad\norm{\operatorname{Im}A_\infty}_\infty\leq\eta,
\end{equation}
and has distinct bounded complete gaps  $G_j$, ordered increasingly so that $\inf G_j\to\infty$, with the corresponding Chern numbers $m_j$.
\end{proposition}
More details on the induction process are presented in the proof of Proposition \ref{prop:chern-sequence}. This concludes the proof of Theorem \ref{thm:chern-main}. The remainder of this section is dedicated to proving Propositions \ref{prop:uniform-seeds}-\ref{prop:chern-sequence}.

\subsection{Proof of Propositions \ref{prop:chern-insert}-\ref{prop:chern-sequence}}


\begin{proof}[Proof of Proposition~\ref{prop:chern-insert}]
The gap estimate is proved by the same argument as in Section~\ref{sec:layer-insertion}. Here we focus on the delicate point, i.e., transfer of the prescribed Chern number to the high-frequency regime by inserting the seed medium. The essential idea is that a spectral bundle between any two gap frequencies in $N\overline{G_1}$ (with a controlled separation distance to the spectrum) have small mass within the layers, which follows by the screening effect of the gapped layer medium as indicated in Section \ref{sec:proof-idea}. In particular, such a weak connection between adjacent cells does not support a nontrivial Chern number. Consequently, any such two gap points in $N\overline{G_1}$ share the same Chern number. To rigorously formulate this intuitive idea, we will apply the estimate in Lemma \ref{lem:layer-decay} and proceed in several steps.

{\color{blue}Step 1.} We first prove that two sufficiently separated gap frequencies in the inner window $N\overline{G_1}$ have the same occupied Chern number; identifying it with the prescribed Chern number of the seed medium is left to the next step. Let $a_N<b_N$ lie in $N\overline{G_1}$, both satisfying that $\inf_{k\in\mathcal B}\dist(E,\spec(M_{D_{N}}(k)))\geq h^{\prime}/N=:g_N$ for $E=a_N,b_N$. In the sequel, we write $w:=w_N$ and $D:=D_N$ for ease of notation. Write the spectral projection to the intermediate interval as $P=P(k)=\one_{[a_N,b_N]}(\widetilde M_{D}(k))$, where the gauged block operator $\widetilde M_{D}(k)$ is introduced in \eqref{eq:guage_transform_2}. We assume $P\neq 0$ without loss of generality, as the claim on constant Chern number follows trivially when $P=0$. In the sequel, we will first prove that the gauged bundle $P$ has zero Chern number, and then transfer this triviality to the original bundle.

We first estimate the mass of the cluster $P(k)$ in the layer $\Sigma_{w/4}$. Let $r:=\operatorname{rank}P\le RN^2$, where the rank bound follows from the counting argument in Lemma~\ref{lem:reference-hole}, uniformly in $C$ and $k$. Let $F_1,\ldots,F_r$ be an orthonormal eigenbasis of $\operatorname{Ran}P(k)$. Since $U_w$ commutes with $\one_{\Sigma_{w/4}}$, the estimate \eqref{eq:mass} applies to $F_j$:
\[
 \|\one_{\Sigma_{w/4}}F_j\|_2\le
 \frac{8\sqrt\beta}{\delta Nw_N}
 e^{-\delta Nw_N/(8\sqrt\beta)}
 =:\epsilon_N=\frac{N^{-5}}{5\log N}.
\]
For $v=\sum_j a_jF_j$, Cauchy--Schwarz gives
$\|\one_{\Sigma_{w/4}}v\|_2\le\epsilon_N\sum_j|a_j|
\le\sqrt r\,\epsilon_N\|v\|_2$. Therefore,
\begin{equation}\label{eq:chern-cluster-mass}
 \|\one_{\Sigma_{w/4}}P(k)\|\le\sqrt r\,\epsilon_N
 =:\zeta_N=O(N^{-4}/\log N).
\end{equation}

{\color{blue}Step 1.1.} To study the Chern number of $P$, we need to control the connection term, given by the momentum-derivative of $P$, i.e., $\partial_{h}P$ (recall the $h$ denotes the direction of differentiation). This is achieved by studying the Sylvester equation satisfied by $\partial_{h}P$. Throughout this step, we will fix the momentum $k$ and abbreviate
\[
 M:=\widetilde M_D(k),\quad G:=\widetilde M_D^{\prime}(k)[h]=\partial_h\widetilde M_D(k).
\]
Note that the differentiation on $\widetilde M_D(k)$ is justified by the fact that $\widetilde M_D(k)$ has common domain and bounded dependence on $k$, as established in Lemma \ref{lem:layer-motion}. We will decompose $M$ by projecting it to $(-\infty,a_N)$, $(b_N,\infty)$ and the intermediate range $[a_N,b_N]$, respectively, which are denoted as
\begin{equation*}
Q_-=\mathbf1_{(-\infty,a_N)}(M),\quad
Q_+=\mathbf1_{(b_N,\infty)}(M),\quad
P=\mathbf1_{[a_N,b_N]}(M)
\end{equation*}
with the ranges $\mathcal H_P:=\operatorname{Ran}P$ and $\mathcal H_\pm:=\operatorname{Ran}Q_\pm$. Now, differentiating the identity $[M,P]=0$ yields 
\[
 [M,\partial_{h}P]=-[G,P].
\]
Multiplying on the left by $Q_\pm$ and on the right by $P$ yields the Sylvester equations
\begin{equation} \label{eq:sylvester}
M_\pm X_\pm-X_\pm M_P=-F_\pm,
 \qquad F_\pm:=Q_\pm GP,
\end{equation}
where
\begin{equation*}
X_\pm:=Q_\pm(\partial_{h}P) P,\quad
M_\pm:=M|_{\mathcal H_\pm},\quad M_P:=M|_{\mathcal H_P}.
\end{equation*}
The restrictions have domains $\operatorname{Dom}(M)\cap\mathcal H_\pm$ and $\mathcal H_P$, respectively. Thanks to the spectral margins of $a_N$ and $b_N$, which are separated from $\spec(M)$ by a distance at least $g_N=h^{\prime}/N$, we know
\begin{equation*}
\sigma(M_-)\subset(-\infty,a_N-g_N],\quad
\sigma(M_P)\subset[a_N+g_N,b_N-g_N],\quad
\sigma(M_+)\subset[b_N+g_N,\infty).
\end{equation*}
Hence, one can check using integration by parts that the Sylvester equations \eqref{eq:sylvester} are solved by the explicit formula
\begin{equation}  \label{eq:sylvester_solution}
\begin{aligned}
 X_-&=\int_0^\infty
       e^{t(M_--a_N)}F_-e^{-t(M_P-a_N)}\,dt,\\
 X_+&=-\int_0^\infty
       e^{-t(M_+-b_N)}F_+e^{t(M_P-b_N)}\,dt,
 \end{aligned}
\end{equation}
where both integrals converge in operator norm.
In particular,
\begin{equation} \label{eq:sylvester_estimate}
\|X_\pm\| \le\int_0^\infty e^{-2g_Nt}\|F_\pm\|\,dt =\frac{\|F_\pm\|}{2g_N} \le\frac{\|F_\pm\|}{g_N}.
\end{equation}
We now recover $\partial_{h}P$ from $X_{\pm}$, by which \eqref{eq:sylvester_estimate} leads to the desired estimate of $\partial_{h}P$. Differentiating the projection identity $P^2=P$ gives
\[
P(\partial_{h}P)+(\partial_{h}P)P=\partial_{h}P.
\]
Multiplying this identity on both sides by $P$, or $I-P$, shows that $\partial_{h}P$ is off-diagonal in the sense that
\[
 P\partial_{h}PP=0,\quad (I-P)\partial_{h}P(I-P)=0.
\]
Moreover, $\partial_{h}P=\partial_{h}P^*$ because $P(k)$ is an orthogonal projector. Therefore, decomposing the Hilbert space as $\mathcal H=\mathcal H_P\oplus\mathcal H_P^\perp$, the derivative $\partial_{h}P$ is written as the following block form
\[
 \partial_{h}P=
 \begin{pmatrix}
  0&X^*\\
  X&0
 \end{pmatrix}, \quad
 \text{with }\, X:=(I-P)\partial_{h}PP=X_-+X_+ .
\]
Consequently, we obtain by \eqref{eq:sylvester_estimate} that
\begin{equation}
\|\partial_{h}P\|=\|X\| \leq \|X_-\|+\|X_+\|
\le\frac{\|Q_-GP\|+\|Q_+GP\|}{g_N} \le\frac{2}{g_N}\|GP\|.
\end{equation}
Combining it with the estimates \eqref{eq:gauge} and \eqref{eq:chern-cluster-mass} gives the desired estimate
\begin{equation}\label{eq:chern-cluster-derivative}
 \norm{\partial_hP}
 \leq\frac2{g_N}\norm{GP}
 \leq\frac{4\sqrt\beta\,|h|\zeta_N}{g_Nw_N}
 =O\bigl(|h|N^{-2}/(\log N)^2\bigr).
\end{equation}

{\color{blue}Step 1.2.} We now prove that the gauged bundle $P$ has zero Chern number by showing that its absolute value is strictly smaller than one. We denote
\begin{equation} \label{eq:gauged_Chern_num}
c(P):=\frac{1}{2\pi\ii}\int_{\mathcal B}
\operatorname{tr}\bigl(P[\partial_{k_1}P,\partial_{k_2}P]\bigr)\,\mathrm dk_1\,\mathrm dk_2.
\end{equation}
We claim that $c(P)$ is integer-valued, which is worth a careful proof since we have applied a gauge. This is done by checking the integrand in \eqref{eq:gauged_Chern_num} is a curvature. Indeed, recalling the boundary conditions of the original bundle
\begin{equation*}
P_{\mathrm{orig}}(k+2\pi e_j)
 =\tau_jP_{\mathrm{orig}}(k)\tau_j^*,
 \quad \text{with}\quad
 P_{\mathrm{orig}}(k):=\one_{[a_N,b_N]}(M_D(k)),\quad \tau_j(x):=e^{-2\pi\ii x_j},
\end{equation*}
and the unitary relation
\begin{equation*}
P(k)=U_w(k)^*P_{\mathrm{orig}}(k)U_w(k),
\end{equation*}
we see that $P=P(k)$ satisfies
\[
 P(k+2\pi e_j)=V_jP(k)V_j^*,\quad \text{with}\,
 V_j:=U_w(k+2\pi e_j)^*\tau_jU_w(k)=e^{-2\pi\ii(x_j+f_w(x_j))}.
\]
The operators $V_j$ commute and are independent of $k$. Hence, supposing that  $k\mapsto s(k)\in\operatorname{Ran}P(k)$ is a smooth local section of the gauged bundle, we know $d(V_js)=V_j\,ds$, so the local connections $s\mapsto P\,ds$ respect the boundary identifications and define a connection on the gauged bundle. Its curvature is $P\,dP\wedge dP\,P$, corresponding exactly to the integrand in \eqref{eq:gauged_Chern_num}. In particular, the integrand is periodic: translating $k$ by $2\pi e_j$ conjugates all its factors by the same $k$-independent unitary $V_j$, which leaves the trace unchanged. This concludes that $c(P)\in \Z$.

To prove $|c(P)|<1$, note that \eqref{eq:chern-cluster-derivative} gives
\[
 L_N:=\max_{j=1,2}\sup_{k\in\mathcal B}
       \norm{\partial_{k_j}P(k)}
 =O\bigl(N^{-2}/(\log N)^2\bigr).
\]
On the other hand, as in Lemma \ref{lem:reference-hole}, the min-max principle yields the rank estimate
$$r=\operatorname{rank}P\leq R N^2,$$
for a fixed $R$ independent of $C$ and $k$. Thus, the integrand in \eqref{eq:gauged_Chern_num} is estimated as
\[
 \left|\operatorname{tr}\bigl(P[\partial_{k_1}P,\partial_{k_2}P]\bigr)\right|
 \leq r\norm{[\partial_{k_1}P,\partial_{k_2}P]}
 \leq 2rL_N^2=O\bigl(N^{-2}/(\log N)^4\bigr).
\]
Consequently, for sufficiently large $N$, we conclude by \eqref{eq:gauged_Chern_num} that $|c(P)|<1$, and therefore $c(P)=0$.

{\color{blue}Step 1.3.} Next, we transfer this topological triviality to the original bundle $P_{\mathrm{orig}}$, and then to the concerned scalar operator $H_{D}$. Note that the identity
\[
 U_w(k+2\pi e_j)V_j=\tau_jU_w(k)
\]
shows that $v\mapsto U_w(k)v$ respects the boundary identifications of both the gauged and original bundles. It therefore defines a continuous unitary bundle isomorphism, with inverse induced by $U_w(k)^*$. Consequently, the original bundle $P_{\mathrm{orig}}$ has the same Chern number as the gauged bundle $P$. Finally, we note that the Chern number of $P_{\mathrm{orig}}$ is equal to its scalar counterpart, i.e., $\one_{[a_N^2,b_N^2]}(H_D(k))$, as checked by using the isomorphism between eigenspaces
\[
 u\longmapsto 2^{-1/2}\bigl(u,B_D(k)H_D(k)^{-1/2}u\bigr), \quad u\in\operatorname{Ran}\one_{[a_N^2,b_N^2]}(H_D(k)).
\]
Hence, we conclude that
\[
 \operatorname{Ch}(D;b_N^2)-\operatorname{Ch}(D;a_N^2)
 =c(P_{\mathrm{orig}})=c(P)=0.
\]

{\color{blue}Step 2.} To show that the common value of $\operatorname{Ch}(D;b_N^2)$ and $\operatorname{Ch}(D;a_N^2)$ is equal to the prescribed Chern number $c_{S}$, we follow a homotopy argument by continuously deforming the background medium $C$ to $S_{N}$. For $t\in [0,1]$, we define
\[
 C_t:=(1-t)C+tS_{N},\quad \text{and}\quad
 D_{t}:=\left\{
 \begin{aligned}
 &S_{N},\quad \text{on }\Sigma_w, \\
 &C_{t},\quad \text{off }\Sigma_w .
 \end{aligned}
 \right.
\]
The counting and band-motion argument in Section~\ref{sec:layer-insertion} gives a constant $h_0>0$, independent of the background, such that for every $t\in[0,1]$ one can choose $\omega_t\in NG_1$ with
\begin{equation}
 \inf_{k\in\mathcal B}\dist(\omega_t,\spec(M_{D_t}(k)))\ge h_0/N.
\end{equation}
Choosing $N$ large enough and writing $c(t):=\operatorname{Ch}(D_t;\omega_t^2)$, the result of Step~1 shows that $c(t)$ does not depend on the choice of $\omega_t$. We claim that $c(t)$ is constant and that $c(1)=c_S$. Indeed, this is justified by noting the following two points:
\begin{itemize}
    \item[(i)] The integer $c(t)$ is locally constant. To see this, we fix a reference time $t_0$ and a frequency $\omega_{t_0}$. With $L:=(\beta-\alpha)/\alpha$, the sesquilinear forms associated with $H_{D_t}$ satisfy
\[
 |q_{D_t,k}[u]-q_{D_{t_0},k}[u]|
 \le L|t-t_0|q_{D_{t_0},k}[u],\qquad u\in H^1(\T^2),
\]
uniformly in $k$. The min--max principle therefore gives, for $\varepsilon:=L|t-t_0|<1$,
\[
 (1-\varepsilon)\lambda_j(D_{t_0};k)
 \le\lambda_j(D_t;k)
 \le(1+\varepsilon)\lambda_j(D_{t_0};k).
\]
Hence, for sufficiently small $|t-t_0|$, $\omega_{t_0}^2$ remains in a common gap of $H_{D_t}(k)$, with frequency separation at least $h_0/(2N)$. The corresponding occupied projections vary continuously in operator norm, uniformly in $k$, and therefore define isomorphic bundles. By Step~1, this proves $c(t)=c(t_0)$.

    \item[(ii)] The gap Chern number at $t=1$ equals the prescribed number $c_{S}$. In fact, at $t=1$, the medium $D_{1}=S_{N}$; hence it suffices to show the rescaled operator $H_{S_N}$ has the same gap Chern number as $H_S$ in the corresponding rescaled energy. Indeed, the folded occupied bundle is the direct image of the seed bundle under $\theta\mapsto N\theta$ on the phase torus. Locally, its summands have phases $\theta_l=(k+2\pi l)/N$, $l\in\{0,\ldots,N-1\}^2$. The trace curvature of the scalar occupied bundle is the sum of the seed curvatures on these branches, with the derivative factor $N^{-2}$. Changing variables tiles the seed torus exactly once, so its integral is unchanged.
\end{itemize}
This completes the proof.
\end{proof}

\begin{proof}[Proof of Proposition~\ref{lem:chern-persistence}]
The persistence of gap under small $L^1$ perturbation has been proved in Lemma~\ref{lem:persistence}. Moreover, the resolvent convergence shown in Lemma~\ref{lem:persistence} indicates that the spectral projection of the occupied bands (below the gap) is continuous in norm operator topology. For a sufficiently small perturbation, the occupied projections are uniformly less than one in operator norm and hence define isomorphic Bloch bundles. Their Chern numbers are therefore equal.
\end{proof}

\begin{proof}[Proof of Proposition~\ref{prop:chern-sequence}]
As in Section \ref{sec:infinite-gap-construction}, the induction starts by a constant real-valued medium $C_0\in\mathcal X^{\mathrm R}_{\alpha,\beta}$, and we set the initial width of layer $w_0=1/8$. At step $j$, we use the seed $S=S_j$ in the insertion formula \eqref{eq:insert} with the relevant constants chosen as in Proposition~\ref{prop:chern-insert}. Then, as in Section~5, at each step $j$, we choose sufficiently large $N_j$ so that 1) $w_j<w_{j-1}/2$, 2)
\begin{equation} \label{eq:Chern_adjacent_difference}
\norm{C_j-C_{j-1}}_1
 <2^{-j}\min(1,r_1,\ldots,r_{j-1}),
\end{equation}
and 3) $N_j^2(\omega_{1,-}^{(j)})^2>j$, where
$G_1^{(j)}=(\omega_{1,-}^{(j)},\omega_{1,+}^{(j)})$
is the inner frequency window chosen for the seed $S_j$. To ensure that the selected intervals belong to distinct gap components, let $\mu_n$ denote the ordered eigenvalues of the periodic Laplacian $-\Delta$ on $\mathbb T^2$. For $j\ge2$, before choosing $N_j$, we choose an index $p_j$ such that $\alpha\mu_{p_j}>\sup I_{j-1}$ and impose the additional condition
\[
 N_j^2(\omega_{1,-}^{(j)})^2>\beta\mu_{p_j}.
\]
Note that all these conditions can be satisfied by increasing $N_j$. After fixing the choice of $N_j$ and the resulting medium $C_j$, we choose a compact interval $I_j$ of positive length inside both the new complete gap supplied by Proposition~\ref{prop:chern-insert} and the energy window $\bigl(N_j^2(\omega_{1,-}^{(j)})^2,\,N_j^2(\omega_{1,+}^{(j)})^2\bigr)$. We then choose the persistence radius $r_j$ for $(C_j,I_j)$ from Proposition~\ref{lem:chern-persistence}. Consequently, the estimate \eqref{eq:Chern_adjacent_difference} gives an $L^1$-limit matrix field $A_\infty$ satisfying $\norm{A_\infty-C_j}_1<r_j$ for every $j$. Thus, every $I_j$ remains inside a complete gap of
$H_{A_\infty}$ and retains its prescribed Chern number $m_j$.

It remains to verify that these intervals $I_j$ belong to distinct gap components. By the common ellipticity bounds and the min-max principle, for every $j\ge2$,
\[
 \sup I_{j-1} <\alpha\mu_{p_j} \leq\lambda_{p_j}(A_\infty;0) \leq\beta\mu_{p_j} <\inf I_j.
\]
Thus, a point of $\sigma(H_{A_\infty})$ separates each pair $I_{j-1},I_j$. Let $G_j$ be the gap component containing $I_j$. The components $G_j$ are therefore distinct and increasingly ordered. Moreover, for $j\ge2$,
\[
 \inf G_j\geq\lambda_{p_j}(A_\infty;0) >\sup I_{j-1}>j-1.
\]
Hence, $\inf G_j\to\infty$, and $\operatorname{Ch}(A_\infty;G_j)=m_j$ for every $j$.

Finally, the estimates \eqref{eq:limiting_medium_bounded} follows by noting that each seed we apply in the induction satisfies the same bounds, which pass to the limit along an almost-everywhere convergent subsequence.
\end{proof}

\subsection{Topological seeds with fixed material bounds: proof of Proposition \ref{prop:uniform-seeds}}\label{sec:uniform-seeds}
\clubpenalty=10000 \widowpenalty=10000

\subsubsection{Roadmap of the proof} \label{sec:chern_roadmap}
The proof of Proposition~\ref{prop:uniform-seeds} is based on a high-frequency homogenization approach. First, we will construct a periodic magnetic operator with the prescribed occupied Chern number $m\in\Z$; see Lemma \ref{lem:seed-magnetic-target}. Next, we show that, by threading the magnetic and scalar potential near a non-degenerate band edge of a fixed real medium (supplied by Lemma \ref{lem:seed-controllable-edge}) using an appropriate scaling, the target operator in Lemma \ref{lem:seed-magnetic-target} is realized in the high-frequency limit. Hence, by showing that the limit is indeed uniform using the Schur complement, we conclude the existence of a single medium with prescribed gap Chern number, as claimed in Proposition~\ref{prop:uniform-seeds}. See the ansatz \eqref{eq:short-seed-coefficient}, Lemma \ref{lem:seed-transfer} and its proof for more details. We note that the idea of using band-edge homogenization to produce an effective Hamiltonian has a long history; see e.g. \cite{PanatiSpohnTeufel2003,shapiro2022tight}.

Throughout this subsection, we write $D:=-\ii\nabla$ and $J=\left(\begin{smallmatrix}0&-1\\1&0\end{smallmatrix}\right)$. For a projection bundle $P$, we write $c(P)$ for its first Chern number, with the convention in \eqref{eq:chern-number}; in particular, $c(\one_{(-\infty,E)}(H_A(\cdot)))=\operatorname{Ch}(A;E)$. We also write $Q_{\rm cell}:=(0,1)^2$.

We now state the three ingredients used for the proof as mentioned above; their proofs are postponed later. First, the existence of a target magnetic operator with prescribed gap Chern number and arbitrary kinetic matrix $M$ is summarized as follows. In fact, this kinetic matrix will be fixed later by the dispersion near the concerned band edge.

\begin{lemma} \label{lem:seed-magnetic-target}
For every real symmetric positive definite $2\times2$ matrix $M$ and $m\in\Z$, there exist smooth real-valued $\Z^2$-periodic fields $b:\R^2\to\R^2$ and $V:\R^2\to\R$ such that the fiber operator
\[
 K(k)=(D+k-b)^{\mathsf T}M(D+k-b)+V
\]
have a common spectral gap containing zero for all $k\in \mathcal{B}$, and
$c(\one_{(-\infty,0)}(K))=m$. Moreover, $b$ and $V$ can be chosen as trigonometric polynomials.
\end{lemma}

The next result provides a real medium with a non-degenerate band edge and sufficient information on the local Bloch eigenmodes such that the local band structure is controllable when applying an imaginary perturbation:
\begin{lemma} \label{lem:seed-controllable-edge}
There exist constants $0<s_-< s_+<\infty$, a smooth real-valued symmetric $\Z^2$-periodic matrix field $S$, two smooth real-valued periodic functions $\beta_1,\beta_2$, and $\lambda_0>0$ with the following properties:
\begin{itemize}
\item[(i)] $s_-I_2\leq S\leq s_+I_2$ holds, and $\lambda_0$ is the upper edge of a complete gap of $H_S$. Moreover, the edge is attained only at the momentum $k_0=(\pi,0)$, modulo the reciprocal lattice vectors.
\item[(ii)] The local band structure near $\lambda_0$ admits the asymptotic expansion
\[
 \lambda(k_0+q)=\lambda_0+q^{\mathsf T}Mq+O(|q|^3),\quad\text{with}\quad M>0.
\]
Moreover, there exist $\kappa_0,\delta_0>0$ such that $\dist(\lambda_0,\spec(H_S(k)))>\delta_0$ whenever $\dist_{\mathcal B}(k,k_0)\geq\kappa_0$, where $\dist_{\mathcal B}(k,k_0):=\inf_{\ell\in\Z^2}|k-k_0+2\pi\ell|$.
\item[(iii)] For small $s$, let $\lambda_{S+s\ii\beta_\ell J}(k)$ be the perturbed band near $\lambda_0$ of $H_{S+s\ii\beta_\ell J}(k)$. It holds that
\[
 \mu_j(\beta_\ell):=
 \left.\partial_{q_j}\partial_s
 \lambda_{S+s\ii\beta_\ell J}(k_0+q)\right|_{q=s=0}
 =\delta_{j\ell},\quad j,\ell\in\{1,2\}.
\]
\end{itemize}
\end{lemma}
From now on, we fix a medium $S$ and periodic functions supplied by Lemma~\ref{lem:seed-controllable-edge}. Let $\phi$ be a normalized real $k_0$-quasi-periodic eigenfunction of $H_0:=D^{\mathsf T}SD$ at $\lambda_0$; equivalently, $e^{-\ii k_0\cdot x}\phi$ is an eigenfunction of the periodic fiber $H_S(k_0)$. We define the reduced inverse
$R:=(H_0-\lambda_0)^{-1}_{\phi^\perp}$, and
\begin{equation} \label{eq:V_C_def}
V_\ell:=D^{\mathsf T}\ii\beta_\ell JD,
 \quad C_{\ell s}:=\langle V_\ell\phi,RV_s\phi\rangle.
\end{equation}
Note that the matrix $C=(C_{\ell s})$ is real symmetric but not necessarily positive-definite. We also fix the real trigonometric polynomial fields $b,V$ supplied by Lemma \ref{lem:seed-magnetic-target}.

To pass from the target operator in Lemma \ref{lem:seed-magnetic-target} to a desired material coefficient $A$ as stated in Proposition \ref{prop:uniform-seeds}, we will study the high-frequency limit of the following ansatz
\begin{equation} \label{eq:short-seed-coefficient}
A_N(X):=S_N(X)+\ii N^{-1}\sum_{\ell=1}^2
 B_\ell(X)\beta_\ell(NX)J+N^{-2}v(X)S(NX)
\end{equation}
with
\begin{equation} \label{eq:short-seed-coefficient_details}
S_N(X)=S(NX),\quad B:=-2Mb,\quad
 v=\lambda_0^{-1}\big(V+b^{\mathsf T}Mb+\sum_{\ell,s}C_{\ell s}B_\ell B_s\big) .
\end{equation}
We take $N$ to be even so that the Bloch mode at $k_0=(\pi,0)$ is periodic after rescaling. The following lemma justifies the claim that the operator in the ansatz \eqref{eq:short-seed-coefficient} inherits the gap and the prescribed Chern number of the target magnetic operator, when $N$ is sufficiently large.
\begin{lemma} \label{lem:seed-transfer}
Let $K(k)=(D+k-b)^{\mathsf T}M(D+k-b)+V$ be the magnetic target operator in Lemma \ref{lem:seed-magnetic-target} and let the coefficient matrix $A_N$ be defined by \eqref{eq:short-seed-coefficient}. There exists $\gamma>0$ such that, for every sufficiently large even $N$,
\[
 (N^2\lambda_0-\gamma,N^2\lambda_0+\gamma)
       \cap\spec(H_{A_N})=\varnothing,
 \quad \operatorname{Ch}(A_N;N^2\lambda_0)=m.
\]
In particular, this is a complete positive gap of $H_{A_N}$ for all sufficiently large $N$.
\end{lemma}

With these preparations, the proof of Proposition~\ref{prop:uniform-seeds} follows directly:

\begin{proof}[Proof of Proposition~\ref{prop:uniform-seeds}]
By our choice of medium/fields, as stated following Lemma \ref{lem:seed-controllable-edge}, we know by Lemma~\ref{lem:seed-transfer} that $H_{A_{N}}$ with $A_{N}$ defined in \eqref{eq:short-seed-coefficient} has a complete positive gap with the prescribed occupied Chern number $m$ for every sufficiently large even $N$. It then suffices to verify that $A_N$ satisfies the bounds \eqref{eq:Chern_seed_bound} for sufficiently large $N$. In fact, since $\|J\|=1$, we have
\begin{equation} \label{eq:A_N_estimate}
\|A_N-S(N\cdot)\|_\infty
 \leq N^{-1}\sum_{\ell=1}^2 \|B_\ell\|_\infty\|\beta_\ell\|_\infty +N^{-2}\|v\|_\infty\|S\|_\infty
 \to 0.
\end{equation}
Choose constants $0<\alpha<\min\{1,s_-\}$ and $\beta>\max\{2,s_+\}$, which depend only on the fixed matrix $S$, and are independent of $m$ and $\eta$. Since $S$ is real, \eqref{eq:A_N_estimate} allows us to select a sufficiently large even $N$ such that
\[
\|A_N-S(N\cdot)\|_\infty \leq\min\{s_--\alpha,\beta-s_+,\eta\}.
\]
Consequently,
\[
\alpha I_2\leq A_N\leq\beta I_2,\quad \|\operatorname{Im}A_N\|_\infty\leq\eta,
\]
which proves \eqref{eq:Chern_seed_bound} hence completes the proof.
\end{proof}

\subsubsection{Proofs of the auxiliary lemmas} \label{sec:chern_auxiliary_proof}

\paragraph{Proof of Lemma~\ref{lem:seed-magnetic-target}}

Fix the kinetic matrix $M$. To construct the target magnetic operator with Chern number $m=r>0$, we start with the Landau operators, whose gap Chern numbers are direct to calculate. Then we remove a small disk from the torus; consequently, thanks to the topological triviality of the punctured domain, we are able to select a smooth section of the complex bundle, which helps to reduce the magnetic boundary condition (of the Landau operator) to the ordinary periodic condition, as required in Lemma~\ref{lem:seed-magnetic-target}. Finally, we fill in the removed disk by a finite positive potential, with the final result being the desired target magnetic operator. Importantly, we will clearly show that the Chern number is preserved in each step.

{\color{blue}Step 1. The Landau operator with kinetic matrix $M$.}

Let $B_0=2\pi$, $D=-\ii\nabla_x$, and $b_0(x)=(0,B_0x_1)$. The Landau operator is imposed on the following domain equipped with the magnetic boundary condition
\begin{equation} \label{eq:Landau_domain}
\mathcal V_L=\{v\in H^1(Q_{\rm cell}):
 v(1,x_2)=e^{\ii B_0x_2}v(0,x_2),\,
 v(x_1,1)=v(x_1,0)\}.
\end{equation}
Here, the boundary identifications are compatible because $e^{\ii B_0}=1$. In particular, this identification defines the magnetic line bundle $L$ over $\T^2$: for $n=(n_1,n_2)\in\Z^2$, we define the transport operator
\[
 T_n(x,z):=(x+n,e^{\ii B_0n_1x_2}z),
\]
which satisfies $T_nT_m=T_{n+m}$ since $B_0=2\pi$. Then the Hermitian line bundle $L$ is defined as
\begin{equation} \label{eq:torus_line_bundle}
L:=(\R^2\times\C)/\Z^2,
 \quad \pi_L([x,z])=x+\Z^2 .
\end{equation}
Its smooth sections are represented by smooth functions $v$ satisfying
\[
 v(x+n)=e^{\ii B_0n_1x_2}v(x).
\]
In particular, their restrictions to $Q_{\rm cell}$ satisfy precisely the magnetic boundary conditions defining $\mathcal V_L$. WIth these notions, the Landau operator (with kinetic matrix $M$) is defined as follows: for each $k$, let $\mathcal L(k)$ be the self-adjoint operator on $L^2(Q_{\rm cell})$ associated with the form
\[
 \ell_k[v]:=\int_{Q_{\rm cell}}((D+k-b_0)v)^*M((D+k-b_0)v)\,\mathrm dx, \quad v\in\mathcal V_L.
\]
It admits the differential expression $\mathcal L(k)=(D+k-b_0)^{\mathsf T}M(D+k-b_0)$, and has compact resolvent. The spectrum of $\mathcal L(k)$ is given by the Landau levels (see e.g. \cite[Eq. (7.12)]{SakamotoTanimura2003})
\[
 \sigma(\mathcal L(k))
 =\bigl\{\Lambda_j:=B_0\sqrt{\det M}\,(2j+1)\bigr\}_{j\geq0}.
\]
Each eigenvalue is simple and independent of $k$. The normalized eigenfunctions associated with $\Lambda_j$, denoted by $v_{j,k}$, form a complete orthonormal basis of $L^2(Q_{\rm cell})$. To choose these eigenfunctions consistently in $k$, we write $D_t=-\ii\partial_t$, $\Pi=(D_t,-B_0t)^{\mathsf T}$, and let $\mathfrak h_M$ be the self-adjoint realization of $\Pi^{\mathsf T}M\Pi$ on $L^2(\R)$. Define the unitary map
\[
 (\mathcal U_kh)(x)
 :=\sum_{n\in\Z}
 e^{2\pi\ii n x_2-\ii k_1(x_1-n)}
 h(x_1-n-k_2/B_0).
\]
In particular, $\mathcal U_kh$ satisfies the magnetic boundary conditions defining $\mathcal V_L$, and direct differentiation gives
\[
 \mathcal L(k)\mathcal U_kh=\mathcal U_k\mathfrak h_Mh,
 \quad h\in\mathcal S(\R).
\]
For a normalized eigenbasis $\{h_j\}_{j\geq0}$ of $\mathfrak h_M$, with $\mathfrak h_Mh_j=\Lambda_jh_j$, we therefore choose $v_{j,k}:=\mathcal U_kh_j$. The Gaussian decay of $h_j$ and its derivatives justify the termwise differentiation in $k$ of the series defining $v_{j,k}$. Writing $\tau_\ell(x):=e^{-2\pi\ii x_\ell}$, the explicit formula for $\mathcal U_k$ gives
\[
 v_{j,k+2\pi e_1}=\tau_1v_{j,k},\quad
 v_{j,k+2\pi e_2}=e^{-\ii k_1}\tau_2v_{j,k}.
\]
Hence, for the connection form $\mathcal A_j=\sum_{\ell=1}^2\langle\partial_{k_\ell}v_{j,k},v_{j,k}\rangle_{L^2(Q_{\rm cell})}\,\mathrm dk_\ell$, we have $\mathcal A_j(k+2\pi e_1)=\mathcal A_j(k)$ and $\mathcal A_j(k+2\pi e_2)=\mathcal A_j(k)-\ii\,\mathrm dk_1$. By Stokes' formula on $\mathcal B$,
\[
 \frac{1}{2\pi\ii}\int_{\mathcal B}d_k\mathcal A_j=1.
\]
Choosing an energy level $E\in(\Lambda_{r-1},\Lambda_r)$ and setting $P(k):=\one_{(-\infty,E)}(\mathcal L(k))$, additivity then shows that $P$ has rank $r$ and Chern number $c(P)=r$.

{\color{blue}Step 2. Removing a disk.}

In this step, we show that removing a small disk from the domain of Landau operator $\mathcal{L}$ by placing the Dirichlet boundary condition on the removed surface will not change the gap Chern number of the resulting operator. In fact, as we will see later in Step 3, removing such a disk gives the convenience of modifying the magnetic boundary condition.

Fix $x_*\in Q_{\rm cell}$. For sufficiently small $\varepsilon>0$, let $\Omega_\varepsilon:=\T^2\setminus \overline{B(x_*,\varepsilon)}$. Define $\mathcal L_\varepsilon(k)$ by restricting $\ell_k$ to functions in $\mathcal V_L$ that vanish almost everywhere on $\overline{B(x_*,\varepsilon)}$, regarded as functions on $\Omega_\varepsilon$. In particular, these functions still satisfy the magnetic boundary conditions in \eqref{eq:Landau_domain}. Writing the eigenvalues of $\mathcal L_\varepsilon(k)$ as $\lambda_j^\varepsilon(k)$, indexed from zero, one knows by the min-max principle that
\begin{equation} \label{eq:removing_disk_proof_1}
    \lambda_j^\varepsilon(k)\ge\Lambda_j ,\quad j\leq r.
\end{equation}
We also require a reverse of \eqref{eq:removing_disk_proof_1} (for $j<r$) to study the gap of the punctured operator $\mathcal L_\varepsilon(k)$. To this end, we introduce the radial cutoff
\[
 \eta_\varepsilon(x):=
 \begin{cases}
 0,&|x-x_*|\le\varepsilon,\\
 \displaystyle\frac{\log(|x-x_*|/\varepsilon)}
 {\log(\varepsilon^{-1/2})},
     &\varepsilon<|x-x_*|<\sqrt\varepsilon,\\
 1,&|x-x_*|\ge\sqrt\varepsilon,
 \end{cases}
\]
which satisfies 1) $\int|\nabla\eta_\varepsilon|^2=4\pi/|\log\varepsilon|$, and 2) $1-\eta_\varepsilon$ is supported in a disk of area $\pi\varepsilon$. Now we define the trial space built on the first $j$ vectors $v_{j,k}$ ($0\leq j\leq r-1$):
\[
E_{j,\varepsilon}(k)
=\operatorname{span}\{\eta_\varepsilon v_{0,k},\ldots,
                      \eta_\varepsilon v_{j,k}\}.
\]
In particular, in the level of quadratic forms, the punctured operator $\mathcal L_\varepsilon(k)$ equals $\mathcal L(k)$ on $E_{j,\varepsilon}(k)$. Moreover, we claim that
\begin{equation} \label{eq:removing_disk_proof_2}
\|T-\operatorname{diag}(\Lambda_0,\ldots,\Lambda_{r-1})\|\le\frac{C_{r,M}}{|\log\varepsilon|},\quad 
\|G-I\|\le C_{r,M}\varepsilon ,
\end{equation}
with the matrices defined by
\begin{equation*}
T_{ab}:=\ell_k[\eta_\varepsilon v_{a,k},\eta_\varepsilon v_{b,k}],\quad 
G_{ab}:=\langle\eta_\varepsilon v_{a,k}, \eta_\varepsilon v_{b,k}\rangle \quad (0\le a,b<r).
\end{equation*}
Indeed, \eqref{eq:removing_disk_proof_2} follows by the $H^1$ estimate of the cutoff function $\eta_{\epsilon}$ and the following identity, which is proved by direct calculation using the eigenvalue equations satisfied by $v_{a,k}$:
\[
 T_{ab}=\frac{\Lambda_a+\Lambda_b}{2}G_{ab}
 +\int_{Q_{\rm cell}}(\nabla\eta_\varepsilon)^{\mathsf T}
 M\nabla\eta_\varepsilon\,
 v_{a,k}\overline{v_{b,k}}\,\mathrm dx.
\]
Consequently, \eqref{eq:removing_disk_proof_2} shows that
\[
\sup_{0\ne w\in E_{j,\varepsilon}(k)}
\frac{\ell_k[w]}{\|w\|^2}
\le \Lambda_j+\frac{C_{r,M}}{|\log\varepsilon|}.
\]
Then, recalling the identification of $\mathcal L_\varepsilon(k)$ and $\mathcal L(k)$ on the trial space $E_{j,\varepsilon}(k)$, we know by the min-max principle that
\begin{equation} \label{eq:removing_disk_proof_3}
0\le\lambda_j^\varepsilon(k)-\Lambda_j \le\frac{C_{r,M}}{|\log\varepsilon|},\quad j<r ,
\end{equation}
which is uniformly in $k$. Together with
\eqref{eq:removing_disk_proof_1} applied for $j=r$, we conclude that the punctured operator $\mathcal L_\varepsilon(k)$ has a spectral gap about $E$ (E is fixed at the end of Step 1), which is uniform in $k$, and has exactly $r$ eigenvalues below $E$ for small $\varepsilon$.

To discuss the gap Chern number, we write $P_{\varepsilon}(k):=\one_{(-\infty,E)}(\mathcal L_{\varepsilon}(k))$, which is extended by zero to $L^2(Q_{\rm cell})$. Note that
\[
 (\Lambda_r-\Lambda_{r-1})
 \operatorname{tr}((I-P)P_\varepsilon)
 \le\sum_{j<r}(\lambda_j^\varepsilon-\Lambda_j),
\]
which is proved by expanding the trace in the complete Landau eigenbasis. Since both projections $P$ and $P_\varepsilon$ have rank $r$ as we have proved, we then obtain
\[
 \|P-P_\varepsilon\|^2
 \le\|P-P_\varepsilon\|_{\mathrm{HS}}^2
 =2\operatorname{tr}((I-P)P_\varepsilon)\to 0, 
\]
where the estimate is uniform in $k$. Hence, for small
$\varepsilon$, the restriction $P_\varepsilon(k)|_{\operatorname{Ran}P(k)}$ is a continuous bundle isomorphism, which leads to the identification of Chern number, i.e., $c(P_\varepsilon)=r$.

{\color{blue}Step 3. Replacing the magnetic boundary conditions by a smooth section.}

Fix $\varepsilon>0$ such that $\mathcal{L}_{\varepsilon}$ is gapped at $E$ and has the same Chern number as $\mathcal{L}$. Remarkably, since the punctured torus has the cohomology group $H^2(\Omega_{\varepsilon/2};\Z)=0$, we know that the restriction of the complex line bundle $L$ (defined in \eqref{eq:torus_line_bundle}) to $\Omega_{\varepsilon/2}$ is trivial. Consequently, we can choose a smooth section $\sigma$ of unit length on $\Omega_{\varepsilon/2}$, which satisfies the magnetic boundary condition
\[
 \sigma(x+e_1)=e^{\ii B_0x_2}\sigma(x),\quad
 \sigma(x+e_2)=\sigma(x) .
\]
Based on this section, we define the following vector field on the punctured torus $\Omega_{\varepsilon/2}$
\[
 b_\Omega:=b_0-\overline\sigma D\sigma.
\]
It is then clear that $b_\Omega$ is smooth, real-valued and satisfies the periodic boundary condition, as the gauge increment of $b_0$ under $x\mapsto x+e_1$ is canceled by that of $\overline\sigma D\sigma$. Importantly, the section $\sigma$ induces a unitary transform whose conjugation keeps the form of the Landau operator: letting $T_\sigma f:=\sigma f$, direct
differentiation gives
\[
 T_\sigma^*(D+k-b_0)T_\sigma=D+k-b_\Omega.
\]
Thus $T_\sigma$ is a unitary map (recall that $\sigma$ has unit length) from scalar periodic functions on $\Omega_\varepsilon$ with zero trace on its boundary to sections with the magnetic boundary conditions and the same Dirichlet condition. Moreover, $T_\sigma$ is independent of $k$ and
$T_\sigma\tau_\ell=\tau_\ell T_\sigma$.

We now use the unitary transform $T_\sigma$ to produce a periodic magnetic operator, which differs from the target operator in Lemma~\ref{lem:seed-magnetic-target} by the removed disk (it will be filled in the next step).

Choose a smooth cutoff $\rho$ supported in $\Omega_{\varepsilon/2}$ and equal to one on a neighborhood of $\overline{\Omega_\varepsilon}$. Extend $b=\rho b_\Omega$ by zero across the smaller disk $B(x_*,\varepsilon/2)$. Then $b\in C^\infty(\T^2;\R^2)$ and $b=b_\Omega$ on $\Omega_\varepsilon$. Let $H_\Omega^b(k)$ be the scalar operator on $L^2(\Omega_\varepsilon)$ with differential expression
\begin{equation*}
H_\Omega^b(k):=(D+k-b)^{\mathsf T}M(D+k-b),
\end{equation*}
which is equipped with periodic outer boundary conditions and the Dirichlet condition on the removed disk. One can check that $H_\Omega^b(k)$ is unitarily equivalent to the punctured operator $\mathcal L_\varepsilon(k)$:
\[
 H_\Omega^b(k)=T_\sigma^*\mathcal L_\varepsilon(k)T_\sigma .
\]
Therefore, the projector $\Pi_D(k)=\one_{(-\infty,E)}(H_\Omega^b(k))$ inherits the rank $r$ and Chern number $r$.

{\color{blue}Step 4. Filling the hole by a large finite barrier potential.}

To move back to the torus geometry, we fill in the hole by a large barrier potential. Choose a non-negative function $\chi\in C^\infty(\T^2)$, which is strictly positive on the interior of $\overline{B(x_*,\varepsilon)}$, and $\chi=0$ elsewhere. On $L^2(\T^2)$ define the family of operator indexed by $t\geq 0$ through its quadratic form on $H^1(\T^2)$:
\[
 S_t(k)=(D+k-b)^{\mathsf T}M(D+k-b)+t\cdot \chi.
\]
Let $J_\Omega:L^2(\Omega_\varepsilon)\to L^2(\T^2)$ be the zero extension. We claim that, as $t\to\infty$,
\begin{equation} \label{eq:removing_disk_proof_4}
\sup_{k\in\mathcal B}\left\|
 (S_t(k)+1)^{-1}
 -J_\Omega(H_\Omega^b(k)+1)^{-1}J_\Omega^*
 \right\|\to 0.
\end{equation}
To prove this, let $t_n\to\infty$, $k_n\to k$, and let $f_n$ be $L^2$-bounded functions, with $f_n\rightharpoonup f$ after passing to a subsequence if necessary. Define $u_n=(S_{t_n}(k_n)+1)^{-1}f_n$. Note the form inequality
\[
 \langle u,S_t(k)u\rangle
 \ge\tfrac12\lambda_{\min}(M)\|\nabla u\|^2
       -C\|u\|^2+t\int\chi|u|^2,\quad \forall  u\in H^1(\mathbb T^2;\mathbb C).
\]
Applying it to $u_n$ together with the fact that $u_n$ has a uniform $H^1$ bound, thanks to the boundedness of resolvent $(S_{t_n}(k_n)+1)^{-1}$, we know that $t_n\int\chi|u_n|^2$ are uniformly bounded. Consequently, passing to a subsequence for which $u_n$ converges strongly in $L^2$ and weakly in $H^1$ by the Rellich compactness, we know that the limiting function $u$ vanishes on $\overline{B(x_*,\varepsilon)}$, by which it follows that
\[
 u=J_\Omega(H_\Omega^b(k)+1)^{-1}J_\Omega^*f.
\]
This proves that the difference in \eqref{eq:removing_disk_proof_4} converges in strong operator topology. The uniformity in both $\|f\|$ and $k\in\mathcal{B}$ is clear from the argument, which then concludes the norm convergence in \eqref{eq:removing_disk_proof_4}.


By \eqref{eq:removing_disk_proof_4}, we know that $S_t(k)$ has a common gap about $E$ for all sufficiently large $t$, and, by the Riesz formula,
\begin{equation}
\label{eq:removing_disk_proof_6}
 \sup_k\left\|
 \one_{(-\infty,E)}(S_t(k))
 -J_\Omega\Pi_D(k)J_\Omega^*
 \right\|\longrightarrow0.
\end{equation}
Hence, for large $t$, these two projections are isomorphic. Fix such a $t$ and set $V:=t\chi-E$. Then we finally obtain the target operator
\begin{equation} \label{eq:removing_disk_proof_5}
K(k):=(D+k-b)^{\mathsf T}M(D+k-b)+V=S_t(k)-E .
\end{equation}
By the results obtained from Steps 1-4, we know that the operator $K$ has a complete gap about zero and occupied Chern number $r>0$.

If the prescribed Chern number is negative, i.e., $m=-r<0$, we replace $b$ by $-b$ and retain $V$. In particular, writing $\mathcal C$ for complex conjugation, we have
\[
 K_{-b,V}(k)=\mathcal C K_{b,V}(-k)\mathcal C.
\]
Hence, using the formula \eqref{eq:chern-number}, one sees the Chern number of $K_{-b,V}$ at zero changes
sign, giving the result $-r$. If the prescribed Chern number is trivial, i.e., $m=0$, we take $b=0$ and $V=1$. This produces the desired target operator for any integer $m$.

Finally, the claim about the trigonometric-polynomial choice of $b$ and $V$ comes from the trigonometric approximation and the stability of gaps and Chern numbers. To be precise, we choose real trigonometric polynomials $b_n,V_n$ with $b_n\to b$ in $C^1$ and $V_n\to V$ in $C^0$. Denoting the corresponding sesquilinear forms as $q_{n,k}$ and $q_k$, we know by direct calculation that
\[
 |q_{n,k}[u]-q_k[u]|
 \le\epsilon_n\bigl(\|\nabla u\|^2+\|u\|^2\bigr),
 \quad\text{as }\, \epsilon_n\to0,
\]
uniformly in $k\in\mathcal B$. Then, the uniform ellipticity and this form estimate imply norm-resolvent convergence of the corresponding magnetic operators (which take the form as in \eqref{eq:removing_disk_proof_5}), uniformly in $k\in\mathcal B$. Again, using the Riesz formula as in \eqref{eq:removing_disk_proof_6}, we know that the zero gap, as well as the Chern number, persists for sufficiently large $n$. This completes the proof.

\paragraph{Proof of Lemma~\ref{lem:seed-controllable-edge}}

To prove the lemma, we first construct a smooth one-dimensional coefficient with a sufficiently large first gap. We use this coefficient to obtain a unique simple upper gap edge in two dimensions. We then compute the two mixed responses against the external perturbation and choose the diagonal perturbations by inverting the corresponding Gram matrix. 

{\color{blue}Step 1. A smooth one-dimensional coefficient.}

The idea of constructing the one-dimensional gapped medium is simple: we start from a constant medium and insert a large periodic barrier. Specifically, let $a^{\text{per},(0)}$ be the $1$-periodic extension of the piecewise constant function
\begin{equation} \label{eq:1d_potential_proof_1}
a^{(0)}(x):=
 \begin{cases}
  1,&0<x<1/16\ \text{or}\ 15/16<x<1,\\
  128,&1/16<x<15/16.
 \end{cases}
\end{equation}
Note that the constants are not chosen judiciously, but nevertheless sufficient to prove the lemma. Denote the $j$\textsuperscript{th} Dirichlet and Neumann eigenvalues of $-(a^{(0)}v')'$ on $(0,1)$, written as $T^{D,(0)}$ and $T^{N,(0)}$, by $\lambda_j^D(a^{(0)})$ and $\lambda_j^N(a^{(0)})$, respectively. We apply the min-max argument as in Lemma \ref{lem:seed} to investigate the first gap of the periodic operator $T^{\text{per},(0)}v:=-(a^{\text{per}}v')'$. Consider the trial function $v(x):=\min\{16x,1,16(1-x)\}$, which satisfies
\[
 \int_0^1a^{(0)}|v'|^2=32,\quad
 \int_0^1|v|^2\geq\frac78.
\]
Hence,
\begin{equation} \label{eq:1d_potential_proof_2}
\lambda_1^D(a^{(0)})\leq\frac{\int_0^1a^{(0)}|v'|^2}{\int_0^1|v|^2}<37 .
\end{equation}
Next, we consider $\lambda_2^N(a^{(0)})$. First, note that the reflection operator about $1/2$ commutes with $T^{N,(0)}$. On the other hand, by Sturm--Liouville theory, the second eigenvalue of $T^{N,(0)}$ is simple, and its eigenfunction $v_{2}^{N}$ has unique zero inside $(0,1)$ (cf. \cite[Theorem~4.3.1]{Zettl2005}); hence $v_{2}^{N}$ vanishes only at $1/2$. If $(v_{2}^{N})^{\prime}(1/2)=0$, the unique extension of solution to ODE implies $v_{2}^{N}$ vanishes identically. Thus, we conclude that $v_{2}^{N}$ is an odd function about $1/2$. Then we claim that
\begin{equation} \label{eq:1d_potential_proof_3}
\lambda=\lambda_2^N(a^{(0)})>500.
\end{equation}
In fact, solving $-\left(a^{(0)}(v_{2}^{N})^{\prime}\right)^{\prime}=\lambda v_{2}^{N}$ on $(0,1/16)$ and $(1/16,1/2)$, with the boundary conditions $(v_{2}^{N})^{\prime}(0)=0$ and $v_{2}^{N}(1/2)=0$, and then matching both values $v_{2}^{N}$ and $(v_{2}^{N})^{\prime}$ at $x=1/16$ give the identity
\[
 \tan\!\left(\frac{\sqrt\lambda}{16}\right)
 =\sqrt{128}\cot\!\left(
       \frac{7\sqrt\lambda}{16\sqrt{128}}\right).
\]
If $0<\lambda\leq500$, one can check that the left-hand side is less than $6$ while the right side is greater than $6$, which leads to a contradiction. This proves \eqref{eq:1d_potential_proof_3}.

By \eqref{eq:1d_potential_proof_2} and \eqref{eq:1d_potential_proof_3}, together with the min-max principle, we know that the periodic operator $T^{\text{per},(0)}$ has a first gap of size at least $500-37=463$. Now we mollify the coefficients of $T^{\text{per},(0)}$. Let $a^{(h)}$ be a periodic mollification of $a^{(0)}$ by a nonnegative smooth kernel such that $1\leq a^{(h)}\leq128$ for all $h>0$ and $(a^{(h)})^{-1}\to(a^{(0)})^{-1}$ in $L^1(0,1)$; we also denote the periodization of $a^{(h)}$ as $a^{\text{per},(h)}$. Then, by the continuity of isolated eigenvalues in the norm resolvent topology, one knows that the Dirichlet/Neumann eigenvalues associated with $a^{(h)}$ converge to that of $a^{(0)}$. In other words, by fixing a sufficiently small $h$ and write $a=a^{(h)}$, we have
\[
 1\leq a\leq128,\qquad \lambda_1^D(a)<40,
 \qquad \lambda_2^N(a)>490.
\]
Let $e_j(\theta)$ be the eigenvalues of $(D_x+\theta)a^{\text{per},(h)}(x)(D_x+\theta)$ on the periodic domain, where $D_x=-\mathrm i\partial_x$ and $\theta\in\mathbb R/(2\pi\mathbb Z)$. We apply the min-max principle again and obtain 
\begin{equation} \label{eq:simple_edge_proof_3}
\begin{gathered}
  0=e_1(0)\leq e_1(\theta)\leq e_1(\pi)=:b_*<40,\\
  e_2(\theta)\geq e_2(\pi)=:a_*>490,\qquad
  a_3:=\min_\theta e_3(\theta)>a_*.
 \end{gathered}
\end{equation}
By periodic/antiperiodic eigenvalue ordering (see \cite[Theorem~5.33]{Teschl2012}) and the strict inequality $b_*<a_*$, the eigenvalue $a_*=e_2(\pi)$ is simple and
\[
 e_2(\theta)\geq a_*
 \quad\text{for every }\theta\in\mathbb R.
\]
To identify the minimizing quasi-momentum and compute the curvature at this minimum, we use the monodromy matrix of the equation $-(au')'=\lambda u$. More precisely, let $M(\lambda)$ be the matrix satisfying
\[
 \binom{u(1)}{a(1)u'(1)}
 =M(\lambda)\binom{u(0)}{a(0)u'(0)}
\]
for every solution, and let $\Delta(\lambda):=\operatorname{tr}M(\lambda)$. Constancy of the Wronskian gives $\det M(\lambda)=1$. Taking $\lambda=e_2(\theta)$, we see that the corresponding Bloch eigenfunction satisfies
\[
 \binom{u(1)}{a(1)u'(1)}
 =e^{\mathrm i\theta}\binom{u(0)}{a(0)u'(0)}.
\]
Thus, $e^{\mathrm i\theta}$ is an eigenvalue of $M(\lambda)$, and its characteristic equation gives
\[
 0=\det\bigl(M(\lambda)-e^{\mathrm i\theta}I_2\bigr)
   =e^{2\mathrm i\theta}
      -\Delta(\lambda)e^{\mathrm i\theta}+1.
\]
Dividing by $e^{\mathrm i\theta}$, we obtain
\[
 \Delta(e_2(\theta))=2\cos\theta.
\]
In particular, $\Delta(a_*)=-2$. Hence, if $e_2(\theta_{min})=a_*$, we know $\theta_{min}=\pi$ modulo $2\pi$. This proves that the minimum is attained only at the quasi-momentum $\theta=\pi$.

To compute the second derivative of dispersion curve at $\theta=\pi$, we write $E(q)=e_2(\pi+q)$. Differentiating the equation $\Delta(E(q))=2\cos(\pi+q)$ yields
\[
 \begin{aligned}
 \Delta'(E(q))E'(q)
   &=-2\sin(\pi+q),\\
 \Delta''(E(q))(E'(q))^2+\Delta'(E(q))E''(q)
   &=-2\cos(\pi+q).
 \end{aligned}
\]
Hence, using the minimum property $E'(0)=0$, we obtain
\[
 \Delta'(a_*)E''(0)=2.
\]
Consequently, the inequality $E''(0)\geq0$ becomes strict. We conclude that
\begin{equation}
\label{eq:simple_edge_proof_7}
 e_2'(\pi)=0,\quad
 e_2''(\pi)=\frac{2}{\Delta'(a_*)}>0.
\end{equation}

{\color{blue}Step 2. A unique simple upper gap edge.}

We now construct a two-dimensional medium based on the one-dimensional medium $a$ found in Step 1. Choose a non-constant $g\in C^\infty(\mathbb R/\mathbb Z;\mathbb R)$ with $\|g\|_\infty\leq1$ and a real $\eta$ with $0<|\eta|\leq1/4$. Let $t(y):=1+\eta g(y)$, and
\[
S(x,y):=\operatorname{diag}\bigl(a(x)t(y),2a(y)\bigr).
\]
Then $1-|\eta|\leq t\leq1+|\eta|$, and $s_-I_2\leq S\leq s_+I_2$ with $s_-:=1-|\eta|>0$ and $s_+:=2\|a\|_{L^\infty(0,1)}$. For $\eta\geq0$, let $\rho_n(\eta,\theta)$ be the ordered eigenvalues of the following operator with periodic boundary conditions on $(0,1)$
\begin{equation} \label{eq:simple_edge_proof_1}
K_\eta(\theta)
 =(D_y+\theta)2a(y)(D_y+\theta)+\eta t(y).
\end{equation}
Notably, $K_\eta$ is linked to the elliptic operator $H_{S}$ by the following relation
\begin{equation} \label{eq:simple_edge_proof_2}
H_S(k_1,k_2)\simeq
 \bigoplus_{j\geq1}K_{e_j(k_1)}(k_2),
\end{equation}
where the unitary equivalence is justified by expanding $H_S(k_1,k_2)$ in its eigenbasis when regarding $H_S(k_1,k_2)$ as an operator in the variable $x$ with the fiber depending on $y$; in particular, every fiber eigenvalue is of the form $\rho_n(e_j(k_1),k_2)$. Hence, we can estimate the eigenvalue of $H_S(k_1,k_2)$ using that of $K_\eta(\theta)$. We now list some estimates of the latter. First, the definition \eqref{eq:simple_edge_proof_1} together with the min-max principle yields
\[
 2e_n(\theta)+(1-|\eta|)\eta
 \leq\rho_n(\eta,\theta)
 \leq2e_n(\theta)+(1+|\eta|)\eta
\]
and, for $\eta_2\geq\eta_1$,
\[
 \rho_1(\eta_2,\theta)
 \geq\rho_1(\eta_1,\theta)+(1-|\eta|)(\eta_2-\eta_1).
\]
On the other hand, the ground state $f_\eta$ of $K_\eta(0)$ can be chosen to be strictly positive and normalized. Then, for every periodic function $z\in H^1(0,1)$, integration by parts gives
\[
 \int_0^1\!\left(2a|(D_y+\theta)z|^2
           +\eta t|z|^2\right)
 -\rho_1(\eta,0)\|z\|_2^2
 =\int_0^1 2a f_\eta^2
       \left|(\partial_y+\mathrm i\theta)
                   \frac{z}{f_\eta}\right|^2.
\]
Thus, $\rho_1(\eta,\theta)\geq\rho_1(\eta,0)$. The equality for a nonzero minimizing function requires $z/f_\eta=c e^{-\mathrm i\theta y}$, which is periodic only when $\theta=0$ modulo $2\pi$.

Now we estimate the eigenvalues of $H_{S}$. Writing $\lambda_0:=\rho_1(a_*,0)$, the estimates obtained above and the decomposition \eqref{eq:simple_edge_proof_2} imply that
\[
 \begin{aligned}
 \rho_1(e_1(k_1),k_2)&\leq(3+|\eta|)b_*,\\
 \rho_n(e_j(k_1),k_2)&\geq2a_* &&(n\geq2),\\
 \rho_1(e_j(k_1),k_2)&\geq
       \lambda_0+(1-|\eta|)(a_3-a_*) &&(j\geq3),\\
 (1-|\eta|)a_*&\leq\lambda_0\leq(1+|\eta|)a_*.
 \end{aligned}
\]
Since $(3+|\eta|)b_*<(1-|\eta|)a_*$ and $(1+|\eta|)a_*<2a_*$ (recall \eqref{eq:simple_edge_proof_3}), the interval $((3+|\eta|)b_*,\lambda_0)$ is disjoint from the spectrum of $H_{S}(k)$ for all $k$. Moreover, $\rho_1(e_2(k_1),k_2)\geq\lambda_0$, with equality only at $k_0=(\pi,0)$ modulo $2\pi\mathbb Z^2$. At this point the eigenvalue is simple, because the second antiperiodic $x$-eigenvalue and the periodic $y$-ground eigenvalue are simple and all other branches satisfy the strict bounds above. This concludes the existence of a gap of $H_{S}$ and the simplicity of eigenvalue at the band edge.

We also record the Bloch eigenfunction of $H_{S}$ at the band edge for later use, which takes the following form according to the above discussion
\begin{equation}
\phi(x,y)=u(x)f(y).
\end{equation}
Here $u$ is the normalized real antiperiodic eigenfunction of $D_xa(x)D_x$ at $a_*$, and $f=f_{a_*}>0$ is the ground state of $K_{a_*}(0)$. In particular, the function $u$ is non-constant by antiperiodicity. The function $f$ is also non-constant; otherwise its eigenvalue equation would give $a_*t(y)=\lambda_0$, contradicting the choice of $g$ and $\eta$.

{\color{blue}Step 3. Curvatures and mixed responses.}

To complete the proof of Lemma~\ref{lem:seed-controllable-edge}, it remains to prove the nondegeneracy and spectral separation in claim (ii), and to construct the two perturbations that diagonalize the mixed response as in claim (iii). The idea is to differentiate the one-dimensional eigenvalue equations underlying the decomposition \eqref{eq:simple_edge_proof_2}. The resulting identities determine both the curvature of the two-dimensional band and its mixed responses to the perturbations.

{\color{blue}Step 3.1. Nondegeneracy of the band edge.}

We first choose eigenfunctions on fixed domains so that their derivatives with respect to quasi-momentum are well defined. On the antiperiodic domain, let $u_q$ be a normalized eigenfunction
of
\begin{equation} \label{eq:simple_edge_proof_6}
\big(L_x(q)-e_2(\pi+q)\big)u_q=0,\quad \text{with}\quad L_x(q):=(D_x+q)a(x)(D_x+q),\quad u_0=u.
\end{equation}
On the periodic domain, let $f_{\eta,q}$ be a normalized ground eigenfunction of $K_\eta(q)$, with $f_{a_*,0}=f$. The simplicity of eigenvalue established in Steps 1-2 permits these eigenfunctions to be chosen smoothly near the specified parameters. We also choose their phases so that 
\begin{equation} \label{eq:simple_edge_proof_4}
\langle u,u_q\rangle,\langle f,f_{\eta,q}\rangle \in\R,\quad \text{for small $q$.}
\end{equation}
Since the coefficient $a$ is real (i.e., the concerned operators are time-reversal symmetric), complex conjugation identifies the eigenfunction at $q$ to that at $-q$ in the sense that
\[
 u_{-q}=\overline{u_q},
 \quad
 f_{\eta,-q}=\overline{f_{\eta,q}}.
\]
Their derivatives at $q=0$ are therefore purely imaginary. Hence, using the phase conditions \eqref{eq:simple_edge_proof_4} and normalization of eigenfunctions, one calculates that
\[
 \left.\partial_q u_q\right|_{q=0}=\mathrm i w_x,
 \qquad
 \left.\partial_q f_{a_*,q}\right|_{q=0}=\mathrm i w_y,
 \qquad
 \int_0^1u w_x=\int_0^1f w_y=0,
\]
where $w_x$ is real and antiperiodic, and $w_y$ is real and periodic. These momentum-derivatives of Bloch eigenfunctions will appear naturally when calculating the second-order derivative of dispersion curves (see \eqref{eq:simple_edge_proof_5}). To see this, we note that, by differentiating \eqref{eq:simple_edge_proof_6} with respect to $q$, and similarly for the eigenvalue equation satisfied by $f_{a_{*},q}$, we obtain
\[
 \begin{aligned}
 (D_xaD_x-a_*)w_x&=a'u+2au',\\
 (K_{a_*}(0)-\lambda_0)w_y&=2a'f+4af',
 \end{aligned}
\]
where we have applied the facts $e_2'(\pi)=0$ and $\partial_\theta\rho_1(a_*,0)=0$, as proved in Step 1-2. Combining these equations with the eigenvalue equations at the band vertex, i.e.,
\[
 -(au')'=a_*u,
 \quad
 -(2af')'+a_*tf=\lambda_0f,
\]
we arrive at
\[
 \begin{aligned}
 \bigl[a(uw_x'-u'w_x)\bigr]'
   &=u(aw_x')'-w_x(au')'=-(au^2)',\\
 \bigl[2a(fw_y'-f'w_y)\bigr]'
   &=f(2aw_y')'-w_y(2af')'=-(2af^2)'.
 \end{aligned}
\]
Thus, the following quantities are constant in $x$ and $y$, respectively:
\begin{equation}
\label{eq:simple_edge_proof_10}
 C_1:=a(uw_x'-u'w_x+u^2),
 \quad
 C_2:=2a(fw_y'-f'w_y+f^2)
\end{equation}
To relate these constants to the band curvatures, we apply the Hellmann-Feynman formula and integrate by parts. Here we omit the details and directly show the results:
\begin{equation} \label{eq:simple_edge_proof_5}
\begin{aligned}
 \frac12e_2''(\pi)
 &=\int_0^1a u^2
     -\int_0^1w_x(a'u+2au')=\int_0^1a(uw_x'-u'w_x+u^2)=C_1,\\
 \frac12\partial_\theta^2\rho_1(a_*,0)
 &=\int_0^1 2a f^2
     -\int_0^1w_y(2a'f+4af')=\int_0^1 2a(fw_y'-f'w_y+f^2)=C_2.
 \end{aligned}
\end{equation}
The boundary terms vanish because $u,w_x$ are both antiperiodic and $f,w_y$ are periodic. By the identity \eqref{eq:simple_edge_proof_7} in Step 1, we know
\begin{equation}
\label{eq:simple_edge_proof_8}
 C_1=\frac12e_2''(\pi)=\frac{1}{\Delta'(a_*)}>0.
\end{equation}
To determine the sign of $C_2$, we use the fact $f>0$ (see Step 2) and rewrite defining identity of $C_2$ in \eqref{eq:simple_edge_proof_10} as
\[
 \left(\frac{w_y}{f}\right)'
 =\frac{C_2}{2af^2}-1.
\]
Since $w_y/f$ is periodic, integration over $(0,1)$ gives
\[
 0=C_2\int_0^1\frac{dy}{2a(y)f(y)^2}-1.
\]
Consequently,
\begin{equation} \label{eq:simple_edge_proof_9}
C_2=\left(\int_0^1\frac{dy}{2a(y)f(y)^2}\right)^{-1}>0.
\end{equation}

We can now compute the curvature of the two-dimensional band in terms of $C_1$ and $C_2$, by which the positivity claimed in (ii) of Lemma~\ref{lem:seed-controllable-edge} will follow by \eqref{eq:simple_edge_proof_8} and \eqref{eq:simple_edge_proof_9}. Note that, by the decomposition\eqref{eq:simple_edge_proof_2}, the local expression of the concerned band is
\[
 \lambda_S(k_0+q)
 =\rho_1\bigl(e_2(\pi+q_1),q_2\bigr),
 \quad \text{with}\quad k_0=(\pi,0).
\]
We claim that the Hessian matrix of $ \lambda_S$, which is exactly the matrix $M$ in Lemma~\ref{lem:seed-controllable-edge}, is diagonal. In fact, using $e_2'(\pi)=0$ and
$\partial_\theta\rho_1(a_*,0)=0$, the chain rule yields
\[
 \begin{aligned}
 \nabla_k\lambda_S(k_0)&=0,\\
 \partial_{k_1}^2\lambda_S(k_0)
 &=\partial_\eta\rho_1(a_*,0)e_2''(\pi)
   =2C_1\int_0^1t f^2,\\
 \partial_{k_1}\partial_{k_2}\lambda_S(k_0)
 &=\partial_\eta\partial_\theta\rho_1(a_*,0)
      e_2'(\pi)=0,\\
 \partial_{k_2}^2\lambda_S(k_0)
 &=\partial_\theta^2\rho_1(a_*,0)=2C_2.
 \end{aligned}
\]
where we have applied the Hellmann-Feynman formula in the parameter $\eta$ to derive
\[
 \partial_\eta\rho_1(a_*,0) =\int_0^1t(y)f(y)^2\,dy>0.
\]
In conclusion,
\[
 \lambda_S(k_0+q)
 =\lambda_0+q^{\mathsf T}Mq+O(|q|^3),
 \qquad
 M=
 \begin{pmatrix}
 C_1\displaystyle\int_0^1t(y)f(y)^2\,dy&0\\
 0&C_2
 \end{pmatrix}>0.
\]
For completeness, we verify the uniform spectral separation away from the vertex $k_0$, which follows by a standard compactness argument. Let $\nu_n(k)$ be the ordered eigenvalues of $H_S(k)$ with $n\geq 1$. First, the ellipticity bound $S\geq s_- I_2$ and the min-max principle imply that $\nu_n(k)\to+\infty$ uniformly in $k$. Hence, we can choose large $N$ so that
\[
 \nu_{N+1}(k)\geq\lambda_0+1
 \quad\text{for every }k\in\mathcal B.
\]
Fix a sufficiently small $\kappa_0>0$, with distance on $\mathcal B$ understood modulo $2\pi\mathbb Z^2$. Then the result from Step 2 shows that $\lambda_0\notin\sigma(H_S(k))$ whenever $k\neq k_0$. By the continuity of the finite collection of dispersion curves $\{\nu_1,\ldots,\nu_N\}$, the compactness of Brillouin zone therefore give
\[
 \gamma:=
 \min_{\substack{k\in\mathcal B,\ 
                 \operatorname{dist}_{\mathcal B}(k,k_0)
                    \geq\kappa_0\\
                 1\leq n\leq N}}
       |\nu_n(k)-\lambda_0|>0.
\]
Taking $\delta_0=\frac12\min\{1,\gamma\}$, we obtain
\[
 \operatorname{dist}\bigl(\lambda_0,\sigma(H_S(k))\bigr)
 >\delta_0
 \quad\text{whenever}\quad
 \operatorname{dist}_{\mathcal B}(k,k_0)\geq\kappa_0.
\]
This completes the proof of claim (ii).

{\color{blue}Step 3.2. Construction of the two mixed responses.}

To prove claim (iii), we first express each mixed response as a linear functional of the perturbation field. We then prove that these two functionals are linearly independent and diagonalize the responses correspondingly.

Recalling that $\beta$ is a smooth real-valued periodic function, and $J=\begin{pmatrix}0&-1\\1&0\end{pmatrix}$, we know that $S+s\mathrm i\beta J$ is Hermitian and satisfies
\[
 S+s\mathrm i\beta J\geq\frac{1}{2}s_- I_2,\quad \text{whenever}\quad |s|\|\beta\|_\infty<s_-/2.
\]
We work on the fixed domain with antiperiodic boundary conditions in $x$ and periodic boundary conditions in $y$; the concerned operator is then
\[
 e^{\ii k_0\cdot x}H_{S+s\ii\beta J}(k_0+q)e^{-\ii k_0\cdot x} =(D+q)^{\mathsf T}(S+s\ii\beta J)(D+q).
\]
By the standard perturbation theory, its simple eigenvalue near $(q,s)=(0,0)$ has a smooth continuation (cf. \cite[Chapter VII]{Kato1995}), denoted by $\lambda_\beta(q,s)$. Hence, it is legal to define
\begin{equation} \label{eq:simple_edge_proof_11}
\mu_j(\beta):=
\left.\partial_{q_j}\partial_s\lambda_\beta(q,s)
 \right|_{q=s=0},
 \quad j=1,2.
\end{equation}
At $s=0$, a normalized eigenfunction of $(D+q)^{\mathsf T}S(D+q)$ associated with the eigenvalue $\lambda_\beta(q,0)$ is
\[
 \phi_q(x,y)
 =u_{q_1}(x)f_{e_2(\pi+q_1),q_2}(y).
\]
With this notion and the Hellmann-Feynman formula, the $s$-derivative in \eqref{eq:simple_edge_proof_11} equals
\begin{equation}
\label{eq:simple_edge_proof_12}
 \partial_s\lambda_\beta(q,0)
 =\int_{Q_{\rm cell}} ((D+q)\phi_q)^*
       \mathrm i\beta J((D+q)\phi_q).
\end{equation}
To further take the $q$-derivative, we first calculate that of $\phi_q$. Note that, when differentiating $\phi_q$ with respect to $q_1$, the derivative of the parameter $e_2(\pi+q_1)$ contributes a factor $e_2'(\pi)=0$. Hence,
\[
 \left.\partial_{q_1}\phi_q\right|_{q=0}
 =\mathrm i w_xf=:\mathrm iW_1,
 \qquad
 \left.\partial_{q_2}\phi_q\right|_{q=0}
 =\mathrm iuw_y=:\mathrm iW_2,
\]
Recall that $\phi_0=\phi=uf$ is real. Hence, one further calculates that
\[
\left.\partial_{q_j}\bigl((D+q)\phi_q\bigr)\right|_{q=0}
 =\nabla W_j+\phi\mathbf e_j,
\]
where $\mathbf e_j$ is the $j$th coordinate vector. Using the facts that $\phi,W_j,\beta$ are real and $J^{\mathsf T}=-J$, one can calculate from \eqref{eq:simple_edge_proof_11} and \eqref{eq:simple_edge_proof_12} that
\begin{equation} \label{eq:simple_edge_proof_13}
\mu_j(\beta)
=2\int_{Q_{\rm cell}}\beta(\nabla W_j+\phi\mathbf e_j)^{\mathsf T}J\nabla\phi
=\int_{Q_{\rm cell}}\beta R_j,
\end{equation}
where
\[
 R_j:=2(\nabla W_j+\phi\mathbf e_j)^{\mathsf T}J\nabla\phi.
\]
Substituting $\phi=uf$, $W_1=w_xf$, and $W_2=uw_y$, we obtain
\[
 \begin{aligned}
 R_1(x,y)
 &=-2f(y)f'(y)
   \bigl(u(x)w_x'(x)-u'(x)w_x(x)+u(x)^2\bigr)=-\frac{2C_1f(y)f'(y)}{a(x)},\\[1mm]
 R_2(x,y)
 &=2u(x)u'(x)
   \bigl(f(y)w_y'(y)-f'(y)w_y(y)+f(y)^2\bigr)=\frac{C_2u(x)u'(x)}{a(y)}.
 \end{aligned}
\]
The last equalities use the constant identities in \eqref{eq:simple_edge_proof_10} from Step 3.1. Both $R_1$ and $R_2$ are smooth real-valued periodic functions; in particular, $uu'$ is periodic because both $u$ and $u'$ are antiperiodic.

We now prove that $R_1$ and $R_2$ are linearly independent, by which we can diagonalize \eqref{eq:simple_edge_proof_13} using the associated Gram matrix. Suppose that $c_1R_1+c_2R_2=0$ for some real constants $c_1,c_2$. Multiplying by $a(x)a(y)$ then gives
\[
 -2c_1C_1a(y)f(y)f'(y)
 +c_2C_2a(x)u(x)u'(x)=0.
\]
If $c_2\neq0$, fixing $y$ shows that $a(x)u(x)u'(x)$ is a constant $c$. Since $u^2$ is periodic, we know that
\[
 0=\int_0^1u(x)u'(x)\,dx
   =c\int_0^1\frac{dx}{a(x)}.
\]
Thus, $c=0$ and $(u^2)'=0$. This is impossible: a nonzero, continuous and real-valued antiperiodic function cannot have a constant square. We conclude that $c_2=0$. Since $f>0$ is nonconstant, $ff'\not\equiv0$, and the remaining identity then gives $c_1=0$. This concludes the linear independence of $R_1$ and $R_2$. Consequently, the real Gram matrix $G=(G_{ij})$ is well-defined and positive definite, where
\[
 G_{ij}:=\int_{Q_{\rm cell}}R_iR_j,\qquad i,j\in\{1,2\} .
\]
Hence, introducing the real-valued smooth periodic function $\beta_\ell:=\sum_{r=1}^2(G^{-1})_{r\ell}R_r$ for $\ell=1,2$, we have
\[
 \mu_j(\beta_\ell)
 =\int_{Q_{\rm cell}}\beta_\ell R_j
 =\sum_{r=1}^2G_{jr}(G^{-1})_{r\ell}
 =\delta_{j\ell}.
\]
This proves claim (iii) and completes the proof of the lemma.

\paragraph{Proof of Lemma~\ref{lem:seed-transfer}}

The idea is to expand the near-$k_0$ part of the (unperturbed) operator $H_{S_N}(k)$ into a two-scale form, which is the product of a slow-varying envelope and a microscopic carrier; see, e.g. \eqref{eq:two_scale_proof_1}. Then, in the high-frequency regime (of order $N^2$), the envelope part effectively dominates the spectral property; combined with the two perturbation terms in \eqref{eq:short-seed-coefficient}, the effective operator as $N\to\infty$ is exactly the target magnetic operator $K$ in Lemma \ref{lem:seed-magnetic-target}. See Steps 3 and 4 in the sequel. The far-$k_0$ part is controlled using the Schur complement. The proof is then complete by recalling that the target magnetic operator is gapped as established in Lemma \ref{lem:seed-magnetic-target}.



Throughout the proof, $N$ tends to infinity through even integers and $\varepsilon:=N^{-1}$. The emerging constants are uniform in $N$ and the momentum $k\in\mathcal B$ unless independently specified. We take $N$ large enough that the matrix $A_N=S_N+O(N^{-1})$ is uniformly positive.

\medskip\noindent
{\color{blue}Step 1. The microscopic waves and their slow envelopes.}

In this step, we introduce the Hilbert space accommodating the slow-varying envelopes and establish the identification between it and the near-$k_0$ part of microscopic Bloch modes. To start, we denote the concerned microscopic operator on $(0,1)^2$ with $k_0$-quasi-periodic boundary conditions as
\begin{equation} \label{eq:two_scale_proof_14}
H(q):=e^{\ii k_0\cdot y}H_S(k_0+q)e^{-\ii k_0\cdot y}=(D_y+q)^{\mathsf T}S(y)(D_y+q).
\end{equation}
Here, the subscript indicating the coefficient $S$ is omitted for ease of notation, as it is fixed throughout the proof. For brevity, we write $\lambda(q)$ for the near-edge eigenvalue of $H(q)$ for $q$ near zero, which admits the asymptotic expansion on Lemma~\ref{lem:seed-controllable-edge}(ii). In particular, we can choose $0<\rho<\pi/4$ and normalized eigenfunctions $\phi_q$, smooth in $q$ with values in the $H^2$ space, such that
\begin{equation}\label{eq:transfer-edge-chart}
 \begin{gathered}
 H(q)\phi_q=\lambda(q)\phi_q,\quad
 \lambda(q)=\lambda_0+q^{\mathsf T}Mq+O(|q|^3),\\
 \lambda(q)-\lambda_0\geq c|q|^2,
 \quad |q|<2\rho.
 \end{gathered}
\end{equation}
The eigenvalue is simple for $|q|<2\rho$. Since $S$ is real and $k_0=-k_0$ modulo $2\pi\Z^2$, we can choose a gauge such that $\phi_0=\phi$ and $\phi_{-q}=\overline{\phi_q}$. 

Now we define the following unitary map, which enlarges the underlying unit cell,
\[
 (\mathcal D_{N,k}f)(y):=e^{\ii\varepsilon k\cdot y}f(\varepsilon y).
\]
It sends $L^2\big((0,1)^2\big)$ to $L^2\big((0,N)^2\big)$ with the quasi-periodic boundary conditions $u(y+Ne_j)=e^{\ii k_j}u(y)$. The enlarged cell is equipped with measure $N^{-2}\,dy$ to ensure the unitarity, and all Sobolev norms on this enlarged cell use the same normalized measure. In particular, by direct substitution, the conjugation of $\mathcal D_{N,k}$ transforms $H_{A_N}$ as follows:
\begin{equation}\label{eq:transfer-dilation}
 \mathcal D_{N,k}H_{A_N}(k)\mathcal D_{N,k}^{-1}
 =N^2H_N(k),\qquad
 H_N=H_{0,N}+\varepsilon V_{1,N}+\varepsilon^2V_{2,N},
\end{equation}
where
\begin{equation} \label{eq:two_scale_proof_2}
H_{0,N}:=D_y^{\mathsf T}S(y)D_y,\quad
 V_{1,N}:=D_y^{\mathsf T}
   \left(\ii\sum_{\ell=1}^2B_\ell(\varepsilon y)\beta_\ell(y)J\right)D_y,\quad
 V_{2,N}:=D_y^{\mathsf T}\bigl(v(\varepsilon y)S(y)\bigr)D_y.
\end{equation}
The dependence on $k$ is now imposed in the boundary conditions.

With the unitary transform $\mathcal D_{N,k}$, the emergence of slow-varying envelope will be natural. To see this, we introduce the following notions, where, in particular, the index set $I_N(k)$ records the band indices which  contribute near the edge:
\[
 p_n(k)=k+2\pi n,\quad q_n(k)=\varepsilon p_n(k),\quad
 I_N(k)=\{n\in\Z^2:|q_n(k)|<\rho\}.
\]
For $n\in I_N(k)$, the following modes satisfy the $k$-quasi-boundary conditions on $(0,N)^2$
\[
 \Psi_{n,N,k}(y):=e^{\ii q_n(k)\cdot y}\phi_{q_n(k)}(y)
\]
as its phase under $y\mapsto y+Ne_j$ is $e^{\ii k_j}e^{\ii Nk_{0,j}}=e^{\ii k_j}$ because $N$ is even. Moreover, one can check that $\Psi_{n,N,k}$ are orthonormal. We let $P_N(k)$ project onto $\operatorname{span}\{\Psi_{n,N,k}:n\in I_N(k)\}$ and set $Q_N:=I-P_N$. Importantly, the image of $\Psi_{n,N,k}$ under $\mathcal D_{N,k}^{-1}$ admits the claimed two-scale form:
\begin{equation} \label{eq:two_scale_proof_1}
\mathcal D_{N,k}^{-1}\Psi_{n,N,k}(X)
 =e^{2\pi\ii n\cdot X}\phi_{q_n(k)}(NX).
\end{equation}
Here, the factor $\phi_{q_n}(NX)$ is the microscopic mode, whereas $e^{2\pi\ii n\cdot X}$ is the slow-varying envelope. We will identify each mode in $\operatorname{span}\{\Psi_{n,N,k}:n\in I_N(k)\}$ with the envelope using the following notions:
\begin{equation}\label{eq:transfer-envelope-space}
 \begin{gathered}
 \mathcal E_N(k):=\operatorname{span}
 \{e^{2\pi\ii n\cdot X}:n\in I_N(k)\},\qquad
 \Pi_N(k):L^2(\T^2)\longrightarrow\mathcal E_N(k),\\
 \mathcal U_N(k)e^{2\pi\ii n\cdot X}=\Psi_{n,N,k}.
 \end{gathered}
\end{equation}
Here, $\Pi_N$ is the orthogonal Fourier projection and $\mathcal U_N$ is unitary from $\mathcal E_N(k)$ onto $\operatorname{Ran}P_N(k)$. As we will see later, the identification with the envelope space $\mathcal E_N(k)$ is convenient for notation: more importantly, the effective operator in the high-frequency regime will be explicitly calculated in $\mathcal E_N(k)$ and clearly shown to be the target operator in Lemma \ref{lem:seed-magnetic-target}; see Step 3 for details.


\medskip\noindent
{\color{blue}Step 2. Elimination of far-$k_0$ component.}

Before studying the effective operator, we first establish the Schur decomposition framework, which is essential for rigorously justifying the limiting effective operator in the later steps. First, note that the spectral isolation established in Lemma \ref{lem:seed-controllable-edge}(ii) and the asymptotics \eqref{eq:transfer-edge-chart} imply that, for some $d>0$,
\begin{equation}\label{eq:transfer-complement-gap}
 \operatorname{dist}\left(\lambda_0,
 \sigma\!\left(Q_NH_{0,N}Q_N\big|_{\operatorname{Ran}Q_N}\right)
 \right)\geq d.
\end{equation}
Thus, it is legal to define the reduced resolvent
\[
 R_N:=Q_N\left(Q_N(H_{0,N}-\lambda_0)Q_N
              \big|_{\operatorname{Ran}Q_N}\right)^{-1}Q_N.
\]
By the elliptic estimates and Bloch decomposition, it is $H^2$ bounded:
\begin{equation}\label{eq:transfer-regularity-bounds}
 \|P_N\|_{L^2\to H^2}+\|R_N\|_{L^2\to H^2}\leq C.
\end{equation}
Note that these estimates are uniform in $N$ because 1) the coefficient $S$ of $H_{0,N}$ is fixed, 2) the spectral separation in \eqref{eq:transfer-complement-gap} is uniform in $N$. To facilitate the Schur decomposition, we need to estimate the perturbation operators $V_{j,N}$ in \eqref{eq:transfer-dilation} when restricted to $\operatorname{Ran}P_N$ and $\operatorname{Ran}Q_N$, respectively. To this end, we first note that, for any real function $h$, the antisymmetry of $J$ implies the identity
\[
 D_y^{\mathsf T}(\ii hJ)D_y
 =\sum_{i,j}J_{ij}(\partial_i h)D_{y_j}.
\]
In other words, $V_{1,N}$ ($V_{2,N}$, resp.) is a first-order (second-order, resp.) differential operator with uniformly bounded coefficients, as checked by definitions \eqref{eq:two_scale_proof_2} and \eqref{eq:short-seed-coefficient_details}. Hence, it follows from \eqref{eq:transfer-regularity-bounds} that
\begin{equation} \label{eq:two_scale_proof_3}
\|V_{j,N}P_N\|+\|P_NV_{j,N}\|+\|V_{j,N}R_N\|\leq C,
 \qquad j=1,2.
\end{equation}
Here, $P_NV_{j,N}$ denotes the bounded extension of the adjoint of
$V_{j,N}P_N$. Unless otherwise specified, operator norms in this proof
are $L^2\to L^2$ norms.

The purpose of Schur decomposition is to study the resolvent of the total operator $H_N$ for spectral parameter of the range $\lambda=\lambda_0+\varepsilon^2z$, with $z$ fixed within a bounded set of complex plane. The key point is first to establish the invertibility of the reduced resolvent restricting on $\operatorname{Ran}Q_N$. We define
\[
 C_N(z):=Q_N(H_N-\lambda_0-\varepsilon^2z)Q_N.
\]
It is indeed invertible for small $\varepsilon$: denoting the reduced perturbation operator as $Z_N(z):=Q_N(\varepsilon V_{1,N}+\varepsilon^2V_{2,N}
-\varepsilon^2z)Q_N$, then, using the relative boundedness $\|Z_N(z)R_N\|\leq C\varepsilon$ and the fact that the unperturbed resolvent $R_N$ is bounded, one concludes by the Neumann series that 
\begin{equation}\label{eq:transfer-complement-inverse}
 \|C_N(z)^{-1}-R_N\|_{L^2\to H^2}\leq C\varepsilon.
\end{equation}
Here and below all the reduced inverse to $\operatorname{Ran}Q_N$ is extended by zero on $\operatorname{Ran}P_N$. With this invertibility, we can now define the Schur complement
\begin{equation} \label{eq:two_scale_proof_4}
\mathscr S_N(z):=P_N(H_N-\lambda_0-\varepsilon^2z)P_N
 -P_NH_NQ_NC_N(z)^{-1}Q_NH_NP_N.
\end{equation}
By the block decomposition adapted to $\operatorname{Ran}P_N$ and $\operatorname{Ran}Q_N$, one can check that
\begin{equation*}
\text{$H_N-\lambda_0-\varepsilon^2z$
is invertible}\quad\Longleftrightarrow \quad
\text{$\mathscr S_N(z)$ is invertible.}
\end{equation*}
Since $P_N$ commutes with $H_{0,N}$, both the two off-diagonal factors $P_NH_NQ_N$ and $Q_NH_NP_N$ in \eqref{eq:two_scale_proof_4} contain at least one perturbation parameter $\varepsilon$. With this observation, together with estimate \eqref{eq:transfer-complement-inverse}, we obtain the following asymptotic expansion of the Schur complement
\begin{equation} \label{eq:transfer-schur-remainder}
 \varepsilon^{-2}\mathcal U_N^*\mathscr S_N(z)\mathcal U_N
 =:F_N(k)-z+\mathfrak r_N(k,z),
\end{equation}
where the selfadjoint operator $F_N(k)$ on $\mathcal E_N(k)$ is defined as
\begin{equation}\label{eq:short-seed-schur}
 \begin{split}
 F_N(k):=\mathcal U_N^*\big[&
 \varepsilon^{-2}P_N(H_{0,N}-\lambda_0)P_N
 +\varepsilon^{-1}P_NV_{1,N}P_N\\
 &+P_NV_{2,N}P_N-P_NV_{1,N}R_NV_{1,N}P_N
 \big]\mathcal U_N,
 \end{split}
\end{equation}
and the remainder satisfies the bound
\begin{equation}
\|\mathfrak r_N(k,z)\|\leq C\varepsilon .
\end{equation}
Note that we have transformed the Schur complement by the unitary $\mathcal U_N$ because, as we have explained, the effective operator as $\varepsilon\to 0$ will be explicitly computed on $\mathcal E_N(k)$; see Steps 3-4 in the following.

\medskip\noindent
{\color{blue}Step 3. Computation of the effective operator on the envelope space.}

In this step, we calculate each term in the leading order term \eqref{eq:short-seed-schur} of the Schur complement in the Fourier basis of $\mathcal E_N(k)$. Remarkably, as one shall see, the limiting operator of $F_N$ as $N\to\infty$ takes exactly the form of the target magnetic operator in Lemma \ref{lem:seed-magnetic-target}; see \eqref{eq:transfer-core-limit}. This explains why we choose the profiles $B,v$ in their specific expressions in \eqref{eq:short-seed-coefficient}.

{\color{blue}Step 3.1.} The leading order term in $F_N$, i.e., $\varepsilon^{-2}\mathcal U_N^*P_N(H_{0,N}-\lambda_0)P_N \mathcal U_N$, is direct to calculate as $H_{0,N}-\lambda_0$ is diagonalized by the Bloch eigenfunctions in $\operatorname{Ran}P_N$. Explicitly, this diagonal matrix is given by
\begin{equation} \label{eq:two_scale_proof_5}
(\varepsilon^{-2}\mathcal U_N^*P_N(H_{0,N}-\lambda_0)P_N \mathcal U_N)_{n^{\prime}n}=\varepsilon^{-2}(\lambda(q_n)-\lambda_0)\delta_{n^{\prime}n}=\big(p_n^{\mathsf T}Mp_n+O(\varepsilon|p_n|^3)\big)\delta_{n^{\prime}n}.
\end{equation}

{\color{blue}Step 3.2.} To calculate the matrix form of the first perturbation term $\mathcal U_N^*P_NV_{1,N}P_N\mathcal U_N$, we introduce the kernel
\begin{equation} \label{eq:two_scale_proof_8}
\mathcal B_\ell(q',q):=\langle \ii\beta_\ell J(D_y+q)\phi_q, (D_y+q')\phi_{q'}
\rangle_{L^2((0,1)^2)}.
\end{equation}
It satisfies the following properties:
\begin{itemize}
    \item[(i)] $\mathcal B_\ell(0,0)=0$, which follows by the reality of $\phi$ and anti-symmetry of $J$.
    \item[(ii)] Self-adjointness of the operator $\ii\beta_\ell J$ and our gauge choice $\phi_{-q}=\overline{\phi_q}$ give
\begin{equation} \label{eq:two_scale_proof_6}
\mathcal B_\ell(q',q)=\overline{\mathcal B_\ell(q,q')},\quad
 \mathcal B_\ell(-q',-q)=-\overline{\mathcal B_\ell(q',q)}.
\end{equation}
    \item[(iii)] Every first-order derivative of $\mathcal B_\ell(q',q)$ at $(0,0)$ is real, which follows by the second identity in \eqref{eq:two_scale_proof_6}. Then, applying the first identity in \eqref{eq:two_scale_proof_6}, we know that the derivatives of $\mathcal B_\ell(q',q)$ in $q'$ and $q$ at $(0,0)$ are equal.
    \item[(iv)] At the diagonal, the Hellmann--Feynman formula and Lemma~\ref{lem:seed-controllable-edge} give
    \begin{equation} \label{eq:two_scale_proof_7}
    \left.\partial_{q_j}\mathcal B_\ell(q,q)\right|_{q=0}
 =\left.\partial_{q_j}\partial_s
 \lambda_{S+s\ii\beta_\ell J}(k_0+q)\right|_{q=s=0}
 =\delta_{j\ell}.
    \end{equation}
\end{itemize}
Using these properties and the Taylor expansion, we conclude that
\begin{equation}\label{eq:transfer-cell-kernel}
 \mathcal B_\ell(q',q)
 =\tfrac12(q'_\ell+q_\ell)+O(|q|^2+|q'|^2),\quad
 |\mathcal B_\ell(q',q)|\leq C(|q|+|q'|).
\end{equation}

Now we can calculate the matrix form of $\mathcal U_N^*P_NV_{1,N}P_N\mathcal U_N$. Since the Hessian matrix $M$ is constant and the profiles $b,V$ are trigonometric (see Lemma \ref{lem:seed-magnetic-target}), we know by the definition \eqref{eq:short-seed-coefficient_details} that we can choose $R_0>0$ so that the Fourier coefficients of $B_1,B_2,v$ vanish for orders $|r|>R_0$. In particular, the intermediate function $B_\ell$ in $V_{1,N}$ is expanded as
\begin{equation} \label{eq:two_scale_proof_12}
B_\ell(X)=\sum_{\substack{r\in\Z^2\\|r|\le R_0}}\widehat B_\ell(r)e^{2\pi\ii r\cdot X}, \quad
 \widehat B_\ell(r) :=\int_{(0,1)^2}B_\ell(X)e^{-2\pi\ii r\cdot X}\,dX.
\end{equation}
To see how these coefficients enter the matrix, recall that $\mathcal U_Ne^{2\pi\ii n\cdot X}=\Psi_{n,N,k}$ and $P_N\Psi_{n,N,k}=\Psi_{n,N,k}$ for $n\in I_N(k)$; for brevity, we write $\Psi_n=\Psi_{n,N,k}$ and $q_n=q_n(k)$ in the sequel. Using the definition \eqref{eq:two_scale_proof_2}, the desired matrix element is expanded as the following volume integral
\[
\begin{aligned}
(\mathcal U_N^*P_NV_{1,N}P_N\mathcal U_N)_{n'n}
= \frac1{N^2}\sum_{\ell=1}^2 \int_{(0,N)^2} B_\ell(\varepsilon y) (D_y\Psi_{n'})^*\,\ii\beta_\ell(y)J\,D_y\Psi_n\,dy.
\end{aligned}
\]
Substituting the Fourier expansion of $B_\ell$ and the identity
\[
 D_y\Psi_n =e^{\ii q_n\cdot y}(D_y+q_n)\phi_{q_n},
\]
therefore, yields
\[
\begin{aligned}
 (\mathcal U_N^*P_NV_{1,N}P_N\mathcal U_N)_{n'n}=
 \sum_{\ell=1}^2\sum_{|r|\le R_0}&\widehat B_\ell(r)
 \frac1{N^2}\int_{(0,N)^2}
 e^{2\pi\ii(n-n'+r)\cdot y/N}\\
 &\times
 \bigl((D_y+q_{n'})\phi_{q_{n'}}(y)\bigr)^*
 \ii\beta_\ell(y)J
 \bigl((D_y+q_n)\phi_{q_n}(y)\bigr)\,dy.
\end{aligned}
\]
The factor following the exponential is $\Z^2$-periodic: $\beta_\ell$ is periodic, and the two Bloch factors have the same $k_0$-quasiperiodicity, whose phases cancel. Hence, splitting the enlarged cell into translations of the unit cell, extracting the common factor and recalling the definition \eqref{eq:two_scale_proof_8}, we obtain
\begin{equation} \label{eq:two_scale_proof_9}
(\mathcal U_N^*P_NV_{1,N}P_N\mathcal U_N)_{n'n}=
 \frac1{N^2}\sum_{\ell=1}^2\sum_{|r|\le R_0}\widehat B_\ell(r)\mathcal B_\ell(q_{n'},q_n)
 \sum_{a\in\{0,\ldots,N-1\}^2}
 e^{2\pi\ii(n-n'+r)\cdot a/N} .
\end{equation}
Note that
\[
 \frac1{N^2}
 \sum_{a\in\{0,\ldots,N-1\}^2}
 e^{2\pi\ii(n-n'+r)\cdot a/N}
 =
 \begin{cases}
  1,&n-n'+r\in N\Z^2,\\
  0,&n-n'+r\notin N\Z^2.
 \end{cases}
\]
On the other hand, for $n,n'\in I_N(k)$, we have $|n'-n|<\rho N/\pi$. Hence, for sufficiently large $N$,
\[
 |n-n'+r|
 <\frac{\rho N}{\pi}+R_0<N.
\]
The only possible vector in $N\Z^2$ is therefore zero. Thus, only the Fourier mode $r=n'-n$ contributes. This, together with \eqref{eq:two_scale_proof_9}, yields the desired matrix form
\begin{equation} \label{eq:two_scale_proof_10}
 (\mathcal U_N^*P_NV_{1,N}P_N\mathcal U_N)_{n'n}
 =
 \sum_{\ell=1}^2
 \widehat B_\ell(n'-n)\,
 \mathcal B_\ell(q_{n'},q_n).
\end{equation}
For fixed $n,n'$, \eqref{eq:transfer-cell-kernel} shows that, multiplying by $\varepsilon^{-1}$, the above matrix element converges to
\[
 \frac12\sum_\ell\widehat B_\ell(n'-n)
                         (p_{n',\ell}+p_{n,\ell}).
\]
These are exactly the Fourier matrix entries of $\frac12\sum_\ell\{D_{X_\ell}+k_\ell,B_\ell\}$, where $\{T,B\}=TB+BT$.

{\color{blue}Step 3.3.} Following the same calculations, the matrix form of $\mathcal U_N^*P_NV_{2,N}P_N\mathcal U_N$ is given by
\begin{equation} \label{eq:two_scale_proof_11}
 (\mathcal U_N^*P_NV_{2,N}P_N\mathcal U_N)_{n'n}
 =\widehat v(n'-n)\mathcal T(q_{n^{\prime}},q_n),
\end{equation}
where
\[
\mathcal T(q',q):=
\langle S(D_y+q)\phi_q,(D_y+q')\phi_{q'}\rangle.
\]
In particular, the limiting entries are  $\lambda_0\widehat v(n'-n)$.

{\color{blue}Step 3.4.}  Next, we calculate the matrix of the last term in \eqref{eq:short-seed-schur}, i.e., $\mathcal U_N^*P_NV_{1,N}R_NV_{1,N}P_N\mathcal U_N$. To do this, we first expand $V_{1,N}\Psi_n$ into microscopic Bloch eigenmode $\phi_q$, then study the action of $R_N$ on $\phi_q$. Note that we keep the abbreviation $\Psi_n=\Psi_{n,N,k}$ and $q_n=q_n(k)$ as in Step 3.2. Since $P_N\Psi_n=\Psi_n$ and $V_{1,N}$ is selfadjoint, the concerned matrix element is written as
\begin{equation} \label{eq:two_scale_proof_15}
 (\mathcal U_N^*P_NV_{1,N}R_NV_{1,N}P_N\mathcal U_N)_{n'n}
 =\langle R_NV_{1,N}\Psi_n,V_{1,N}\Psi_{n'}\rangle_N,
\end{equation}
where $\langle\cdot,\cdot\rangle_N$ is the enlarged-cell inner
product with normalized measure $N^{-2}\,dy$. 

First, substituting the Fourier expansion of $B_\ell$ in \eqref{eq:two_scale_proof_12} into the definition \eqref{eq:two_scale_proof_2} gives
\begin{equation} \label{eq:two_scale_proof_13}
 V_{1,N}\Psi_n
 =\sum_{\ell=1}^2\sum_{|r|\le R_0}
 \widehat B_\ell(r)e^{\ii q_{n+r}\cdot y}
 (D_y+q_{n+r})^{\mathsf T}
 \ii\beta_\ell J(D_y+q_n)\phi_{q_n}.
\end{equation}
Note that the operator $(D_y+q_{n+r})^{\mathsf T}\ii\beta_\ell J(D_y+q_n)$ in \eqref{eq:two_scale_proof_13} is the momentum shift of the operator $V_\ell$ introduced in \eqref{eq:V_C_def}, with the momentum in the outer derivative being $q_{n+r}=q_n+2\pi\varepsilon r$. For fixed index $n$, by the Fourier support bound from Step 3.2, the intermediate indices in the expansions of $V_{1,N}\Psi_n$ belong to $\{a=n+r:|r|\le R_0\}$; hence, as $N\to\infty$, we have $q_a(k)\to0$ uniformly in $k\in\mathcal B$. Thus, all indices appearing in the sum belong to $I_N(k)$ for sufficiently large $N$. In particular, this observation indicates that the application of $H_{0,N}$ and $P_N$ to the vector in \eqref{eq:two_scale_proof_13} is identified with the application of $H(q_a)$ (see \eqref{eq:two_scale_proof_14}) and $P(q_a)$, with the latter being the orthogonal projection onto $\phi_{q_a}$. Transferring this property to the resolvent, we have
\begin{equation} \label{eq:two_scale_proof_16}
\begin{aligned}
&R_N\bigl(e^{\ii q_{n+r}\cdot y}(D_y+q_{n+r})^{\mathsf T}\ii\beta_\ell J(D_y+q_n)\phi_{q_n}\bigr) \\
&=e^{\ii q_{n+r}\cdot y}\mathcal R(q_{n+r})\big((D_y+q_{n+r})^{\mathsf T} \ii\beta_\ell J(D_y+q_n)\phi_{q_n}\big),
\end{aligned}
\end{equation}
where
\begin{equation*}
\mathcal R(q):=Q(q)\left((H(q)-\lambda_0)\big|_{\operatorname{Ran}Q(q)}\right)^{-1}Q(q),\quad
Q(q):=I-P(q) .
\end{equation*}
Consequently, \eqref{eq:two_scale_proof_15}-\eqref{eq:two_scale_proof_16} yield the desired matrix form
\begin{equation} \label{eq:two_scale_proof_17}
\begin{aligned}
 &(\mathcal U_N^*P_NV_{1,N}R_NV_{1,N}P_N\mathcal U_N)_{n'n}\\
 &\quad=\sum_{\ell,s=1}^2
 \sum_{\substack{a\in\Z^2\\|a-n|\le R_0,\ |n'-a|\le R_0}}
 \widehat B_\ell(n'-a)\widehat B_s(a-n)\\
 &\qquad\times\Big\langle
 \mathcal R(q_a)(D_y+q_a)^{\mathsf T}
 \ii\beta_sJ(D_y+q_n)\phi_{q_n},
 (D_y+q_a)^{\mathsf T}
 \ii\beta_\ell J(D_y+q_{n'})\phi_{q_{n'}}
 \Big\rangle_{L^2((0,1)^2)}.
\end{aligned}
\end{equation}
Here we have applied 1) the unit-cell orthogonality calculation in Step 3.2, which forces the two intermediate indices to agree modulo $N\Z^2$ and hence equal exactly as these indices belong to fixed finite sets, and 2) the identity $\overline{\widehat B_\ell(a-n')}=\widehat B_\ell(n'-a)$. To calculate the limit of \eqref{eq:two_scale_proof_17} as $N\to\infty$, we note the following facts:
\begin{itemize}
    \item[(i)] $\mathcal R(q)\to R$ in $L^2\to H^2$ norm, with the operator $R:=(H_0-\lambda_0)^{-1}_{\phi^\perp}$ defined above \eqref{eq:V_C_def} and extended by zero on $\operatorname{span}\{\phi\}$. Indeed, $H(0)-\lambda_0+P(0)$ is invertible from the fixed $k_0$-quasiperiodic $H^2$ space to $L^2$, and $H(q)-\lambda_0+P(q)$ converges to it in the $H^2\to L^2$ norm. Then, the continuity of inversion together with the identity $\mathcal R(q)=Q(q)\bigl(H(q)-\lambda_0+P(q)\bigr)^{-1}Q(q)$ leads to the conclusion.
    \item[(ii)] For each $a$ in the summation of \eqref{eq:two_scale_proof_17}, the following convergence holds in $L^2$ by the smooth $H^2$ dependence of $\phi_q$,
    \begin{equation*}
    (D_y+q_a)^{\mathsf T}\ii\beta_sJ(D_y+q_n)\phi_{q_n}\to V_s\phi,\quad
    (D_y+q_a)^{\mathsf T}\ii\beta_\ell J(D_y+q_{n'})\phi_{q_{n'}}\to V_\ell\phi,
    \end{equation*}
    which are uniformly in $k$.
\end{itemize}
As a result, the inner product in \eqref{eq:two_scale_proof_17} has the following limit
\[
 \langle RV_s\phi,V_\ell\phi\rangle
 =\overline{\langle V_\ell\phi,RV_s\phi\rangle}
 =\overline{C_{\ell s}}=C_{\ell s},
\]
where the real coefficients $C_{\ell s}$ are introduced in \eqref{eq:V_C_def}, and $RV_s\phi$, $RV_{\ell}\phi$ are well-defined because $V_s\phi\in\phi^\perp$ as proved by $\langle\phi,V_s\phi\rangle=0$. In conclusion, using the Fourier convolution formula, we have shown that \eqref{eq:two_scale_proof_17} admits the following limit
\begin{equation} \label{eq:two_scale_proof_18}
\begin{aligned}
(\mathcal U_N^*P_NV_{1,N}R_NV_{1,N}P_N\mathcal U_N)_{n'n}
&\to\sum_{\ell,s=1}^2C_{\ell s}\sum_{a\in\Z^2}\widehat B_\ell(n'-a)\widehat B_s(a-n) \\
&=\sum_{\ell,s=1}^2C_{\ell s}\widehat{B_\ell B_s}(n'-n).
\end{aligned}
\end{equation}

{\color{blue}Step 3.5.} Here we summarize the results obtained from Step 3.2-3.4. First, we have seen that the matrix elements of each term in \eqref{eq:short-seed-schur} are compactly supported. Explicitly, one can check that
\[
 (F_N(k))_{n'n}=0
 \qquad\hbox{whenever }|n'-n|>2R_0.
\]
Hence, as all the computed limits in Steps 3.2-3.4 are uniform in $k$ for any fixed finite set of indices, we arrived at the following strong convergence with $u$ being any finite Fourier sum
\begin{equation}\label{eq:transfer-core-limit}
 \sup_{k\in\mathcal B}\|F_N(k)u-K(k)u\|_{L^2}\longrightarrow0.
\end{equation}
Indeed, using \eqref{eq:two_scale_proof_10}, \eqref{eq:two_scale_proof_11} and \eqref{eq:two_scale_proof_18}, the limiting operator of $F_N$ as $N\to\infty$ equals
\[
 (D_X+k)^{\mathsf T}M(D_X+k)
 +\frac12\sum_\ell\{D_{X_\ell}+k_\ell,B_\ell\}
 +\lambda_0v-\sum_{\ell,s}C_{\ell s}B_\ell B_s.
\]
Substituting $B=-2Mb$ and $\lambda_0v=V+b^{\mathsf T}Mb+\sum C_{\ell s}B_\ell B_s$ reduces this expression exactly to $(D_X+k-b)^{\mathsf T}M(D_X+k-b)+V=K(k)$.

The strong convergence in \eqref{eq:transfer-core-limit} is not sufficient to justify the resolvent convergence in the next step, for which we need to control the difference in \eqref{eq:transfer-core-limit} as the Fourier support of $u$ grows. To this end, we record the following uniform bound
\begin{equation} \label{eq:transfer-coercivity}
 \langle u,F_N(k)u\rangle
 \geq c_1\|(D_X+k)u\|^2-C_1\|u\|^2,
 \qquad u\in\mathcal E_N(k).
\end{equation}
The same inequality, with possibly different constants, holds for $K(k)$ by its magnetic quadratic form. Indeed, the leading order term $P_N(H_{0,N}-\lambda_0)P_N$ in \eqref{eq:short-seed-schur} is bounded below by $c\|(D_X+k)u\|^2$ using \eqref{eq:transfer-edge-chart}. On the other hand, the first perturbation term in \eqref{eq:short-seed-schur} is estimated as follows
\[
\left|\langle\mathcal U_Nu,\varepsilon^{-1}P_NV_{1,N}P_N\mathcal U_Nu\rangle\right|
\leq C\|(D_X+k)u\|\|u\|+C\|u\|^2,
\]
which is justified by \eqref{eq:transfer-cell-kernel} together with the finite Fourier support. The other two perturbation terms in $F_N$ are bounded in norm as shown in Step 2. This completes the proof of \eqref{eq:transfer-coercivity}.

\medskip\noindent
{\color{blue}Step 4. Resolvent convergence and the complete gap.}

In this step, we first improve the strong convergence \eqref{eq:transfer-core-limit} to the norm resolvent convergence on the envelope space, using the coercive estimate \eqref{eq:transfer-coercivity}; then, by transferring this result back to the physical space using the Schur complement and recalling that the target magnetic operator $K$ is gapped, we will conclude the existence of a gap of $H_{A_N}$ as claimed in Lemma \ref{lem:seed-transfer}. To start, we claim that
\begin{equation}\label{eq:transfer-effective-resolvent}
 \sup_k\|(F_N(k)-z)^{-1}\Pi_N(k)-(K(k)-z)^{-1}\|
 \to 0
\end{equation}
uniformly for $z$ in any compact subset of the resolvent set $\bigcap_{k\in\mathcal{B}}\rho(K(k))$, where $\Pi_N$ is introduced in \eqref{eq:transfer-envelope-space}. We first prove \eqref{eq:transfer-effective-resolvent} for a fixed spectral parameter and then extend it to the whole resolvent set by continuity. Choose $z_*\in\R$ strictly below the uniform lower bounds in \eqref{eq:transfer-coercivity}, and write $T_N(k):=(F_N(k)-z_*)^{-1}\Pi_N(k)$ and $T(k):=(K(k)-z_*)^{-1}$. Then the coercive estimate \eqref{eq:transfer-coercivity} gives the bound
\begin{equation*}
\|T_N(k)\|_{L^2\to H^1}+\|T(k)\|_{L^2\to H^1}\leq C.
\end{equation*}
Suppose \eqref{eq:transfer-effective-resolvent} fails at $z_*$. Then we can choose $k_N\to k$, unit vectors $f_N\rightharpoonup f$, and $a>0$ such that
\begin{equation}
\label{eq:two_scale_proof_19}
 \|(T_N(k_N)-T(k_N))f_N\|\geq a.
\end{equation}
By the $H^1$ bounds, the compactness of embedding $H^1(\T^2)\hookrightarrow L^2$, and selecting a subsequence if necessary, $T_N(k_N)f_N$ and $T(k_N)f_N$ converge strongly in $L^2$. Writing $w:=\lim_{N\to\infty}T_N(k_N)f_N$, for every finite Fourier sum $u$, the strong convergence \eqref{eq:transfer-core-limit} and self-adjointness give
\[
 \begin{split}
 \langle(K(k)-z_*)u,w\rangle
 =\lim_N\langle(F_N(k_N)-z_*)u,T_N(k_N)f_N\rangle=\lim_N\langle u,f_N\rangle=\langle u,f\rangle.
 \end{split}
\]
Since the operator $K(k)$ has domain $H^2(\T^2)$ (recall that its kinetic matrix is constant positive definite and its lower-order coefficients are smooth), the finite Fourier sums form an operator core of $K(k)$. Hence, the above identity implies that $w=T(k)f$. Applying the same argument to $T(k_N)f_N$ gives the same limit, which consequently contradicts \eqref{eq:two_scale_proof_19}. This proves \eqref{eq:transfer-effective-resolvent} for $z=z_*$. To prove \eqref{eq:transfer-effective-resolvent} for general $z$, note that
\[
 (F_N(k)-z)^{-1}\Pi_N
 =T_N(k)[I-(z-z_*)T_N(k)]^{-1}.
\]
The limiting inverse at the right side exists and is uniformly bounded on the stated compact sets. Thus, by the justified norm convergence at $z=z_*$ and the Neumann series, we know \eqref{eq:transfer-effective-resolvent} holds within a neighborhood of $z_*$. A finite cover of the concerned compact complex set then completes the proof. In particular, we point out that \eqref{eq:transfer-effective-resolvent} holds by incorporating the perturbation term in \eqref{eq:transfer-schur-remainder}, i.e., replacing $F_N(k)$ with $F_N(k)+\mathfrak r_N(k,z)$, using the estimate $\|\mathfrak r_N(k,z)\|\leq C\varepsilon$.

We next transfer the resolvent convergence \eqref{eq:transfer-effective-resolvent} back to the physical space. We will present the results with a smooth momentum cutoff map, which replaces its discontinuous counterpart $\mathcal D_{N,k}^{-1}\mathcal U_N(k)\Pi_N(k)$ and is introduced mainly for later use: it will give a continuous isomorphism between eigenbundles that identify the gap Chern numbers. To this end, we choose a real-valued function $\chi\in C_c^\infty(\{|q|<\rho\})$ with $0\leq\chi\leq1$ and $\chi(q)=1$ for $|q|\leq a$, where $a>0$. Then the cutoff map is defined as
\begin{equation}\label{eq:transfer-smooth-lift}
\mathcal J_N(k)f(X):=\sum_{n\in\Z^2}\widehat f(n)\chi(q_n(k))e^{2\pi\ii n\cdot X}\phi_{q_n(k)}(NX).
\end{equation}
or equivalently,
\begin{equation*}
\mathcal J_N=\mathcal D_{N,k}^{-1}\mathcal U_N
\chi(\varepsilon(D_X+k))
\end{equation*}
with the Fourier multiplier restricted to its support. In particular, using the properties of $\mathcal D_{N,k}$ and $\mathcal U_N$ from Step 1, we have
\begin{equation}\label{eq:transfer-lift-metric}
 \mathcal J_N(k)^*\mathcal J_N(k)
 =\chi(\varepsilon(D_X+k))^2,\quad \|\mathcal J_N(k)\|\leq1.
\end{equation}
For fixed $N$ and near a fixed $k$, only finitely many indices occur in the sum of \eqref{eq:transfer-smooth-lift}; hence, by the fact that $\chi\phi_q$ is smooth and vanishes outside the support of $\chi$, we know that the map $k\mapsto \mathcal J_N(k)$ is smooth in operator norm.

With this notion and using the Schur complement adapted to $\operatorname{Ran}P_N\oplus \operatorname{Ran}Q_N$ (recall the result from Step 2, in particular the definition \eqref{eq:two_scale_proof_4}), the inverse of $H_N-\lambda_0-\varepsilon^2 z$ admits the following block form
\[
 \begin{pmatrix}
 \mathscr S_N^{-1}&-\mathscr S_N^{-1}B_NC_N^{-1}\\
 -C_N^{-1}D_N\mathscr S_N^{-1}&
 C_N^{-1}+C_N^{-1}D_N\mathscr S_N^{-1}B_NC_N^{-1}
 \end{pmatrix}.
\]
Here $B_N:=P_NH_NQ_N$, $D_N:=Q_NH_NP_N$, which satisfy the estimates $\|B_N\|+\|D_N\|=O(\varepsilon)$. Note also by \eqref{eq:transfer-schur-remainder} and \eqref{eq:transfer-effective-resolvent} that $\|\mathscr S_N^{-1}\|=O(\varepsilon^{-2})$.
The four blocks therefore have orders $O(\varepsilon^{-2})$, $O(\varepsilon^{-1})$, $O(\varepsilon^{-1})$, and $O(1)$ in norm, respectively. By \eqref{eq:transfer-dilation}, the resolvent of the physical operator $H_{A_N}(k)$ is their unitary conjugate multiplied by $N^{-2}=\varepsilon^2$. This implies that only the first block, which consists only of the Schur complement, contributes to the limit of the resolvent of $H_{A_N}(k)$. Moreover, in that block, inserting the smooth cutoff does not change the limit. Indeed, $1-\chi(q_n)\ne0$ implies $|p_n|\geq a/\varepsilon$, and the $L^2\to H^1$ estimate for the effective resolvent gives
\[
 \|[I-\chi(\varepsilon(D_X+k))](K(k)-z)^{-1}\|
 \leq C\varepsilon.
\]
The right cutoff yields the same bound by taking adjoints and replacing $z$ by $\overline z$. In conclusion, using equation \eqref{eq:transfer-effective-resolvent}, we have proved the desired estimate
\begin{equation}\label{eq:transfer-full-resolvent}
\sup_k\big\|(H_{A_N}(k)-N^2\lambda_0-z)^{-1}-\mathcal J_N(k)(K(k)-z)^{-1}\mathcal J_N(k)^*\big\| \to 0,
\end{equation}
uniformly on compact subsets of the common resolvent set of $K(k)$. This immediately gives the existence of spectral gap of $H_{A_N}$. In fact, by Lemma \ref{lem:seed-magnetic-target}, we can choose $\gamma>0$ with $[-2\gamma,2\gamma]\cap\sigma(K(k))=\varnothing$ for every $k$. For $z\in[-\gamma,\gamma]$, both $C_N(z)$ and its Schur complement are invertible for large $N$, by Steps 2 and 4. Therefore,
\[
 (N^2\lambda_0-\gamma,N^2\lambda_0+\gamma)
 \cap\sigma(H_{A_N})=\varnothing,
\]
which justifies the claim of gap existence in Lemma \ref{lem:seed-transfer}.

For later use, we record in the following how the map $\mathcal J_N$ identifies the Bloch modes. For $U_\ell f(X)=e^{-2\pi\ii\ell\cdot X}f(X)$, reindexing \eqref{eq:transfer-smooth-lift} gives
\begin{equation}\label{eq:transfer-lift-equivariance}
 \mathcal J_N(k+2\pi\ell)U_\ell=U_\ell\mathcal J_N(k),
 \qquad \ell\in\Z^2.
\end{equation}
Thus, $\mathcal J_N$ defines a map between families with the same Bloch identifications on the boundary of $\mathcal B$.

\medskip\noindent
{\color{blue}Step 5. Identification of the Chern number}

Since the resolvent convergence \eqref{eq:transfer-full-resolvent} holds on any compact subset of the common resolvent set of $K(k)$, it, together with a fixed contour integral, identifies the spectral projection of $H_{A_N}(k)$ on an interval of fixed length, i.e., $\one_{[N^2\lambda_0-L_0,N^2\lambda_0]}(H_{A_N}(k))$ for any $L_0>0$ providing that $N^2\lambda_0-L$ is a gap point. This does not identify the full occupied spectral bundle below $N^2\lambda_0$. To solve this, we first fix a gap point $\lambda_0^{\prime}\notin \spec(H_{S})$ and show that the occupied bundle below $N^2\lambda_0^{\prime}$ is topologically trivial. Physically, this triviality follows by the facts that 1) the target magnetic operator controls only the Chern number of the near-edge band group, on which it is threaded to the host operator $H_{S_N}$, and 2) the host operator $H_{S_N}$ is topologically trivial since $S_N$ is real-valued. Then, showing that the bands between $N^2\lambda_0^{\prime}$ and $N^2\lambda_0$ have exactly the target Chern number $m$, we conclude the proof of Lemma \ref{lem:seed-transfer}.

{\color{blue}Step 5.1.} Select $\lambda_0^{\prime}\notin \spec(H_{S})$, which lies inside the first gap of $H_{S}$ immediately below $\lambda_0$, and consider the homotopic family
\[
 A_{N,t}:=S_N+t(A_N-S_N),\quad 0\leq t\leq 1.
\]
Then, the uniform ellipticity and the estimate $\|A_N-S_N\|_\infty\leq C\varepsilon$ (recall \eqref{eq:short-seed-coefficient} and $\varepsilon=N^{-1}$) imply, in the sense of sesquilinear form,
\[
 (1-C\varepsilon)H_{S_N}(k)
 \leq H_{A_{N,t}}(k)
 \leq(1+C\varepsilon)H_{S_N}(k).
\]
By the min-max principle, the same inequalities hold for every ordered Bloch eigenvalue. For sufficiently large $N$, this implies that $N^2\lambda_0^{\prime}\notin \spec(H_{A_{N,t}}(k))$ for every $k,t$: otherwise, \(H_S(\theta)\) has an eigenvalue for some microscopic momentum \(\theta\in\mathcal B\), which is separated from $\lambda_0^{\prime}$ only by the order $1/N$, contradicting the uniform isolation of $\lambda_0^{\prime}$. Thus, the projection $P_{N,d,t}(k):=\one_{(-\infty,N^2\lambda_0^{\prime})}(H_{A_{N,t}}(k))$ is well-defined and is a continuous homotopy of spectral bundles. Hence, we conclude that
\begin{equation}\label{eq:transfer-lower-chern}
 c(P_{N,d,1})=c(P_{N,d,0})=0,
\end{equation}
since $H_{S_N}$ is topologically trivial, recalling that the coefficient matrix $S_N$ is real-valued. Moreover, corresponding to the argument at the beginning of Step 5, we claim that for a constant $L_0$ independent of $N$,
\begin{equation}\label{eq:transfer-lower-exclusion}
 (N^2\lambda_0^{\prime},N^2\lambda_0-L_0)
 \cap\sigma(H_{A_N}(k))=\varnothing
 \quad\hbox{for every }k.
\end{equation}
To prove this exclusion, we use the Schur complement near $\lambda_0$ on the remaining compact part of the host gap. To be precise, we choose $s_0<d/2$ with $\lambda_0-s_0>\lambda_0^{\prime}$. For \(0\le s\le s_0\), the restriction of \(Q_N(H_N(k)-\lambda_0+s)Q_N\) to \(\operatorname{Ran}Q_N\) has a uniformly bounded inverse, by \eqref{eq:transfer-complement-gap} and the Neumann-series argument of Step 2. By introducing the notation
\[
 [qu]:=\big(\sum_{n\in I_N(k)}|q_n(k)|^2|\widehat u(n)|^2\big)^{\frac{1}{2}},
\]
we know by the unperturbed band estimate \(\lambda(q)-\lambda_0\ge c|q|^2\) together with the  first-order estimate in Step 3 that the Schur complement satisfies the estimate
\begin{equation*}
\langle\mathcal U_Nu,\mathscr S_N(-s/\varepsilon^2)\mathcal U_Nu\rangle
 \geq c[qu]^2-C\varepsilon[qu]\|u\| +(s-C\varepsilon^2)\|u\|^2.
\end{equation*}
In particular, completing the square gives a constant $L_0>0$, independent of $N$ and $k$, such that the right side is bounded below by $(s-L_0\varepsilon^2)\|u\|^2$. Hence, the Schur complement $\mathscr S_N(-s/\varepsilon^2)$ is invertible whenever $L_0\varepsilon^2<s<s_0$, and therefore
\[
 (\lambda_0-s_0,\lambda_0-L_0\varepsilon^2) \cap\sigma(H_N(k))=\varnothing.
\]
To cover the remaining compact interval $J_0:=[\lambda_0^{\prime},\lambda_0-s_0]$, we note that it is contained in the host gap, i.e.,
\[
 \operatorname{dist}(J_0,\sigma(H_S))>0.
\]
Hence, the form comparison and the min-max principle show that an eigenvalue of $H_N(k)$ in $J_0$ would require an eigenvalue of $H_{0,N}(k)$ within an $O(\varepsilon)$ neighborhood of $J_0$. By band folding, every such unperturbed eigenvalue belongs to $\sigma(H_S)$, which leads to a contradiction. Thus $J_0\cap\sigma(H_N(k))=\varnothing$. Combining the two exclusions and applying \eqref{eq:transfer-dilation} proves \eqref{eq:transfer-lower-exclusion}.


{\color{blue}Step 5.2.} To identify the occupied Chern number by a contour integral and the Riesz formula, we first choose the lower endpoint of the near-edge spectral interval. Let $L_0$ be the constant \eqref{eq:transfer-lower-exclusion}. Since $K(k)$ is uniformly bounded below, we may fix
\[
 L>\max\left\{
 L_0+1,\,
 1-\inf_{k\in\mathcal B}\inf\sigma(K(k))
 \right\}.
\]
Then $-L$ lies strictly below the spectrum of $K(k)$ for every $k$. For all sufficiently large $N$, $N^2\lambda_0-L$ lies strictly inside the interval excluded by \eqref{eq:transfer-lower-exclusion}. The upper endpoints $0$ and $N^2\lambda_0$ lie in the respective complete gaps established in Step 4.

Let $\Gamma$ be the counterclockwise boundary of the rectangle with vertical sides $\operatorname{Re}z=-L,0$ and horizontal sides $\operatorname{Im}z=\pm1$. It lies in the common resolvent sets of $K(k)$ and $H_{A_N}(k)-N^2\lambda_0$, and encloses all negative eigenvalues of $K(k)$. Therefore, the Riesz formula gives
\begin{equation*}
\begin{aligned}
&P_-(k):=\one_{(-\infty,0)}(K(k))=-\frac{1}{2\pi\ii}\int_\Gamma(K(k)-z)^{-1}\,dz, \\
&P_{N,e}(k):=\one_{(N^2\lambda_0-L,N^2\lambda_0)}(H_{A_N}(k))=-\frac{1}{2\pi\ii}\int_\Gamma(H_{A_N}(k)-N^2\lambda_0-z)^{-1}\,dz .
\end{aligned}
\end{equation*}
These formulas are well-defined for all sufficiently large $N$ by the choice of $\Gamma$ above. Thus, the resolvent convergence \eqref{eq:transfer-full-resolvent} implies that
\begin{equation}\label{eq:transfer-projection-comparison}
 \sup_k\|P_{N,e}(k)-\mathcal J_N(k)P_-(k)\mathcal J_N(k)^*\|
 \longrightarrow0.
\end{equation}
However, $\mathcal J_N(k)P_-(k)\mathcal J_N(k)^*$ may not be a projection because $\mathcal J_N$ contains a cutoff. To remedy this, we now introduce a projection-valued normalization of $\mathcal J_N(k)P_-(k)\mathcal J_N(k)^*$. First, the coercive estimate for $K$, restricted to its negative spectral subspace, gives $\|P_-(k)\|_{L^2\to H^1}\leq C$. This, together with \eqref{eq:transfer-lift-metric} and the Fourier tail estimate
in Step 4, implies that
\begin{equation} \label{eq:two_scale_proof_20}
\sup_k\|(\mathcal J_N(k)^*\mathcal J_N(k)-I)P_-(k)\| \to 0.
\end{equation}
To perform the normalization on a fixed Hilbert space, define
\[
 G_N(k):=P_-(k)\mathcal J_N(k)^*\mathcal J_N(k)P_-(k)+I-P_-(k).
\]
Thanks to \eqref{eq:two_scale_proof_20}, we know $\sup_k\|G_N(k)-I\|\to 0$; therefore, $G_N$ is positive and invertible for large $N$ and $G_N^{-1/2}\to I$ uniformly. Consequently, defining that
\[
 W_N(k):=\mathcal J_N(k)G_N(k)^{-1/2}P_-(k),
\]
we know $G_N$ commutes with $P_-$, and $W_N^*W_N=P_-$. The desired normalization is hence defined as its range projection
\begin{equation*}
\widetilde P_N:=W_NW_N^* ,
\end{equation*}
which, by \eqref{eq:transfer-projection-comparison}, satisfies
\[
 \sup_k\|\widetilde P_N(k)-P_{N,e}(k)\|\longrightarrow0.
\]
For large $N$, this norm is less than one. Consequently, the two bundles are isomorphic by similar arguments in Step 2 of Lemma~\ref{lem:seed-magnetic-target}. Thus,
\[
 P_{N,e}(k)W_N(k):\operatorname{Ran}P_-(k)
                  \longrightarrow\operatorname{Ran}P_{N,e}(k)
\]
is an isomorphism, that is continuous in $k$ and obeys the transition rule \eqref{eq:transfer-lift-equivariance}. In conclusion, $P_{N,e}(k)W_N(k)$ defines an isomorphism of Bloch bundles, which proves the identification of Chern number $c(P_{N,e})=c(P_-)=m$.

Finally, recalling \eqref{eq:transfer-lower-chern} and \eqref{eq:transfer-lower-exclusion}, we conclude that for all sufficiently large $N$,
\[
\operatorname{Ch}(A_N;N^2\lambda_0) =c(P_{N,d,1})+c(P_{N,e})=0+m=m.
\]
This completes the proof of Lemma \ref{lem:seed-transfer}.

\section{Concluding remarks}\label{sec:conclusion}

In this paper, we developed a thin-layer induction framework for producing infinitely many complete spectral gaps for periodic divergence-form operators on $\R^2$. The main result, Theorem~\ref{thm:main}, shows the existence of a single coefficient matrix whose associated operator possesses infinitely many spectral gaps. Moreover, coefficients with this property form a dense $G_\delta$ subset of the specified real symmetric ellipticity class in the $L^1$ metric. Therefore, we obtain infinitely many counterexamples to the Bethe-Sommerfeld conjecture for the class of divergence-form operators.

Importantly, we show in the second main result, Theorem~\ref{thm:chern-main} that the same thin-layer mechanism also controls the spectral topology: one can realize any prescribed integer sequence as the occupied Chern numbers of selected gap components tending to infinity in a single operator. The ellipticity bounds of the constructed operator remain fixed, and the imaginary part of the coefficient can be arbitrarily small. Thus, this result shows that one can realize arbitrary topological complexity without increasing the coefficient bounds or the strength of the time-reversal breaking field, at the cost of increasingly fine spatial structure.

As an on-going work, we will address the BSC of the full three-dimensional Maxwell operator in a second paper, incorporating the divergence constraint and a seed with a complete Maxwell gap. Another question concerns the role of coefficient regularity as the limiting media constructed here may be discontinuous. Therefore, the Maxwell BSC with continuous coefficients as formulated in
Section~\ref{sec:perspective} remains conjectural. We leave it for future study.

\appendix
\section{Reduction of Maxwell equations to divergence form}
\label{app:maxwell-reduction}
\setcounter{equation}{0}
\setcounter{subsection}{0}
\setcounter{theorem}{0}
\renewcommand{\theequation}{A.\arabic{equation}}
\renewcommand{\theHequation}{appendix.A.\arabic{equation}}
\renewcommand{\thesubsection}{A.\arabic{subsection}}
\renewcommand{\thetheorem}{A.\arabic{theorem}}

We explain how the operator $H_A$ arises from a planar polarization of Maxwell's equations, following \cite[Appendix~A.2--A.4]{lee2019elliptic}. Write $x=(x_1,x_2)$ and let $x_3$ denote the longitudinal coordinate. We use a lossless, frequency-independent constitutive model $\mathbf D=\varepsilon\mathbf E$, $\mathbf B=\mathbf H$, with
\[
 \varepsilon(x)=
 \begin{pmatrix}\varepsilon_\perp(x)&0\\0&1\end{pmatrix},
 \quad
 0<cI_2\leq\varepsilon_\perp(x)
       =\varepsilon_\perp(x)^*\leq CI_2<\infty.
\]
In particular, the coefficients are independent of $x_3$, which is necessary for a planar reduction. In units in which the vacuum wave speed is one, the harmonic fields with time dependence $e^{-\ii\omega t}$ satisfy the time-independent Maxwell equations
\begin{equation}\label{eq:app-maxwell-system}
 \operatorname{curl}\mathbf E=\ii\omega\mathbf H, \quad
 \operatorname{curl}\mathbf H=-\ii\omega\varepsilon\mathbf E, \quad
 \operatorname{div}(\varepsilon\mathbf E) =\operatorname{div}\mathbf H=0.
\end{equation}
We consider fields independent of $x_3$, or equivalently the zero longitudinal Fourier parameter, with transverse electric (TE) polarization
\[
 \mathbf H(x)=(0,0,u(x)),\quad \mathbf E(x)=(E_1(x),E_2(x),0).
\]
To obtain the scalar equation under this polarization, we first eliminate the electric field. For $\omega\ne0$, the first two equations in \eqref{eq:app-maxwell-system} give
\[
 \mathbf E=\frac{\ii}{\omega}\varepsilon^{-1}
                  \operatorname{curl}\mathbf H,
 \quad
 \operatorname{curl}\bigl(\varepsilon^{-1}
                  \operatorname{curl}\mathbf H\bigr)
       =\omega^2\mathbf H.
\]
Let $J=\left(\begin{smallmatrix}0&-1\\1&0\end{smallmatrix}\right)$, and write $\nabla=(\partial_1,\partial_2)^{\mathsf T}$. The identities
\[
 \operatorname{curl}(u e_3)=(-J\nabla u,0),\quad
 \operatorname{curl}(v_1,v_2,0)=(0,0,-\diver(Jv))
\]
then imply
\begin{equation}\label{eq:app-maxwell-reduction}
\operatorname{curl}\bigl(\varepsilon^{-1}\operatorname{curl}(u e_3)\bigr)=(H_Au)e_3,\quad
\text{with}\quad A:=J^{\mathsf T}\varepsilon_\perp^{-1}J.
\end{equation}
Thus, the longitudinal magnetic component satisfies $H_Au=-\diver(A\nabla u)=\omega^2u$. Conversely, a solution of this scalar equation determines the TE field by
\begin{equation}\label{eq:app-maxwell-reconstruction}
 \mathbf H=u e_3,\quad \mathbf E=-\frac{\ii}{\omega}(\varepsilon_\perp^{-1}J\nabla u,0).
\end{equation}
Both divergence constraints hold: $\partial_3u=0$ and
$\varepsilon\mathbf E=-(\ii/\omega)(J\nabla u,0)$, while
$\diver(J\nabla u)=0$. Note that the reduction also holds for bounded measurable coefficients. Indeed, the magnetic energy per unit longitudinal length satisfies
\begin{equation}\label{eq:app-maxwell-form}
 \int_{\R^2}
  \bigl(\varepsilon^{-1}\operatorname{curl}(u e_3)\bigr)
       \cdot\overline{\operatorname{curl}(v e_3)}\dd
 =\int_{\R^2}(A\nabla u)\cdot\overline{\nabla v}\dd,
 \qquad u,v\in H^1(\R^2).
\end{equation}
The right side is precisely the sesquilinear form that defines $H_A$. Consequently, \eqref{eq:app-maxwell-reduction} is an identity of the associated operators in the weak sense; in particular, no derivatives of the material coefficients are required.

\section{Proofs of the auxiliary Lemmas for Proposition~\ref{prop:insert}}
\label{app:prop-insert-auxiliary-lemmas}

\begin{proof}[Proof of Lemma~\ref{lem:layer-block}]
We regard $B_C(k)$ as an operator from $L^2(\Omega)$ to $L^2(\Omega;\C^2)$ with domain $H_{\mathrm{per}}^1(\Omega)$. Uniform ellipticity and $\nabla=\nabla_k-\ii k$ give
\[
 \|u\|_2^2+\|B_C(k)u\|_2^2\asymp\|u\|_{H^1(\Omega)}^2,
 \qquad u\in H_{\mathrm{per}}^1(\Omega).
\]
Thus, $B_C(k)$ is closed and densely defined. Distributional integration by parts on $\T^2$ gives
\[
 B_C(k)^*v=-(\nabla+\ii k)\cdot(C^{1/2}v),
 \qquad
 \Dom (B_C(k)^*)=C^{-1/2}H(\diver;\T^2).
\]
By the block-operator construction, $M_C(k)$ is self-adjoint on the $k$-independent domain $\Dom (B_C(k))\oplus\Dom (B_C(k)^*)$ in
\eqref{eq:domain}. On this common domain, $M_C(k)-M_C(0)$ has off-diagonal entries $B_C(k)-B_C(0)=\ii C^{1/2}k$ and its adjoint. Both are bounded multiplication operators linear in $k$, so $M_C(k)$ depends affinely on $k$.

Since $H_C(k)=B_C(k)^*B_C(k)$, the standard spectral correspondence for self-adjoint block operators \cite[Theorem~A.3(i), (iii)]{GesztesyEtAl2012} gives \eqref{eq:block-positive-spectrum}. For each positive eigenvalue $\lambda$ of $H_C(k)$, the map $u\mapsto (u,\lambda^{-1/2}B_C(k)u)$ identifies its eigenspace with that of $M_C(k)$ at $\sqrt{\lambda}$, with projection onto the first component as its inverse. Thus the multiplicities agree.
\end{proof}

\begin{proof}[Proof of Lemma~\ref{lem:reference-hole}]
{\color{blue}Step 1.} \emph{Counting eigenfrequencies.}
By Lemma~\ref{lem:layer-block}, we can bound the number of spectral points of $M_C(0)$ in $N\overline{G_1} [N\omega_{1,-},N\omega_{1,+}]$ by counting eigenvalues of $H_C(0)$. Let $0=\lambda_1\le\lambda_2\le\cdots$ and $0=\nu_1\le\nu_2\le\cdots$ be the eigenvalues of $H_C(0)$ and the periodic Laplacian, respectively, counted with multiplicity. Uniform ellipticity gives
\[
 q_{C,0}[u]\ge\alpha\int_\Omega|\nabla u|^2\,\mathrm dx,
 \qquad u\in H_{\mathrm{per}}^1(\Omega).
\]
By the min--max principle on the common form domain, $\lambda_j\ge\alpha\nu_j$ for every $j\ge1$. Since the periodic Laplacian has eigenvalues $4\pi^2|n|^2$, $n\in\Z^2$, we obtain, for $R\ge0$,
\[
 \#\{j:\lambda_j\le R^2\}
 \le
 \#\left\{n\in\Z^2:
 |n|\le\frac{R}{2\pi\sqrt{\alpha}}\right\}
 \le\left(1+\frac{R}{\pi\sqrt{\alpha}}\right)^2.
\]
The last inequality follows by counting lattice points in the enclosing square. Taking $R=N\omega_{1,+}$ and applying Lemma~\ref{lem:layer-block}, we obtain
\begin{equation}\label{eq:reference-hole_proof_1}
 \begin{aligned}
 \#\bigl(\sigma(M_C(0))\cap N\overline{G_1}\bigr)
 \le\#\{j:\lambda_j\le N^2\omega_{1,+}^2\}
 \le\left(1+\frac{N\omega_{1,+}}{\pi\sqrt{\alpha}}\right)^2
 \le C_{\mathrm{count}}N^2.
 \end{aligned}
\end{equation}

{\color{blue}Step 2.} \emph{Finding a reference gap.}
We now use the counting estimate \eqref{eq:reference-hole_proof_1} and the pigeonhole principle to find a reference gap. Let $q$ be the number of distinct spectral points of $M_C(0)$ in $N\overline{G_1}$. By \eqref{eq:reference-hole_proof_1}, $q\le C_{\mathrm{count}}N^2$. The spectral points in $NG_1$ divide it into at most $q+1$ open intervals disjoint from $\sigma(M_C(0))$. Their lengths sum to $N|G_1|$, so at least one such interval $(l,r)$ satisfies
\[
 \ell:=r-l
 \ge\frac{N|G_1|}{q+1}
 \ge\frac{N|G_1|}{C_{\mathrm{count}}N^2+1}
 \ge\frac{|G_1|}{C_{\mathrm{count}}+1}\frac1N
 =\frac cN.
\]
This gives \eqref{eq:reference-hole-gap}. To obtain a spectral margin, take its central half $J_N=[l+\ell/4,r-\ell/4]$. Since $(l,r)$ contains no spectrum, every $\omega\in J_N$ satisfies
\[
 \operatorname{dist}\bigl(\omega,\sigma(M_C(0))\bigr)
 \ge\min\{\omega-l,r-\omega\}
 \ge\frac{\ell}{4}.
\]
Taking the infimum over $\omega\in J_N$ and using $\ell\ge c/N$ gives \eqref{eq:reference-hole-separation}.
\end{proof}

\begin{proof}[Proof of Lemma~\ref{lem:layer-decay}]

Note that \eqref{eq:CT} follows by \eqref{eq:seedgap} using the Combes-Thomas argument (e.g. \cite[Lemma 12]{figotin1996localization}). The essential idea underlying \eqref{eq:seedgap} is the scaling covariance of the operator $M_{S}$ (see Remark \ref{rmk:scaling_covariance} for an intuitive illustration). 

{\color{blue}Step 1.} \emph{Scaling the seed gap.} We decompose the fibers of $S_N$ on the original unit cell by grouping Fourier frequencies modulo $N$. Let $\mathcal I_N:=\{0,\ldots,N-1\}^2$. For $\ell\in\mathcal I_N$, let $\mathcal H_{\ell,N}$ be the closed subspace
of $L^2(\T^2)$ with Fourier frequencies in $\ell+N\Z^2$, and define
\[
 (U_{\ell,N}f)(x):=e^{2\pi\ii\ell\cdot x}f(Nx).
\]
Since $U_{\ell,N}e^{2\pi\ii n\cdot x} =e^{2\pi\ii(\ell+Nn)\cdot x}$, this map is unitary from $L^2(\T^2)$ onto $\mathcal H_{\ell,N}$.
The frequency classes partition $\Z^2$, and $1+|\ell+Nn|^2\asymp1+|n|^2$ for fixed $N,\ell$. Thus the Fourier characterization of $H^1$ gives
\[
 L^2(\T^2)=\bigoplus_{\ell\in\mathcal I_N}\mathcal H_{\ell,N},
 \qquad
 U_{\ell,N}H^1(\T^2)=\mathcal H_{\ell,N}\cap H^1(\T^2).
\]
We extend $U_{\ell,N}$ componentwise to vector-valued functions. Using $S_N^{1/2}(x)=S^{1/2}(Nx)$, the product and chain rules yield
\[
 B_{S_N}(k)U_{\ell,N}
 =N U_{\ell,N}B_S\!\left(\frac{k+2\pi\ell}{N}\right)
 \qquad\text{on }H^1(\T^2).
\]
These identities and the domain correspondence give a unitary direct-sum decomposition of $B_{S_N}(k)$. Taking adjoints gives the corresponding decomposition of
$B_{S_N}(k)^*$, so
\begin{equation}\label{eq:seed-fiber-dilation}
 M_{S_N}(k)\simeq
 \bigoplus_{\ell\in\mathcal I_N}
 N M_S\!\left(\frac{k+2\pi\ell}{N}\right),
\end{equation}
where $\simeq$ denotes unitary equivalence. The seed gap, \eqref{eq:block-positive-spectrum}, and the periodicity of the fiber spectra give $G\cap\spec(M_S(q))=\varnothing$ for every $q\in\R^2$. By spectral scaling and the spectral formula for finite direct sums, \eqref{eq:seed-fiber-dilation} yields
\[
 \begin{aligned}
 \sigma(M_{S_N}(k))
 =\bigcup_{\ell\in\mathcal I_N}
 \sigma\!\left(
 N M_S\!\left(\frac{k+2\pi\ell}{N}\right)
 \right)
 =\bigcup_{\ell\in\mathcal I_N}
 N\,\sigma\!\left(
 M_S\!\left(\frac{k+2\pi\ell}{N}\right)
 \right)
\subset\R\setminus NG.
 \end{aligned}
\]
Consequently, for $k\in\mathcal B$ and $\omega\in N\overline{G_0}$,
\[
 \dist\bigl(\omega,\spec(M_{S_N}(k))\bigr)
 \ge \dist\bigl(\omega,\R\setminus NG\bigr)
 \ge N\delta.
\]
This establishes the spectral separation in \eqref{eq:seedgap}, uniformly for $k\in\mathcal B$ and $\omega\in N\overline{G_0}$.

{\color{blue}Step 2.} \emph{Local mass estimate.} We now apply \eqref{eq:CT} to control the mass of a mixed-medium eigenmode in $\Sigma_{w/4}$. Let $F=(u,v)\in\Dom M_D(k)$ be a normalized eigenvector satisfying
\[
 M_D(k)F=\omega F,\qquad
 \omega\in N\overline{G_0}.
\]
Define the periodic Lipschitz cutoff
\[
 \chi(x):=
 \begin{cases}
  1,&d(x)\leq w/2,\\
  3-4d(x)/w,&w/2<d(x)<3w/4,\\
  0,&d(x)\geq3w/4.
 \end{cases}
\] 
The transition region $E:=\operatorname{supp}\nabla\chi$ is contained in $\{w/2\le d\le3w/4\}$. Since $d$ is $1$-Lipschitz,
\[
 \norm{\nabla\chi}_\infty\le\frac4w,
 \qquad
 \dist_{\T^2}(\Sigma_{w/4},E)\ge\frac w4.
\]
Multiplication by the Lipschitz function $\chi$ preserves $H^1(\T^2)$ and $H(\diver;\T^2)$. Since $F=(u,v)\in\Dom M_D(k)$, we obtain $\chi u\in H^1(\T^2)$ and $\chi D^{1/2}v\in H(\diver;\T^2)$. Moreover, $\operatorname{supp}\chi\subset\Sigma_w$, where $D=S_N$,
so $S_N^{1/2}(\chi v)=\chi D^{1/2}v$. Thus $\chi F$ belongs to the domains of both $M_D(k)$ and $M_{S_N}(k)$. The product rule and the eigenvalue equation yield
\[
 (M_{S_N}(k)-\omega)(\chi F)
 =(M_D(k)-\omega)(\chi F)
 =[M_D(k),\chi]F=:g.
\]
The commutator is a multiplication operator supported in $E$, with pointwise operator norm at most $\sqrt\beta\,|\nabla\chi|$. Consequently,
\[
 g=\one_Eg,\qquad
 \norm{g}_2
 \le\sqrt\beta\,\norm{\nabla\chi}_\infty\norm{F}_2
 \le\frac{4\sqrt\beta}{w}.
\]
By \eqref{eq:seedgap}, the localized equation gives $\chi F=(M_{S_N}(k)-\omega)^{-1}g$. Since $\chi=1$ on $\Sigma_{w/4}$, we obtain
\[
 \one_{\Sigma_{w/4}}F
 =\one_{\Sigma_{w/4}}(M_{S_N}(k)-\omega)^{-1}\one_Eg.
\]
Applying \eqref{eq:CT} with $X=\Sigma_{w/4}$ and $Y=E$, we obtain
\[
 \begin{aligned}
 \norm{\one_{\Sigma_{w/4}}F}_2
 \le\frac{2}{\delta N}
 \exp\!\left(
 -\frac{\delta N}{2\sqrt\beta}
 \dist_{\T^2}(\Sigma_{w/4},E)
 \right)\norm{g}_2
 \le\frac{8\sqrt\beta}{\delta Nw}
 \exp\!\left(-\frac{\delta Nw}{8\sqrt\beta}\right).
 \end{aligned}
\]
This establishes the local mass estimate in \eqref{eq:mass}.
\end{proof}

\begin{remark} \label{rmk:scaling_covariance}
The scaling property \eqref{eq:seed-fiber-dilation} is essential to our thin-layer construction, which enables us to insert increasingly high-frequency gaps by rescaling the spatial pattern of a fixed seed medium. Importantly, the change of length scale preserves the material bounds:
\[
 \alpha I_2\leq S_N(x)\leq\beta I_2,
\]
which is fundamental to construct a limiting operator. This is in sharp contrast with the case of Schrödinger operators, in which BSC is proved for general potentials \cite{kuchment2016overview}. We now elaborate in more detail on this distinction. Under the unitary dilation $y=Nx$, the second-order operator under consideration satisfies
\[
-\diver_x\bigl(S(Nx)\nabla_x\bigr)
\simeq N^2\bigl[-\diver_y\bigl(S(y)\nabla_y\bigr)\bigr].
\]
Thus, a prescribed spectral gap $(A^2,B^2)$ of $H_{S}$ scales to be a gap $(N^2A^2,N^2B^2)$ of $H_{S_{N}}$, during which the material coefficient matrix of the scaled operator stays in the class $\mathcal X^{\mathrm H}_{\alpha,\beta}$. On the contrary, for a Schr\"odinger operator, spatial rescaling with unchanged
potential amplitude instead gives
\[
-\Delta_x+V(Nx)
\simeq N^2\bigl[-\Delta_y+N^{-2}V(y)\bigr].
\]
The operator inside the brackets is therefore not the original seed. To transport its spectrum exactly by the factor $N^2$, one must use instead
\[
 -\Delta_x+N^2V(Nx) \simeq N^2[-\Delta_y+V(y)],
 \quad \text{with }\, \norm{N^2V(N\cdot)}_\infty=N^2\norm{V}_\infty.
\]
This choice of seed potential has unbounded amplitudes as $N\to\infty$, and consequently, it is impossible to construct a limiting operator using our thin-layer induction.
\end{remark}

\begin{proof}[Proof of Lemma~\ref{lem:layer-motion}]
{\color{blue}Step 1.} \emph{Localizing the phase dependence.} We localize the phase dependence in $\Sigma_{w/4}$ so that the local mass estimate can be used to control eigenfrequency motion. For $h\in\R^2$, let $ b_{w,h}(x):= \bigl(h_1q_w(x_1),h_2q_w(x_2)\bigr)^{\mathsf T}$. The product rule and $f_w'=q_w-1$ give
\[
 \widetilde M_C(k)
 =
 \begin{pmatrix}
  0&\widetilde B_C(k)^*\\
  \widetilde B_C(k)&0
 \end{pmatrix},
 \qquad
 \widetilde B_C(k):=C^{1/2}(\nabla+\ii b_{w,k}).
\]
Differentiating in the direction $h$, we obtain
\[
 \widetilde M_C'(k)[h]
 =
 \begin{pmatrix}
  0&-\ii b_{w,h}^{\mathsf T}C^{1/2}\\
  \ii C^{1/2}b_{w,h}&0
 \end{pmatrix}.
\]
This expression is independent of $k$, and
\[
 \widetilde M_C(k)
 =M_C(0)+\widetilde M_C'(0)[k],
\]
so the family is affine. By the definition of $q_w$,
\[
 b_{w,h}=0
 \quad\text{a.e. on }\T^2\setminus\Sigma_{w/4},
 \qquad
 \norm{b_{w,h}}_{L^\infty}\le\frac{2|h|}{w}.
\]
Consequently,
\[
 \widetilde M_C'(k)[h]
 =
 \one_{\Sigma_{w/4}}
 \bigl(\widetilde M_C'(k)[h]\bigr)
 \one_{\Sigma_{w/4}},
\]
and $C\le\beta I_2$ gives
\[
 \norm{\widetilde M_C'(k)[h]}
 =
 \norm{C^{1/2}b_{w,h}}_{L^\infty}
 \le\sqrt\beta\,\norm{b_{w,h}}_{L^\infty}
 \le\frac{2\sqrt\beta\,|h|}{w}.
\]
This proves the derivative estimate \eqref{eq:gauge}.


{\color{blue}Step 2.} \emph{Controlling eigenfrequency motion.} Let $D:=D^S_{N,w}[C]$ and fix $k\in\mathcal B$. Along $t\mapsto tk$,
\[
 \frac{\mathrm d}{\mathrm dt}\widetilde M_D(tk)
 =\widetilde M_D'(tk)[k]
 =\widetilde M_D'(0)[k].
\]
Since $U_w(tk)$ is unitary and commutes with $\one_{\Sigma_{w/4}}$, estimate \eqref{eq:mass} gives
\[
 \norm{\one_{\Sigma_{w/4}}\widetilde F}_2
 \le
 \frac{8\sqrt\beta}{\delta Nw}
 \exp\!\left(-\frac{\delta Nw}{8\sqrt\beta}\right)
 =:\varepsilon_{N,w}
\]
for every normalized eigenvector $\widetilde F$ of
$\widetilde M_D(tk)$ with eigenfrequency in $N\overline{G_0}$.

By Lemma~\ref{lem:layer-block} and compactness of the resolvent of $H_D(tk)$, the positive eigenfrequencies are isolated and have finite multiplicity. Along a simple eigenfrequency branch, the Hellmann--Feynman formula \cite[Chapter~VII, \S3.4]{Kato1995} and
\eqref{eq:gauge} yield
\[
 \begin{aligned}
 |\omega'(t)|
 =
 \left|
 \left\langle
 \widetilde M_D'(tk)[k]\widetilde F(t),
 \widetilde F(t)
 \right\rangle
 \right|
 \le
 \frac{2\sqrt\beta\,|k|}{w}
 \norm{\one_{\Sigma_{w/4}}\widetilde F(t)}_2^2
 \le
 \frac{2\pi\sqrt{2\beta}}{w}\,\varepsilon_{N,w}^2
 =v_{N,w},
 \end{aligned}
\]
where $|k|\le\pi\sqrt2$. At a multiple eigenfrequency, let $P$ be the orthogonal projection onto its eigenspace. Finite-dimensional spectral reduction identifies the branch derivatives with the eigenvalues of
\[
 \left.P\widetilde M_D'(tk)[k]P\right|_{\operatorname{Ran}P};
\]
see \cite[Chapter~II, Theorem~5.4]{Kato1995}. Since the same mass estimate holds for every unit norm vector in $\operatorname{Ran}P$, self-adjointness gives
\[
 \begin{aligned}
 \norm{P\widetilde M_D'(tk)[k]P}
 =
 \sup_{\substack{\widetilde F\in\operatorname{Ran}P
                 \norm{\widetilde F}_2=1}}
 \left|
 \left\langle
 \widetilde M_D'(tk)[k]\widetilde F,\widetilde F
 \right\rangle
 \right|
 \le
 \frac{2\sqrt\beta\,|k|}{w}\,\varepsilon_{N,w}^2
 \le v_{N,w}.
 \end{aligned}
\]
Thus, \eqref{eq:velocity} holds for every local real-analytic eigenfrequency branch, including at crossings.

Finally, let $\omega_*$ satisfy the stated margin condition, and take a local analytic eigenfrequency branch with $\omega(1)=\omega_*$. Continue this branch to the left while it remains in $NG_0$, and let $t_0\in[0,1)$ be the left endpoint of its maximal continuation within $[0,1]$. For $t\in(t_0,1]$,
\[
 |\omega(t)-\omega_*|
 \le\int_t^1|\omega'(s)|\,\mathrm ds
 \le(1-t)v_{N,w}.
\]
Hence the branch remains in the compact interval
\[
 [\omega_*-v_{N,w},\,\omega_*+v_{N,w}]
 \Subset NG_0.
\]
The velocity bound makes $\omega$ Lipschitz on $(t_0,1]$, so
\[
 \omega_0:=\lim_{t\downarrow t_0}\omega(t)
 \in[\omega_*-v_{N,w},\,\omega_*+v_{N,w}].
\]
The spectral theorem and affine dependence give
\[
 \begin{aligned}
 \dist\bigl(
 \omega(t),\sigma(\widetilde M_D(t_0k))
 \bigr)
 \le
 \norm{\widetilde M_D(tk)-\widetilde M_D(t_0k)}
 =
 |t-t_0|\,\norm{\widetilde M_D'(0)[k]}
 \longrightarrow0.
 \end{aligned}
\]
Since $\omega(t)\to\omega_0$ and the spectrum is closed, $\omega_0\in\sigma(\widetilde M_D(t_0k))$. By Lemma~\ref{lem:layer-block} and compactness of the resolvent of $H_D(t_0k)$, this positive spectral point is isolated and has finite multiplicity. Local analytic spectral reduction therefore extends the branch through $t_0$, with values still in $NG_0$ near $t_0$; see \cite[Chapter~VII, \S3.1--\S3.2]{Kato1995}.
If $t_0>0$, this contradicts maximality. Thus $t_0=0$, and the branch extends analytically to this endpoint. The displacement bound consequently holds for all $t\in[0,1]$.

\end{proof}


\bigskip
\noindent\textbf{Conflict of interest.} The authors declare no conflict of interest.

\bigskip
\noindent\textbf{Data and code availability.} No experimental or observational data were generated. 

\bigskip
\noindent\textbf{AI usage.} During the preparation of this manuscript, the authors used ChatGPT Pro (GPT-6 Astra) to assist in exploring and critically examining mathematical arguments and literature, especially the proof of Proposition \ref{prop:uniform-seeds}. The authors reviewed and revised all AI-assisted material. The authors assume full responsibility for the content of the manuscript.

\bibliographystyle{amsplain}
\bibliography{ref}

\bigskip
\begingroup
\small
\raggedright

\noindent\textsc{Habib Ammari}\\
Department of Mathematics, ETH Z\"urich,
R\"amistrasse 101, 8092 Z\"urich, Switzerland\\
Email: \texttt{habib.ammari@math.ethz.ch}

\medskip
\noindent\textsc{Yu Gao}\\
Department of Mathematics,
The Hong Kong University of Science and Technology,
Clear Water Bay, Kowloon, Hong Kong\\
Email: \texttt{mayugao@ust.hk}

\medskip
\noindent\textsc{Jiayu Qiu} (corresponding author)\\
Department of Mathematics, ETH Z\"urich,
R\"amistrasse 101, 8092 Z\"urich, Switzerland\\
Email: \texttt{jiayu.qiu@math.ethz.ch}

\par
\endgroup

\end{document}